\documentclass[english]{iopart}

\usepackage{graphicx}
\usepackage{amssymb}
\newcounter{fig}
\begin{document}

\title[]{\Large On the diagonals of rational functions: the minimal number of variables (unabridged version)}

\author{S. Hassani$^\ddag$, J.M. Maillard$^\pounds$, N. Zenine$^\S$  
}
\address{$^\ddag$  Retired: Centre de Recherche Nucl\'eaire d'Alger, Algeria}
\address{$^\pounds$ LPTMC, UMR 7600 CNRS, 
Universit\'e de Paris, Tour 23,  5\`eme \'etage, case 121, 
 4 Place Jussieu, 75252 Paris Cedex 05, France} 
\address{\S  Centre de Recherche Nucl\'eaire d'Alger, 
2 Bd. Frantz Fanon, BP 399, 16000 Alger, Algeria}

\begin{abstract}

From some observations on the linear differential operators occurring in the Lattice Green function of the $d$-dimensional  
face centred and simple cubic lattices, and on the linear differential operators occurring in the $n$-particle contributions
$\chi^{(n)}$  to the magnetic susceptibility of the Ising model, we forward some conjectures on the diagonals of rational 
functions. These conjectures are also in agreement with exact results we obtain  for many Calabi-Yau operators, and many 
other examples related, or not related to physics. 

Consider a globally bounded power series  which is the diagonal of  rational functions of a certain number of variables,
annihilated by an irreducible minimal order linear differential operator homomorphic to its adjoint.
Among the logarithmic formal series solutions, at the origin, of this operator, call $\,n$ 
the highest power of the logarithm.
We conjecture that this diagonal series can be represented as a diagonal of a rational function
with a minimal number of variables 
$ \, N_v$ related to this highest power  $\, n$ by the relation $\, N_v = \, n +2$.

Since the operator is homomorphic to its adjoint, its differential Galois group is
symplectic or orthogonal.
We also conjecture that the symplectic or orthogonal character of the differential Galois group is related to 
the parity of the highest power $\,n$, namely  symplectic for  $\,n$ odd and orthogonal for  $\,n$ even.

We also sketch the case where the denominator of the rational function is not irreducible and is
the product of, for instance, two polynomials.
We recall that the linear differential operators occurring in the $n$-particle contributions
$\, \chi^{(n)}$  to the magnetic susceptibility of the square Ising model factorize in a large number
of direct sums and products of factors.
The analysis of the  linear differential operators annihilating the diagonal of rational function
where the denominator is the product of two polynomials, sheds some 
light on the emergence of such mixture of direct sums and products of factors.
The conjecture  $\, N_v = \, n +2$ still holds for such reducible linear differential operators.

\vskip .4cm



\vskip .3cm


\end{abstract} 

\today

\section{Introduction}
\label{Introd}

Diagonals of rational functions have been seen to occur
naturally~\cite{2013-rationality-integrality-ising} for $\, n$-fold integrals in
physics, field theory, enumerative combinatorics, etc. 
On many examples and cases, striking properties emerged that are worthy to be
understood. 

When we seek for a characterization of the diagonal of  rational functions
representation of a D-finite {\em globally bounded} power series\footnote[1]{A diagonal
of a rational function is  necessarily~\cite{2013-rationality-integrality-ising}
a D-finite {\em globally bounded} power series. Conversely, according to
Christol's conjecture~\cite{Chrisconj},
a D-finite {\em globally bounded} power series
should be the diagonal of a rational function.}, one may 
think of the {\em least number of variables}  $ \, N_{v}$
occurring in the rational function\footnote[2]{It is obvious, however,
that the rational function, whose diagonal gives  a given {\em globally bounded} series,
is far from unique. This series can be 
the diagonal of many rational functions, even with  different numbers of variables.}.

Given a diagonal of a rational function $\,R_1$ of $\, N$ variables
(which is necessarily~\cite{2013-rationality-integrality-ising}  a D-finite {\em globally bounded}
power series)
 there is a rational function $\,R$ with a
{\em minimal number} $ \, N_{v}$ of variables. One aim of this paper 
amounts to showing that this number of variables $ \, N_{v} \le N $
is simply related to the logarithmic singular behavior, at the origin,
of the formal series of the (minimal order) linear differential operator
annihilating the diagonal.

We illustrate our conjecture with the analysis of 
the lattice Green function (LGF) of the $\, d$-dimensional simple cubic (s.c) and  face centred cubic (f.c.c)
lattices~\cite{joyce-1998,guttmann-2009,guttmann-2010,broadhurst-2009,koutschan-2013,2015-LGF-fcc7,2016-LGF-fcc8to12},
as well as results obtained for many Calabi-Yau operators~\cite{TablesCalabi-2010}, and
an accumulation of other examples related, or not related, to physics,  displayed (or not displayed) in this paper.

The linear differential operators for LGF of the simple cubic (s.c) and face centred cubic (f.c.c)
lattices are {\em irreducible}. These linear differential operators  are
{\em homomorphic to their adjoint} and, consequently, their differential Galois groups
are (included in) symplectic or orthogonal groups. 
All these lattice Green functions can, obviously, be cast into diagonal of rational functions.

This irreducibility is in sharp contrast with the linear differential operators of 
the $n$-particle contributions $\chi^{(n)}$\footnote[5]{The $\, \chi^{(n)}$ have a very
 convoluted form of algebraic fractions as integrands.
They are shown to be diagonal of rational  functions~\cite{2013-rationality-integrality-ising}.}
 to the magnetic susceptibility of the
Ising model~\cite{wu-mc-tr-ba-76,2004-chi3,2005-chi4,2008-experimental-mathematics-chi,2009-chi5,2010-chi5-exact,2010-chi6},
which  have a large set of factors.
Here, the symplectic, or orthogonal, character of the differential Galois group concerns the {\em factors}
occurring in the factorization of the linear differential operators annihiliting the $ \, \chi^{(n)}$'s.
Furthermore, we observe, for the $\, n$-particle contributions $ \, \chi^{(n)}$
of the susceptibility of the square Ising model, that, for each block of factors in the linear differential operator
which has a unique factorization, (e.g. for a block of three factors $ \, L_n \cdot \,  L_p \cdot \, L_q \,  $),
we have {\em alternately} orthogonal and symplectic groups.

\vskip 0.1cm

We will show that these characteristics can be seen on the diagonals of rational functions,
with simple enough expressions, that may lead to a better understanding of their occurrences.

\vskip 0.1cm

With $\, P$ and $\, Q$ multivariate polynomials (with $\, Q(0, ..., 0) \ne \, 0$), the formal series 
of the (minimal order) linear differential operator annihilating
the diagonal of the rational functions $\,P/Q^r$ (with $ \, r$ an integer),
correspond to a {\em finite dimensional} vectorial space 
related\footnote[9]{We are thankful to  P. Lairez for having clarified this point.},
as shown by Christol~\cite{christol-83,christol-85,christol-EMS}, to the de Rham cohomology. 
There is a {\em homomorphism} between the (minimal order) linear differential operators for the 
diagonal of $\, P/Q^k$ and  for the  diagonal of $\, 1/Q$. 
Therefore, without too much loss of generality (see section 3 of~\cite{Plea}),
we will basically restrict ourselves, in this paper, to rational functions in the form $\,R = \, 1/Q$, 
where $\, Q$ is an irreducible multivariate polynomial,  with $ \, Q(0, ..., 0) \ne \, 0$. 
We will also  consider, for pedagogical reason, the case where the denominator $\, Q$
factors in only {\em two} polynomials, $\, Q = \, 1/Q_1/Q_2$, 
where at least one of the $ \, Q_j$'s depends on {\em all} the
variables\footnote[5]{The case where $ \, Q_1$ and $ \, Q_2$
depend on  different sets of variables corresponds to
a Hadamard product and will not be considered in this paper.}.

\vskip 0.1cm

In the sequel, we notice that, generically, for irreducible $\, Q$ 
over the rationals, the resulting (minimal order)  linear differential operator, 
annihilating the diagonal of $\, 1/Q$, seems to be systematically irreducible\footnote[1]{For
 examples of irreducible $\, Q$ and non irreducible operators see \ref{split}.}. In contrast,
a factorization of the linear  differential operator occurs for factorizable $\, Q$.

\vskip 0.1cm

\subsection{The formal solutions}
\label{formsolu}

The linear differential operators annihilating  diagonals of  rational functions are very selected
linear differential operators~\cite{Plea}. The formal solutions, at the origin,
of such linear differential operators (call it $L_q$) can be 
organized as the union of different sets (\ref{theS01}), (\ref{theS02}), ...
of formal series which makes the monodromy at $\, t \, = \, 0$ crystal clear:
\begin{eqnarray}
\label{theS01}
&&  \hspace{-0.98in}  \quad \quad  \,\,
S_0,  \quad \quad   \,\,
   \,\,  S_0 \cdot \, \ln(t) \,\,  + S_{1,0},  \quad  \quad   \,\,
  \,  S_0 \cdot \,  { \ln(t)^2 \over 2!}  \,\,  +S_{1,0}\cdot \, \ln(t) \, \,  + S_{2,0},  
 \nonumber \\
 &&  \hspace{-0.98in}  \quad  \quad  \,\, \cdots,
 \nonumber \\
&&  \hspace{-0.98in}  \quad  \quad  \,\,
S_0 \cdot \,   {\ln(t)^n \over n!}  \,\,\,  +S_{1,0} \cdot \, {\ln(t)^{n-1} \over (n-1)!} 
 \,\,\, + S_{2,0} \cdot \, {\ln(t)^{n-2} \over (n-2)!} \,  \,\,\, + \, \cdots \, \,   + S_{n,0}, 
\end{eqnarray}
\begin{eqnarray}
\label{theS02}
&&  \hspace{-0.98in}  \quad \quad  \,\,
T_0,  \quad \quad  \,\, 
   \,\,  T_0 \cdot \, \ln(t) \,\,  + T_{1,0},  \quad  \quad   \,\,
  \,  T_0 \cdot \,  { \ln(t)^2 \over 2!}  \,\,  +T_{1,0}\cdot, \, \ln(t) \, \,  + T_{2,0},  
 \nonumber \\
 &&  \hspace{-0.98in}  \quad  \quad  \,\, \cdots,
 \nonumber \\
  &&  \hspace{-0.98in}  \quad  \quad  \,\,
T_0 \cdot \,   {\ln(t)^m \over m!}  \,\,\,  +T_{1,0} \cdot \, {\ln(t)^{m-1} \over (m-1)!} 
 \,\,\, + T_{2,0} \cdot \, {\ln(t)^{m-2} \over (m-2)!} \,  \,\,\, + \, \cdots \, \,   + T_{m,0},  
\end{eqnarray}
etc. \

In each of these ``sets'' of formal solutions, the series (up to a $\, t^{\alpha}$ overall,
$\, \alpha \, = \, 0, \, 1/2, ...$) are analytical at $\,t=\,0$.
One of the  series $\, S_0$, $\, T_0$, ... is the diagonal of the rational function
annihilated by the (minimal order) linear differential operator $\, L_q$. It is therefore
a globally bounded series. 
The other analytic solution series have, at first sight, no reason to be globally bounded series. 

\vskip 0.1cm

\subsection{Conjecture on the number of variables}
\label{conj1}

Our main conjecture corresponds to the {\em exact value} of the {\em minimal number } $\, N_v$
of variables of the {\em rational} functions
required to represent
a given diagonal of rational function series.
Consider a rational function $\,R_1$ with $\,N$ variables, and the (minimal order)
linear differential operator annihilating
$\,{\rm Diag} (R_1)\,$ (i.e. the diagonal of $\, R_1 \, $),
{\em assumed to be homomorphic to its adjoint}\footnote[2]{It has been noticed
in several of our papers~\cite{Plea, 2014-Ising-SG}, that diagonal of rational (or algebraic) functions
almost systematically yield linear differential operators which are {\em homomorphic to their adjoint}.
The rare cases breaking this ``self-adjoint duality'' were seen to correspond to
candidates to rule out Christol's conjecture~\cite{Chrisconj,christol-EMS}. The question of this
self-adjoint symmetry breaking is adressed in section 5.2 of~\cite{Plea}.}
(i.e. its differential Galois group is included either in a symplectic,
or in an orthogonal group).
Among the formal solutions, at the origin, of this linear differential operator,
there is one formal solution which has the {\em highest} log-power, $\, n$,
i.e. behaves as $\, \ln(t)^n$.
We conjecture that
the diagonal of  $\, R_1 \, $ identifies with the diagonal of a
rational function $\, R$ which  depends  on
 a minimal number of variables
 $ \, N_v$, where $ \, N_v$  is simply related
 to the {\em highest} log exponent\footnote[5]{The crucial role played by
  the highest log exponent corresponds to the concept of monodromy filtration (see paragraph 4.2
  page 40 of~\cite{christol-EMS}).} 
$\, n$ by the following relation\footnote[8]{We do not have  a conjecture
 for the minimum number of variables $\, N_v$, for diagonals of {\em algebraic} functions. 
An open question is to see if we could actually have
$\,  N_v = \,n\, +1\, $ in the diagonal of {\em algebraic} functions case.}
 (with $ \, N_v \, \le \, N$): 
\begin{eqnarray}
\label{conjmain}
\quad \quad \quad \quad \quad \, 
N_v\, \, = \,\,\, n \, +2. 
\end{eqnarray}

\vskip 0.1cm

Note that the existence of such rational functions with a minimal
number of variables $\, N_v$  ({\em regardless of the number of monomials} and degrees),
{\em does not prevent} the existence of other rational functions with more variables, giving
the same diagonal. For instance, when the general term of the series writes as nested sums
of binomials~\cite{2013-rationality-integrality-ising,lairez-2014},
it is straightforward to obtain a first rational
function~\cite{2013-rationality-integrality-ising}. This first rational function
has often more variables than this minimal number $\, N_v$.
Also note that a balanced ratio of factorials coefficients can be written
in various binomial forms, thus yielding many rational functions.

\vskip 0.1cm

\subsection{Some remarks}
\label{someRq}
Let us give some remarks on the definition of what we call the rational functions
and on the number of variables occurring there.

 We have seen in previous papers (see for instance section 4.3 and appendix G in ~\cite{Diag2F1}),
that an exact result on the diagonal of a rational function of some variables $\, x$, $\, y$, ..., 
depending on {\em parameters},
can straightforwardly be generalised to the same rational function, but where the {\em parameters}
become arbitrary (rational or algebraic)  {\em functions}
of the product $\, t \, = \, x \, y \, z \, \cdots$
of the variables of the rational function. 
Consequently we extend the definition of rational functions to rational functions where
the variables are rescaled\footnote[1]{See for instance equations (\ref{agreement2})
and (\ref{wherealpha}) below.} by arbitrary (rational or algebraic) 
functions of the product $\, t \, = \, x \, y \, z \, \cdots$

 Let us, now, consider rational functions of some variables $\, x$, $\, y$, ... {\em but also}
$\, N$-th root of other variables ($u^{1/4}, \, v^{1/6}$, ...). The calculation of the diagonal of such
a function, is in fact equivalent to the  calculation of the diagonal of this function
where some $\, M$-th power of these variables and $\, N$-th root of variables, has been taken,
so that we do not have $\, N$-th root of variables any longer. Some examples are given below
(see for instance, the many examples in Table \ref{Ta:1}). 
In the following we will say, by abuse of language, in such cases, that we have a rational function
{\em even if it contains}
$\, N$-th {\em root of some variables} ($u^{1/4}, \, v^{1/6}$, $\, \cdots$). 

 Finally,  we have seen examples of rational functions where some monomials or variables in the rational function
do not contribute to the diagonal, meaning that the rational function has in fact lesser number of variables. 
We used, for this situation, the term ``effective'' number of variables in section 2.4 of~\cite{Plea}. 
Consider the rational function $\, 1/Q_1$ where $ \, Q_1 = \, 1 \, -x^2-y \, -x\,z\, $ depends on three variables. 
The diagonal of $1/Q_1\, $ is "blind"\footnote{Changing the monomial $\,x^2\,$ into $\, \mu \,x^2$, the resulting diagonal
will not depend on the parameter $\,\mu$.} on the monomial $\,x^2$. Furthermore, once the monomial $\,x^2$ is not taken into
account, the variables $\,x$ and $\,z$ in the monomial $\,x \,z$ stands for only one variable. As far as
the diagonal of the rational function is concerned, the polynomial $\,Q_1$ is in fact equivalent to
the polynomial $ \, {\tilde Q}_1 = \, 1 \, -x-y$ depending on two variables.
As a second example, let us consider the polynomial $\, Q_2= \, 1 \,-y-z-x\,z - x \,u -x\,z\,u- x\,y\,u$,
which depends on four variables. The diagonal of $\,1/Q_2\, $
is blind on the monomial $\,x\,z$ and the product $\,x\,u$ stands for one variable, reducing $\, Q_2$ to
$\, {\tilde Q}_2 =\,  1 \, -y-z- v \,  - y\,v \, -z\,v$ which depends only on three variables.
We should note that we have not seen this situation in our examples coming from physics or geometry. 

\vskip 0.1cm
         
\subsection{Complements and additional speculations on conjecture (\ref{conjmain})}
\label{cavetas}

Among the different sets (\ref{theS01}), (\ref{theS02}), ... of formal series solutions
of the (minimal order) linear differential operator, one set, for instance (\ref{theS01}),
corresponds to the highest power of the logarithms, namely $\, n$. We also conjecture that
the series with no logarithm in this set, here $\, S_0$, {\em will necessarily be
a globally bounded series}, and, thus, according to Christol's conjecture~\cite{Chrisconj},
will be a diagonal of a rational function, with, according to the conjecture (\ref{conjmain}),
the minimal number of variables $\, N_v \, = \, n +2$.
The other sets, for instance (\ref{theS02}), will, in general, have a power of  logarithms
$\, m$ which is less than the highest power $\, n$: $\, m \, < \, n$. In this $\, m \, < \, n$ case 
we also conjecture that the power series  (up to a $\, t^{\alpha}\, $
overall, $\, \alpha \, = \, 0, \, 1/2, \, \cdots$)
with no logarithm in this set, here $\, T_0$ for (\ref{theS02}),
{\em cannot be a  globally bounded series}.
Consequently, {\em it cannot be a diagonal of a rational function}.

In the rare cases\footnote[9]{See for instance equations (\ref{un}) and (\ref{deux}) below.}
where another set (\ref{theS02}), is such that its power of  logarithms, $\, m$,
is actually equal to the highest
power of logarithm $\, n$,  we  conjecture that the power series
(up to a $\, t^{\alpha}$ overall, $\, \alpha \, = \, 0, \, 1/2, \, \cdots $)
with no logarithm in this set, here $\, T_0$,  {\em will necessarily also be a globally bounded series},
and, thus, according to Christol's conjecture~\cite{Chrisconj}, 
will be a diagonal of a rational function, with, according to the conjecture
(\ref{conjmain}),  the minimal number of  variables $\, N_v \, = \, n+2$.

Let us recall that the definition of diagonal of rational functions is based on multi-Taylor
expansions around the origin~\cite{2013-rationality-integrality-ising}. Therefore, 
the seeking of the maximum exponent of the 
logarithm in the formal solutions {\em is also made at the origin}. The maximum exponent of the logarithms
in the formal solutions around the other singularities can also be considered, and this is checked
for many of our examples below.

\vskip 0.1cm

\subsection{Conjecture on the parity of the number of variables and the differential Galois groups}
\label{conj2}

Another conjecture is related to the symplectic, or  orthogonal,
character of the differential Galois
group\footnote[1]{An irreducible linear differential
  operator $\, L_q$, of order $\, q$,  has, generically, a symmetric square ($sym^2(L_q)$) of 
order $ \,N_s =\, q \,(q+1)/2$ and an exterior square ($ext^2(L_q)$) of order $\,N_e = \, q \, (q-1)/2$.
If $\, sym^2(L_q)$ (resp. $\, ext^2(L_q)$) annihilates a rational solution, or is of order $\, N_s\, -1$
(resp. $\, N_e\, -1$), the linear  differential operator  $\, L_q$ is included in  
the {\em orthogonal group} $\, SO(q, \,\mathbb{C})$ (resp.  {\em symplectic group} $\, Sp(q, \,\mathbb{C})$) 
that admits an invariant quadratic (resp. alternating) form.}
of the irreducible linear differential operator annihilating the 
diagonal of the rational function with $\,N_v$ variables.

\vskip 0.1cm

Since we assume that the  (minimal order) irreducible linear differential operator 
annihilating the diagonal of the rational function
is {\em homomorphic to its adjoint}, its  differential Galois group is necessarily
symplectic or orthogonal~\cite{2014-Ising-SG,2014-DiffAlg-LGFCY}.

We forward a conjecture stating that the parity of the "minimal" number $ \, N_v$ of variables 
in the irreducible denominator $\, Q$ dictates the character either symplectic $ \,Sp$ ($N_v$ is an odd number) 
or orthogonal $ \, SO$ ($N_v$ is an even number) of the differential Galois group of the (minimal order) 
linear differential operator annihilating  the diagonal of the rational function.
This conjecture is that the group is orthogonal for $\, N_v\,$ even and symplectic for $\, N_v\,$ odd: 
\begin{eqnarray}
\label{conjadditio}
\quad
 N_v\,\, {\rm even}\,\, ({\rm resp. \,\, odd})\,\,\,\, \quad  \longrightarrow \,\,\,\, \quad  \quad 
    SO\, ( {\rm resp.}\,\, Sp) 
\end{eqnarray}

\vskip 0.1cm

The illustrative examples displayed in this paper in favour of these two conjectures
(\ref{conjmain}), (\ref{conjadditio}),
are chosen for pedagogical reasons, but also for their interest {\em per se}. 
We have two different kinds of examples: the ones corresponding to (irreducible)
linear differential operators which have
{\em Maximum Unipotent Monodromy}\footnote[5]{Maximum unipotent monodromy: the critical
  exponents at the origin are all equal.} (MUM),
and the ones that do not have MUM.
In the MUM case, we have a simple relation $\, n\, = \, q\, -1$
between the power of logarithm $\, n$ in (\ref{conjmain}),
and the order $\, q$ of the linear differential operator $\, L_q$. The conjecture
(\ref{conjmain}) thus becomes, in the MUM case,  a simple
relation $\, N_v \, = \, q\, +1\, $ between the number of variables and the order of the operator.
For the non-MUM examples, the number of variables
is {\em not} related to the order of the linear differential operators,  but to the highest power
of the logarithms $\, n$, in other words, {\em it is related to the monodromy matrix at the origin}.

\vskip 0.1cm

The paper is organized as follows.
We recall in section \ref{formalsolexpan} the definition of
the diagonal of a rational function, and in section \ref{LGF}
we recall the results of the Lattice Green Functions of the $d$-dimentional
(face centred cubic, simple cubic and diamond)
lattices, to illustrate  the  two conjectures (\ref{conjmain}), (\ref{conjadditio}).
Section \ref{irreduc} presents some illustrative examples for non
factorizable multivariate polynomials $ \, Q$. 
We give, in section \ref{CalabiYau}, the polynomials $ \, Q \, $
for some Calabi-Yau equations~\cite{TablesCalabi-2010} 
of geometric origin, obtaining, for each case, the polynomial $\, Q$ with the minimum
number of variables $\,N_v \, =\, 3 \, +2 \, =\, 5$.
Section \ref{facto} deals with the situation where the denominator polynomial $\, Q \, $
factorizes as $ \, Q = \,  Q_1 Q_2$,
giving either a direct sum, or a unique factorization, of the (minimal order)
linear differential operator
annihilating the diagonal of $ \, 1/Q$. 
In section \ref{powers},  we give examples of diagonal of  algebraic functions which are $\, N$-th root
of rational functions and equivalent rational functions of the form $ \, 1/Q$,
giving the same diagonals. Finally, in section \ref{assumption},
we discuss the homomorphisms-to-adjoint 
assumption on a  counter-example  candidate to Christol's conjecture~\cite{Chrisconj}. 

\section{Diagonals and multinomial expansion}
\label{formalsolexpan}

The diagonal of a rational function $\,R$, dependent on (for example) three variables,
is obtained by (multi-Taylor) expanding $\,R$  around the origin
\begin{eqnarray}
  \label{defdiag}
&&  \hspace{-0.98in}  \quad  \quad  \quad \quad   \quad  \quad  \quad 
  R(x, y, z) \, \, \, = \, \, \,\,
  \sum_{m}  \sum_{n}  \sum_{l} \, a_{m,n,l} \cdot \, x^m \, y^n \, z^l, 
\end{eqnarray}
keeping only the terms such that $\, m = n = l$. The diagonal reads, with $ \, t = \,  xyz$: 
\begin{eqnarray}
  \label{defdiag2}
&&  \hspace{-0.98in}  \quad  \quad  \quad \quad  \quad  \quad  \quad 
 {\rm  Diag}\Bigl(R(x, y, z)\Bigr)  \, \, \, = \, \, \,   \,   \sum_{m} \, a_{m,m,m} \cdot \, t^m. 
\end{eqnarray}

In most of the examples of this paper the rational function will have the form $\, 1/Q\, $ with 
\begin{eqnarray}
  \quad   \quad   \quad   
  Q  \, \,\,  = \, \, \, \, \,
  1 \, \, \,  \, - \left( T_1 \, + T_2 \, \, + \cdots \, +T_n \right), 
\end{eqnarray}
where the $\, T_j$'s are monomials. The (multinomial) expansion of the rational function $1/Q\, $ reads:
\begin{eqnarray}
&&  \hspace{-0.98in}  \quad  \quad  \quad \quad  \quad  \quad 
{{1} \over {Q}} \,  \, = \, \,\,\,
\sum \,\,   {\frac{(k_1 +k_2 + \, \cdots \, + k_n)!}{k_1! \,  k_2! \, \cdots \, k_n!}}
\cdot  \, T_1^{k_1} \,T_2^{k_2} \,\cdots \, T_n^{k_n}. 
\end{eqnarray}
Calculating the diagonal of $\,1/Q\, $ amounts to distributing the powers $\, k_j$
on the variables occurring in the monomials, 
and putting equal the exponents of each variable.

\vskip 0.1cm

Consider the polynomial $\,\, Q_1\,=  \, 1 \, \, -x\, -y \,z\, -z^2\,+x\,y\,\,$ depending on
three variables. The multi-Taylor expansion, around the origin,
of the rational function $\,1/Q_1\, $  reads:
\begin{eqnarray}
  \label{87expli1}
&&  \hspace{-0.99in}    
 {{1} \over { Q_1}} \, \, = \, \,\,
\sum_{k_1=0}^{\infty} \, \sum_{k_2=0}^{\infty} \,
     \sum_{k_3=0}^{\infty} \, \sum_{k_4=0}^{\infty} \,
     {{ (k_1\, +k_2\, +k_3\,+k_4)! } \over { k_1! \, k_2! \, k_3!\,k_4! }}
     \cdot \, x^{k_1} \cdot \, (y\,z)^{k_2} \cdot \, z^{2\, k_3} \cdot \, (-x\,y)^{ k_4}. 
\end{eqnarray}
The diagonal will extract the terms with, only, the same power $\,p$ for $\, x$, $\, y$ and $\, z$,
i.e. $\, k_1+k_4 \, \, =\,  k_2+k_4 \, = \, k_2+2\,k_3 \, \,=\,
\,p$, namely (with $\, t \, = x \, y\, z\, $ and $\, p \, -2\,k_3 \ge \, 0$ ):
\begin{eqnarray}
\label{87expli}
  &&  \hspace{-0.98in}  \quad     \quad \, \, \, \,   \quad \quad  \quad       
{\rm Diag} \left( {{1} \over { Q_1}} \right)  \,  \, = \, \,  \, \, 
\, \sum_{p=0}^{\infty} \,
 \Bigl( \sum_{k_3=0} \, {{ (2\, p\, -k_3)! } \over { k_3! \,(2\, k_3)!  \, (p-2\,k_3)!^2 }} \Bigr) \cdot \, t^{ p}.
\end{eqnarray}

\vskip 0.1cm

With the example\footnote[5]{For "polynomials" with $\, N$-th root of variables,
  see, for instance, (\ref{u1sur2v1sur5})  in section \ref{CalabiYau}. }
$\,\, Q_2\,=\,1 \, \, -x\, -y \, -z \, -u^{1/2} \, \, -v^{1/5}$,
depending on five variables, or $\, n$-th root of variables, the expansion  of $\,1/Q_2$,
around the origin, reads:
\begin{eqnarray}
  \label{87expli2}
&&  \hspace{-0.98in}  \quad   
     {{1} \over { Q_2}}   
\,  \, \, = \, \,\,
   \sum_{k_i=0, \, i \, =1, \cdots \, 5}^{\infty} \,
 {{ (k_1\, +k_2\, +k_3\, +k_4\, +k_5)! } \over { k_1! \, k_2! \, k_3! \, k_4! \, k_5! }}
 \cdot \, x^{k_1} \,  \, y^{k_2} \,\,  z^{2 \, k_3} \, \, u^{ k_4/2}\,\,  v^{ k_5/5}. 
\end{eqnarray}
The  diagonal will extract the terms with the same power. Introducing the integer $\, p$,
such that $\, k_1\, = k_2\, = k_3 \, = k_4/2 =  k_5/5 \, = \, p$,
the only terms contributing to the diagonal are
\begin{eqnarray}
\label{87expli2contrib}
&&  \hspace{-0.98in}  \quad \quad  
   \sum_{p=0}^{\infty} \,\,
{{ (p\, +p\, +p \, +2 \, p\, +5\, p)! } \over { p! \, \, p! \,\,  p! \, \, (2\, p)! \, (5\, p)! }}
 \cdot \, x^{p} \, y^{p} \, z^{p} \, u^{ p}\, v^{ p}
   \, \, = \, \, \, \,\,
   \sum_{p=0}^{\infty}  \, {{(10\, p)!} \over {  p!^3 \, (2\, p)! \, (5\,  p)! }} \cdot \, t^{ p}.
\end{eqnarray}

\vskip 0.1cm

Let us mention how we obtain the (minimal order) linear differential operator annihilating
the diagonal.
Once a long series is obtained, we use the guessing method to obtain\footnote[9]{Alternately,
  we can use the creative telescoping~\cite{Plea}  to get this linear differential operator
as a telescoper. This method often requires more computing time. }
the linear ODE. 
We make use of the "ODE formula" forwarded 
in section 3.1 of~\cite{2008-experimental-mathematics-chi} (and with details in section 1.2 of~\cite{2010-chi6})
to ensure that we 
actually deal with {\em the minimal order} differential equation. For the linear
differential operators of high order, which are not irreducible,
the factorization is obtained using the 
method of factorization of linear differential operators modulo primes (see Section 4 in~\cite{2009-chi5}
and Remark 6 in~\cite{2014-Ising-SG}).

\section{The lattice Green function of the  simple cubic, diamond  and  face centred cubic lattices}
\label{LGF}
  
For the lattice Green functions (LGF) of the {\em simple cubic} lattice of dimension $ \, d$,
the rational function have $ \, N_v= \, d \, +1$ variables. The corresponding linear
differential operators $ \, L_n$, ($n = \, d$, see \ref{LGFsc})
annihilating the LGF
have Maximum Unipotent Monodromy (MUM), 
and thus, the formal solution with the highest log-power at the origin corresponds to
$ \, \ln^{n-1}(t)$. The relation of conjecture (\ref{conjmain}) is again satisfied.

\vskip 0.1cm

Similarly, one may consider the LGF of the {\em diamond lattice} (\ref{LGFdiam})
in 3, 4 and 5 dimensions,
where the rational functions depend
on respectively 4, 5 and 6 variables, and check that the  formal solutions
with the highest log-power, at the origin, are respectively 
in $ \, \ln(t)^2$, $ \, \ln(t)^3$ and $ \, \ln(t)^4$,  in agreement with conjecture (\ref{conjmain}).
All these linear differential operators also have MUM. 

\vskip 0.1cm

The lattice Green functions (LGF) of the {\em face-centred cubic} lattice of dimension $\, d$
are diagonals of rational functions of the form $ \, 1/Q$
for the dimension $\, d < 7$, see~\cite{joyce-1998,guttmann-2009,guttmann-2010,broadhurst-2009,koutschan-2013},
and we have produced the corresponding (minimal order)
linear differential operators~\cite{2015-LGF-fcc7, 2016-LGF-fcc8to12}
for $ \, d = \,\,  7, \, 8,\,  \cdots, \, 12$. These linear differential operators {\em no longer\footnote[2]{Except
for $\, d=2, \, 3, \, 4$ see Table A.1 in \ref{LGFAppend}. } have MUM}.
According to the parity of the dimension $\, d$, which is related to the number of variables by $ \, N_v = \, d +1$,
the differential Galois group of these linear differential operators are included in $\, Sp$ ($N_v$ odd) or $ \, SO$ ($N_v$ even)
in agreement with our conjecture (\ref{conjadditio}).
We have checked (see \ref{LGFfcc}) that, among the formal solutions at the origin of these linear
differential operators, one solution is in front of
$ \, \ln(t)^n$, where $\,n$ is the highest log-power. This exponent is in agreement with
the relation $ \, N_v =  \, n \, +2 \,$ of the conjecture (\ref{conjmain}).
For the exponent of the logarithmic formal solutions at all the other singularities see \ref{LGFfcc}.

\vskip 0.1cm

Beyond this "LGF laboratory", let us give some examples with irreducible
(non factorizable) denominator $\,Q$.

\vskip 0.1cm

\section{Examples with non factorizable denominator $\, Q$}
\label{irreduc}

\subsection{A LGF-like non-MUM example}
\label{lattGutt}

In the examples with the lattice Green functions, one may imagine that the occurrence of the
differential Galois groups $ \, Sp$ or $ \, SO$ is related 
to the dimension of the lattice. Let us go beyond this relation by considering the structure function
\begin{eqnarray}
\label{lattgutt}
\fl \qquad \qquad
  \lambda \, \,=\,\, \,  c_1 c_2 c_3  +c_1 c_2  c_4 +c_1 c_3 c_4 + c_2 c_3 c_4 ,
  \quad \qquad c_j = \, \cos{\phi_j}, 
\end{eqnarray}
considered by Guttmann in~\cite{guttmann-2010}, {\em which does not correspond to any known obvious lattice}.
The (minimal order)  linear differential operator annihilating
\begin{eqnarray}
  \qquad
  G(t) \, \, = \,\,\,\,
  {1 \over (2\pi)^4} \, \int {\frac{d\phi_1 \, d\phi_2 \,  d\phi_3 \,  d\phi_4 }{1 \, \, -t \cdot \, \lambda}}, 
\end{eqnarray}
is of order 8 and is irreducible. Its exterior square is of order 27,
instead of the generic order $28 \, = \, (8 \times 7)/2$. Its
differential Galois group
is thus included in the simplectic group $ \, Sp(8, \, \mathbb{C} )$.
The lattice Green function $ \,G(t)$ is the diagonal of the rational function $ \, 1/Q$,
depending on 5 variables $ \, (z_0, z_1, z_2, z_3, z_4)$ 
\begin{eqnarray}
  \qquad \qquad \quad
Q  \, \, =\, \, \, \,  1 \,\, \, \, - t \cdot \, \lambda, 
\end{eqnarray}
such that $ \, c_j = \,  z_j +1/z_j \,$ $\, j \, = \, 1, \, \cdots \, 4\, \,$
and  $ \, \, t = \, z_0 z_1  z_2  z_3  z_4$.

\vskip 0.1cm

The eight formal solutions, at the origin, of the order-eight  linear differential operator read
\begin{eqnarray}
  \label{display}
  &&  \hspace{-0.98in} \qquad  \quad 
  S^{(0)},  \,  \,  \quad \quad  S^{(0)} \cdot \, \ln(t)\,\,  + S_{2,0}, \quad \quad  \,   \, 
 S^{(0)}  \cdot\, {{\ln(t)^2} \over {2!}} \, \, + S_{3,1} \cdot \,\ln(t) \,\, +  S_{3,0},
   \nonumber \\
   &&  \hspace{-0.98in}\qquad \quad 
   S^{(0)} \cdot \, {{\ln(t)^3} \over {3!}}  \, \, + S_{4,2} \cdot \, {{\ln(t)^2} \over {2!}}
  \,\,  + S_{4,1} \cdot \,\ln(t) \,\,  + S_{4,0},
  \nonumber \\
  &&  \hspace{-0.98in}  \qquad \quad 
  t^{1/2} \cdot \,  S^{(1/2)}
  \qquad  \quad \quad \,\,
   t^{1/2} \cdot \, S^{(1/2)} \cdot \, \ln(t) \,\,  +  t^{1/2}\cdot \, S_{6,0},
  \nonumber \\
  &&  \hspace{-0.98in} \qquad  \quad 
  t^{1/3} \cdot \, S^{(1/3)}, \quad \qquad \quad \,\, \,\,  t^{2/3} \cdot \, S^{(2/3)},  \quad
\end{eqnarray}
where the $\, S_{i, j}$'s are analytical series at $ \, t= \, 0$, and
where the other series begin as $  \,1 \, +\cdots \,$
The series $S^{(0)}$ is {\em globally bounded}, and corresponds
to the diagonal of the rational function. 
All the other analytical series $\, S^{(1/2)}$, $\, S^{(1/3)}$, $\, S^{(2/3)}$, $\, S_{i, j}$ in (\ref{display})
are {\em not globally bounded}.

The diagonal $\,S^{(0)} \, = \, {\rm Diag}(1/Q)\, $ is the diagonal of a rational function
of $\, N_v =\,  3 \, +2\, = \, 5$ variables, in agreement with conjecture  (\ref{conjmain}).

Around all the other singularities, the highest log-power is in $ \, \ln(t)^1$,
and each formal series, in front of $ \, \ln(t)^1$, is {\em not globally bounded}.

\vskip 0.1cm

The generalization of (\ref{lattgutt}) to 6 variables amounts to considering: 
\begin{eqnarray}
&&  \hspace{-0.98in} \quad
   \lambda \, \, = \,\,
 c_1 c_2 c_3 c_4 + c_1 c_2 c_3 c_5 +c_1 c_2 c_4 c_5 +c_1 c_3 c_4 c_5 + c_2 c_3 c_4 c_5,
  \quad \, \, \, \,  \,  c_j = \, \cos{\phi_j}. 
\end{eqnarray}
The corresponding linear differential operator is expected to have its differential Galois group
included in the orthogonal  group.
The linear differential operator is of order 9. Its differential Galois group
is indeed  (included) in  $ \, SO(9, \, \mathbb{C})$,
(its symmetric square being of order 44, instead of the generic order $\, 45 \, = \, (9 \times 10)/2$).

\vskip 0.1cm

The formal solution of this order-nine linear differential operator with the
highest log-power at the origin, corresponds to
$ \, \ln(t)^4$, giving,  according to the conjecture (\ref{conjmain}),
$ \, N_v = \, 4 \, +2 = \, 6\, $ as the minimal number of variables occurring in the rational function.

\vskip .1cm 

\subsection{3-D fcc example: reduction to three variables}

The diagonal of $ \, 1/Q \, $ where
\begin{eqnarray}
  \label{Q3Dfcc}
&&  \hspace{-0.98in} 
Q \,  = \, \, 
 1 \, \, - x y z u \cdot  \, \left( (x+{1 \over x}) \cdot \, (y+{1 \over y})
     +(x+{1 \over x}) \cdot \, (z+{1 \over z}) +(y+{1 \over y}) \cdot \, (z+{1 \over z})  \right), 
\end{eqnarray}
is a polynomial depending on {\em four variables} $\, x, \, y, \, z, \, u$, reproduces
the $\, 3$-dimensional face-centred cubic lattice Green function. 
The (minimal order) linear differential operator annihilating
this diagonal is of order three, and its differential Galois group
is included in $ \, SO(3, \, \mathbb{C})$. 
The most singular formal solution is in $ \,\ln(t)^2$, in agreement with $ \, 2 \, +2 \,$
variables for the rational function, in agreement with conjecture (\ref{conjmain}).

\vskip 0.1cm

Let us reduce the number of variables of the polynomial $ \, Q$, given in (\ref{Q3Dfcc}), to
 {\em three variables}, by {\em fixing} $\, u= \, 1$. The polynomial $\, Q$ 
becomes: 
\begin{eqnarray}
&&  \hspace{-0.98in} 
  Q_1  \, =\, \, 
  1 \, \,  - x y z \cdot \,
  \left( (x+{1 \over x})\cdot \, (y+{1 \over y}) 
  +(x+{1 \over x}) \cdot \, (z+{1 \over z}) +(y+{1 \over y})\cdot \, (z+{1 \over z})  \right). 
\end{eqnarray}
The (minimal order) linear differential operator, annihilating
this diagonal, is of order six, and its differential Galois group
is included in $\, Sp(6, \, \mathbb{C})$.
The  formal solution  with the highest log-power at the origin
is in $\,\ln(t)^{1}$, in agreement with $ \,1 \,+2 \,= \, 3\, $
variables for the rational function.

\vskip .1cm

\subsection{Another diagonal representation of the 3-D fcc LGF \\}
\label{anotherdiag}
The LGF of the three-dimensional fcc lattice {\em can also be seen}
as the diagonal of $\, 1/Q$, where the polynomial denominator $\, Q$ depends on four variables:
\begin{eqnarray}
  \label{Q18}
&&  \hspace{-0.98in} \, \, \quad \quad  \quad \quad \quad  \quad   \, \,  \, 
   Q  \, \,  =\, \, \,  \,  \,
   1 \, \, \,  \,  - x^2 y z u  \cdot \, (1 \, + 4 \, x y z u) \,  \,\,   -(1 \, +u) \cdot \, (y + z). 
\end{eqnarray}
The diagonal of $ \, 1/Q$, where $\, Q$ depends on four variables, gives (with $ \, t = \, x y z u$):
\begin{eqnarray}
\label{On4vars}
&&  \hspace{-0.98in}  \quad  \quad  \quad  \quad  \quad 
  {\rm Diag}_{(xyzu)} \Bigl({{1} \over {Q}} \Bigr)  \, \,  =  \, \,  \,\,
   {}_3F_2\left( [{1 \over 2}, {1 \over 3}, {2 \over 3}], \, [1, 1],
      \,   \, 108 \cdot \, t^2 \cdot \, (1 \, +4 t) \right).
\end{eqnarray}
The  formal solution of the corresponding linear differential operator,
with the highest log-power at the origin,  is
in $ \, \ln(t)^2$,  in agreement with the minimal number of variables of
$ \,  4\, = \, 2 \, +2$, according to conjecture (\ref{conjmain}).

\vskip 0.1cm

Taking the diagonal of $ \, 1/Q$ with $\, Q$ given in (\ref{Q18})
on {\em only the 3 variables} $\, (x,\, y, \, z)$, which means that the
variable $\, u$ {\em is seen as a parameter}, one obtains (with $ \,s = \, x y z$):
\begin{eqnarray}
\label{On3vars}
&&  \hspace{-0.98in}  \quad  \quad  \quad 
  {\rm Diag}_{(xyz)} \Bigl({{1} \over {Q}} \Bigr)  \,  \, \,= \, \,\, \,
   {}_2F_1\left( [{1 \over 3}, {2 \over 3}],[1], \,\, 
      27 \cdot \, u \cdot \, (1 \, +u)^2 \cdot \, (1 \, +4\,  u s) \cdot \, s^2) \right). 
\end{eqnarray}
The  formal solution of the corresponding linear differential operator
with the highest log-power at the $\, s \, =\, 0$ origin, is in $ \, \ln(s)^{1}$,
in agreement with $ \, 3 \, = \, 1 \, +2$ variables,
according to conjecture (\ref{conjmain}).

\vskip 0.1cm

Here, for the {\em same} rational function $\, 1/Q$, we clearly see that the fact that the diagonal
is taken on 4 or 3 variables, produces
a singular formal solution behaving in $ \, \ln(t)^2$ or $ \, \ln(s)$ .

\vskip 0.1cm

{\bf Remark 4.1.} 
Introducing $\, t \, = \, s \, u$, one can write the expansion of  (\ref{On3vars}) as:
\begin{eqnarray}
\label{On3varsexp}
&&  \hspace{-0.98in}  \quad  \quad \quad \quad \quad \quad
   {}_2F_1\left( [{1 \over 3}, {2 \over 3}],[1], \, \,
   27 \cdot \, \, u \cdot \, (1 \, +u)^2 \cdot \, (1 \, +4\,  u s) \cdot \, s^2 \right)
\nonumber \\
&&  \hspace{-0.98in}  \quad  \qquad  \qquad \quad \quad \, 
 \, \, =\, \, \, \, 
 \sum_{n=0}^{\infty}  \,  \sum_{i=0}^{2\, n} \,
   {(3\, n)! \over n!^3} \cdot \,  {2\,n \choose i}  \cdot \,
   u^{n+i} \cdot \, s^{2\, n} \cdot \, (1\, +4\, t)^n,
\end{eqnarray}
where the binomial expansion of $ (1 \, +u)^{2\, n}$ has been used.
Let us (dare to) take the diagonal of the bi-variate {\em transcendental} power series (\ref{On3varsexp}),
which amounts to extracting the coefficients with the same power in $\, u$ and $\, s$.
From (\ref{On3varsexp}) one obtains the following expansion (with $\, t \, = \, s \, u$):
\begin{eqnarray}
\label{DiagOn3varsexp}
&&  \hspace{-0.98in}  \quad  \qquad \quad \quad 
   {\rm Diag}_{(u\, s)} \Bigl(\, {}_2F_1\left( [{1 \over 3}, {2 \over 3}],[1], \,
 \,  27  \cdot \, u \cdot \, (1 \, +u)^2 \cdot \, (1 \, +4\,  u s) \cdot \, s^2) \right)\Bigr)
\nonumber \\
&&  \hspace{-0.98in}  \quad  \qquad  \qquad \quad \quad 
 \, \, =\, \, \,\, 
 \Bigl(\sum_{n=0}^{\infty}  \,  
   {(3\, n)! \over n!^3} \cdot \,  {2\,n \choose n}  \cdot \,
   u^{2\, n} \cdot \, s^{2\, n} \cdot \, (1\, +4\, t)^n \Bigr)_{|t \, = \, u\, s}
\nonumber \\
&&  \hspace{-0.98in}  \quad  \qquad  \qquad \quad \quad 
\, \, =\, \, \,\, 
\sum_{n=0}^{\infty}  \,  
   {(3\, n)! \, (2\, n)! \over n!^5} \cdot \,  
   t^{2\, n} \cdot \, (1\, +4\, t)^n
\nonumber \\
&&  \hspace{-0.98in}  \quad  \qquad  \qquad \quad \quad 
\, \, = \, \, \,\, 
 {}_3F_2\left( [{1 \over 2}, {1 \over 3}, {2 \over 3}], \, [1, 1],
   \, 108 \cdot \, t^2 \cdot \, (1 \, +4 t) \right).
\end{eqnarray}
We thus see that the diagonal on the variables $\,u$ and $\,s$, of
the {\em transcendental} $\, _2F_1$ function (\ref{On3vars}),
{\em actually reproduces} (\ref{On4vars}).

\subsection{A reflexive polytope example}
\label{polytop}

From the 210 linear differential operators arising from reflexive
$\, 4$-polytopes~\cite{lairez-2014,batyrev-kreuzer-2010} (Laurent polynomials of 4 variables)
with symplectic differential Galois group $ \, Sp$, let us consider the
example\footnote[5]{Polytope v18.10805, topology 28 in~\cite{lairez-2014,batyrev-kreuzer-2010}.},
depending on 5 variables, which corresponds to  the  rational function $ \, 1/Q\, $
with polynomial denominator
\begin{eqnarray}
  \label{QS}
  \quad \quad \quad 
Q \, \,  \, = \,  \, \,   \, 1 \, \, \,  - x \, y\, z\, u\, v \cdot \, S, 
\end{eqnarray}
where:
\begin{eqnarray}
 \label{SQ}
 &&  \hspace{-0.98in}  \quad  \quad  \quad \quad  \quad  
S \,  \,= \,  \,  \, \, 
    x+y+z+u \, \,  \,
    +{{1} \over {x}} +{{1} \over {y}} +{{1} \over {z}}   +{{1} \over {u}}
    \nonumber \\
&&  \hspace{-0.98in}  \quad  \quad  \quad  \quad   \quad \quad   \quad \, 
   +yz\, +{\frac {x}{z}} +{\frac {u}{z}} +{\frac {x}{yz}} +{\frac {z}{xu}}
 + {\frac {u}{yz}} +{\frac {xu}{yz}} +{\frac {yz}{x}} +{\frac {yz}{u}} +{\frac {xu}{{z}^{2}y}}.
\end{eqnarray}
The diagonal of $ \, 1/Q$ is annihilated by an irreducible order-six linear
differential operator whose differential Galois group is symplectic (included in 
$ \, Sp(6, \, \mathbb{C})$). The formal solution, at the origin, of this order-six
linear differential operator, with the highest log-power, behaves as $ \, \ln(t)^3$.
This is in agreement with the number of variables $\, N_v= \,  3 \, +2 \, = \, 5\, $ occurring in
the polynomial $ \, Q$, according to conjecture (\ref{conjmain}). 

Around all the other singularities, the maximum power of the log
in the formal solutions is 1, and the series are {\em non-globally bounded}.

\vskip .1cm 

\subsection{An Apery generalization example}
\label{Apery}

Let us recall the series with Apery numbers\footnote[1]{The sequence is related
 to Apery's proof on the irrationality of $ \, \zeta(3)$.}:
\begin{eqnarray}
  \label{Apery}
&&  \hspace{-0.98in}  \quad  \quad  \quad  \quad     \,  \,    \,   
  \sum_{n=0}^{\infty}\sum_{k=0}^{n}\,  { n \choose k}^2 \, { n+k \choose k}^2 \cdot \, t^n
  \,\,\, = \, \, \, \,\,
  \sum_{n=0}^{\infty}\sum_{k=0}^{n}\,   \left(  {\frac{(n+k)!}{k!^2 \, (n-k)! }}  \right)^2  \cdot \,t^n. 
\end{eqnarray}
This series is actually the diagonal of the rational function $ \, 1/Q$ with four variables~\cite{christol-EMS}:
\begin{eqnarray}
 \label{xxx}
  Q \, \,=\, \,  \, \,
  (1 \, \, -x_1  \, -x_2) \cdot \,  (1 \, \, -x_3  \, -x_4) \, \,  \, -x_1 \,x_2\, x_3 \, x_4.
\end{eqnarray}
More generally, let us consider the series given by (with $ \, m$ a positive integer):
\begin{eqnarray}
  \label{xx}
&&  \hspace{-0.98in}  \quad  \quad  \quad  \,   \,   
   \sum_{n=0}^{\infty}\sum_{k=0}^{n}\,  { n \choose k}^m { n+k \choose k}^m \cdot \, t^n
 \,\,\, = \, \, \, \,\,
  \sum_{n=0}^{\infty}\sum_{k=0}^{n}\,   \left(  {\frac{(n+k)!}{k!^2 \,  (n-k)! }}  \right)^m  \cdot \,t^n. 
\end{eqnarray}
This series is the diagonal of the rational function $ \, 1/Q_m$,
with\footnote[5]{Use the multinomial expansion, then equate the exponents of the variables.}
\begin{eqnarray}
 \label{xxx}
&&  \hspace{-0.98in}  \quad  \quad   \quad \qquad 
  Q_m \, \,=\, \,  \, \, \,
  1 \, \,  \, \,
  - \, \prod_{j=0}^{m-1} \, \left( x_{2j+1} \,+x_{2j+2} \,+x_{2j+1} \cdot \, x_{2j+2} \right), 
\end{eqnarray}
depending on $ \, 2 m$ variables.
For any value of $ \, m$, the number of variables being even, the
linear differential operator,
corresponding to the diagonal of $ \, 1/Q_m$, has a differential Galois group
which should be included in the orthogonal group $\, SO$, according to conjecture (\ref{conjadditio}).

\vskip 0.1cm

For $ \, m= \, 3$, we have 6 variables in (\ref{xxx}),
\begin{eqnarray}
\label{xxx3}
&&  \hspace{-0.98in}  \quad  \quad   \quad \, \, 
  Q_3 \, \,=\, \,  \, \,
  1 \, \,  \, \,
  - \, \left( x_{1}+x_{2}+x_{1} x_{2} \right) \cdot \,
  \left( x_{3}+x_{4}+x_{3} x_{4} \right) \cdot \, \left( x_{5}+x_{6}+x_{5} x_{6} \right), 
\end{eqnarray}
and the diagonal of the rational function $ \, 1/Q_3 \, $ reads:
\begin{eqnarray}
 \label{seriesm3}
&&  \hspace{-0.98in}  \quad  \quad \quad   \quad    \quad   \,  \,  \,  \,  \, 
   \sum_{n=0}^{\infty} \sum_{k=0}^{n} \, { n \choose k}^3 { n+k \choose k}^3 \cdot \, t^n
   \,  \,  =  \,  \, \,  \,  \,
   \sum_{n=0}^{\infty} \sum_{k=0}^{n} \, \left(  {\frac{(n+k)!}{k!^2  \,  (n-k)! }}  \right)^3 \cdot \,t^n
   \nonumber \\
&&  \hspace{-0.98in}  \quad  \quad   \quad   \quad   \quad   \quad   \quad      \quad   \quad      \quad   \quad     
  \, = \,  \,   \,  \,  \,
   1 \,  \,   \, +9\,t \,  \,  \,  +433\,{t}^{2}\,  \,  +36729\,{t}^{3}
   \, \,  \,   \, \,  + \, \cdots 
\end{eqnarray}
This series (\ref{seriesm3}) is annihilated by an order-nine linear differential operator
$ \, L_{9}$. The symmetric square of $ \, L_{9}$ is of order 44, instead 
of the generic order $\,45 \, = \, (9  \times \, 10)/2$. The differential Galois group
of this order-nine linear differential
operator $\, L_{9}$ is (included)
in the orthogonal group $ \, SO(9, \, \mathbb{C})$.
The formal solution of $\, L_{9}$ with the highest log-power, at the origin,
behaves as $\, \ln(t)^4$, in agreement with the conjecture
(\ref{conjmain}): $\, N_v \, = \, 4 \, +2 \, = \,6 $.

The maximum exponent of the log's in the formal solutions around all the other singularities
$\,t = \, t_j\, $ (with the exception of $\,t = \, \infty$) is 2, and the series, 
in front of $\,\ln(t-t_j)^2$, are {\em not globally bounded}. 

For the singularity $ \, t = \,\infty \,= \, 1/s$, the highest log-power is {\em also} 4,
i.e. the same value as the maximum log-power around 
$ \, t= \, 0$, so we expect the corresponding series in  front of $\, \ln(s)^4 \, $
to be {\em globally bounded}. The first terms of the series read
\begin{eqnarray}
  \label{ser_s}
&&  \hspace{-0.98in}  \quad  \quad   \quad \quad \quad \quad \quad   \quad  
   s \, \cdot \,
   \Bigl( 1\, \,   +9\,s \,  \, \,   +433\,{s}^{2} \,   \, \,  +36729\,{s}^{3}
   \, \,   \,  + \, \cdots \, \Bigr). 
\end{eqnarray}
Morever, this series (\ref{ser_s}) identifies with (\ref{seriesm3}) up to a multiplicative variable $\, s$, 
which is a consequence of the symmetry $\, t \rightarrow \, 1/t \, $
in the  linear differential operator $\,L_9$
\begin{eqnarray}
&&  \hspace{-0.98in}  \quad  \quad   \quad \quad \quad \quad \quad  \quad \quad   \quad  
   L_9 \left( {1 \over t} \right) \, \,= \, \, \, \, 
      t \cdot \, L_9 \left( t \right) \cdot \, {1 \over t}.
\end{eqnarray}
For the rational function $\,1/Q_3$, the change $\, x_j \, \rightarrow\,  1/x_j\, $ appears as
\begin{eqnarray}
\hspace{-0.98in}    \quad \quad \quad \quad \quad   \quad  
{\rm Diag} \left( {t \over Q_3} \right) \,  \,  \,= \,\, \, \, 
{\rm Diag} \left( - {1 \over {\tilde Q}_3} \right)
\quad \quad \quad  \quad \quad \quad  \hbox{with:} 
 \quad \\
\hspace{-0.98in}  \quad  \quad   \quad \quad \quad\quad
  {\tilde Q}_3 \,\,    =\,  \, \,  \,
1 \,\,  \, \,  - \, {1 \over t} \cdot
  \left(1+  x_{1}+x_{2} \right) \cdot \,  \left(1+ x_{3}+x_{4} \right) \cdot \, \left( 1+x_{5}+x_{6} \right).
 \nonumber 
\end{eqnarray}

\vskip 0.1cm

Similarly, for $ \, m= \, 4$, we have 8 variables in (\ref{xxx}).
The linear differential operator, annihilating the diagonal of the rational function $\, 1/Q_4$,
is of order 15. Its differential Galois group is (included in) the orthogonal group
$\, SO(15, \, \mathbb{C})$. The 
formal series solution, at the origin, of the order fifteen linear differential operator,
with the highest log-power, behaves as $\, \ln(t)^6$
in agreement with the conjecture (\ref{conjmain}):
$\, N_v \, =   \, 6 \, +2 \, = \, 8$.

\vskip .1cm 


\section{Some Calabi-Yau examples}
\label{CalabiYau}

We consider, in this section, some examples of Calabi-Yau equations from~\cite{TablesCalabi-2010},
where the expressions of the general term of the series are known in closed form. 
These order-four linear differential operators are irreducible, their differential Galois groups are
(included in) symplectic groups $ \, Sp(4, \, \mathbb{C} )$. They all have the MUM property~\cite{TablesCalabi-2010},
and their formal solutions with the highest log-power behave as $\, \ln(t)^3$. 
Therefore, according  to conjecture (\ref{conjmain}), these linear differential operators should annihilate
 rational functions of  $\, N_v \, = \, 3 \, +2 \,= \, 5 \, $ variables.

\vskip .1cm 

\subsection{The first 19 Calabi-Yau operators in Almkvist et al. Table~\cite{TablesCalabi-2010}}
\label{first19}

We have considered the first 19 Calabi-Yau linear differential operators of~\cite{TablesCalabi-2010}.
These Calabi-Yau equations (except $\#$9) have a geometric
origin~\cite{morrison-1992, klemm-theisen-1992, klemm-theisen-1993,libgober-teitelbaum-1993,batyrev-stranten-1995}.
The general term of the series are (or can be) written
{\em as nested sums of products of binomials}~\cite{lairez-2014}, known to correspond to diagonals of rational
functions~\cite{2013-rationality-integrality-ising,lairez-2014}. The aim, here, is to give
the rational function {\em with only 5} variables.

\vskip 0.1cm

Assuming some $\, 1/Q \, $  form for the rational function, the polynomials $\, Q \, $ are
obtained from closed formulae (given in~\cite{TablesCalabi-2010}) written as {\em ratio of factorials}, 
using a ``guessing'' procedure sketched in \ref{binomialVSfactorial}.
The procedure in \ref{binomialVSfactorial}, amounts, from the general term given in \cite{TablesCalabi-2010}, to 
going back up to the expression coming out from a {\em multinomial} expansion.
Eventually one finds that all the polynomials $\, Q \, $ can be written with five variables.
We give the corresponding multivariate polynomials 
$\, Q \, $ obtained from this guessing procedure, in Table 1.

Note that, for some $ \, Q$'s, we still say that  $\, Q$ is ``polynomial'' even if it contains
$\, N$-th root of some variables ($u^{1/4}, \, v^{1/6}$, ...). 
Section \ref{formalsolexpan} shows that the diagonal of the rational function with $\, N$-th root of variables is actually
a power series (not a Puiseux series). In the following, instead of some "polynomials" containing
$\, N$-th root of variables,  we give equivalent polynomials still with 5 variables.

\vskip 0.1cm

\begin{table}[htp]
\caption{Rational functions $ \, 1/Q$ for some Calabi-Yau series $\sum \, A_n\, t^n$, where $\, A_n$
is the general term of the series given in \cite{TablesCalabi-2010}. We follow the numbering $\#$ of \cite{TablesCalabi-2010}.
 }

\label{Ta:1}
\begin{center}
\begin{tabular}{|c|c|c|c|}\hline
    $\#$   &  $A_n$    &  $ 1-Q $ &  $N_v$        \\ \hline 
\hline
    1     & ${\frac{(5n)!}{n!^5}}$               & $x + y + z + u + v$                                 &  5     \\
    2     & ${\frac{(10n)!}{n!^3 (2n)! (5n)!}}$  & $x + y + z + u^{1/2} + v^{1/5} $                    &  5     \\
    3     & ${\frac{(2n)!^4}{n!^8}}$             & $ (x+z)(1+y) (1+v) (1+u)$           &  5     \\
    4     & $\left({\frac{(3n)!}{n!^3}}\right)^2$             & $ \left(x + yu + zv\right)\left(y + xz + uv\right) $        &  5     \\
    5     & ${\frac{(2n)!^2 (3n)!}{n!^7}}$             & $ x + y + z + u +v (x+y) (z+u)$          &  5     \\
    6     & ${\frac{(2n)! (4n)!}{n!^6}}$             & $ x + y + z + u + v ( z + u ) $                   &  5     \\
    7     & ${\frac{(8n)!}{n!^4 (4n)!}}$             & $ x+y+z+u+v^{1/4} $        &  5     \\
    8     & ${\frac{(6n)!}{n!^4 (2n)!}}$             & $ x + y + z + u + v^{1/2} $              &  5     \\
    9     & ${\frac{(2n)!(12n)!}{n!^4 (4n)! (6n)!}}$       & $ x + y + z (x+y) + u^{1/4} + v^{1/6} $  & 5      \\
   10     & $\left( {\frac{ (4n)!}{n!^2 (2n)!}} \right)^2$       & $ \left( x + y u + (z v)^{1/2} \right)\left( y + x z + (u v)^{1/2} \right) $  & 5      \\
   11     & $ {\frac{ (3n)! (4n)!}{n!^5 (2n)!}} $       & $ x + y + z + u + v (y + z + u) $        & 5      \\
   12     & $ {\frac{ (4n)! (6n)!}{n!^3 (2n)!^2(3n)!}} $       & $ x + y + (1+z) (u^{1/2}+v^{1/2}) $   &  5     \\
   13     & $ \left( {\frac{ (6n)!}{n! (2n)!(3n)!}} \right)^2 $       & $ \left( x + (y u)^{1/2} + (z v)^{1/3} \right)\left( y + (x z)^{1/2} + (u v)^{1/3} \right) $   &  5     \\
   14     & $  {\frac{ (2n)!(6n)!}{n!^5 (3n)!}}  $       & $ x + y + z + u (x+y) +v^{1/3} $   &  5     \\
   15     & ${\frac{(3n)!}{n!^3}} {n \choose k}^3$  & $ x+y+z \left( x+v \right) +u \left( y+v \right) $                               &  5      \\
   16     & ${2n \choose n}{n \choose k}^2{2k \choose k}{2n-2k \choose n-k}$  & $x + y + z +u+v(xyz+xyu+xzu+yzu)$  &  5      \\
   17     & $\left( {\frac{n!}{j! k! (n-j-k)!}} \right)^3$  & $\left( x+y+z \right) \left( v+x u + y z \right) \left( u v + z v + x y u \right)$  &  5      \\
   18     & ${ 2 n \choose n } { n \choose k }^4$  & $u + \left( x+y \right) \left( x+ v \right) \left(  z+ v \right) \left(  y+ z \right)$  &  5      \\
   19     & ${ n \choose k }^3 { n+k \choose n } { 2 n- k \choose n }$  & $\left( x+y \right)  \left( u+x \right)  \left( u+v \right)  \left(y+z \right)  \left( v+z+y+u \right)$  &  5      \\
\hline
 \end{tabular}
\end{center}
\end{table}

\vskip 0.1cm

\subsection{The $\, Q$'s are  multivariate polynomials of variables and $\, N$-th root of variables}
\label{trick}

For some $ \, Q$'s which are  polynomials with $\, N$-roots of variables,
a straighforward change of variables, such 
$\, (x,\,  y, \, \cdots)\,  \rightarrow \, (x^n, \, y^n, \, \cdots)$, may be introduced.

\subsubsection{Calabi-Yau number 2 \\}
\label{trickcY2}

For instance, for Calabi-Yau number 2 (that we denote $ \, CY_2$),
\begin{eqnarray}
 \label{CY2} 
&&  \hspace{-0.98in}  \quad  \, \, \,  \, \, 
  CY_{2} \, \,= \,\, \, \,
   \sum_{n=0}^{\infty} \,  {\frac{(10n)! }{n!^3 \, (2n)! \,  (5n)! }} \cdot \, t^n
\,\, \,  = \,\, \, \, 
{}_4F_3\left( [{1 \over 10}, {3 \over 10}, {7 \over 10}, {9 \over 10}],[1, 1, 1], \,  2^8 \,5^5 \cdot  \,t \right)
\nonumber \\
 &&  \hspace{-0.98in}  \, \,   \quad \quad \quad \, \, \,
=   \, \,   \, \,\,
1\,  \,\,  +15120\,t \, \,  \,  +3491888400\,{t}^{2}  \, \, \, +1304290155168000 \,{t}^{3}
  \,  \, \, \,\, + \, \,\cdots
\end{eqnarray}
which is actually the diagonal of $\, 1/Q$, with  
\begin{eqnarray}
  \label{u1sur2v1sur5}
&&  \hspace{-0.98in}  \quad  \quad   \quad   \quad  \quad   \quad  \quad  \quad 
Q \, \,=\, \,\, \,  \, 1  \,\, \,  \,- (x + y + z \, \,  \, + u^{1/2} \, + v^{1/5}), 
\end{eqnarray}
the series is in $\, t =\,  x y z u v$.
If one wants to get rid of the $\, N$-th roots $\,u^{1/2}$ and  $\, v^{1/5}$,
one can rather consider the polynomial
\begin{eqnarray}
 \label{x5y5z5}
&&  \hspace{-0.98in}  \quad \quad \quad     \quad   \quad  \quad  \quad  \quad 
Q \, \, = \, \,\, \, \,
 1  \,\, \, \, \, - (x^{10} + y^{10} + z^{10} \, \,  \, + u^{5} \, + v^{2}), 
\end{eqnarray}
yielding for the diagonal of $\, 1/Q$ where $\, Q$ is given by (\ref{x5y5z5}):
\begin{eqnarray}
 \label{diagx4y4u4v5}
&&  \hspace{-0.98in}  \quad   \quad\,  \, \, \, \, \, 
{\rm Diag}\Bigl( {{1} \over {Q}} \Bigr)
\, \, = \, \, \, 
{}_4F_3\left( [{1 \over 10}, {3 \over 10}, {7 \over 10}, {9 \over 10}],\, [1, 1, 1], \,  2^8 \, 5^5 \cdot \, t^{10} \right) 
 \\
&&  \hspace{-0.98in}  \quad  \, \, \, \quad  \quad  \quad    \quad  
 = \, \, \,  \, \,
1 \, \, \,   +15120  \,t^{10}\, \, +3491888400  \,{t}^{20} \, \, +1304290155168000 \,{t}^{30}
\, \,  \, \,   +  \cdots
\nonumber 
\end{eqnarray}
Other polynomials with five variables can be found. With the polynomial  
\begin{eqnarray}
  \label{x4y4u4v5}
&&  \hspace{-0.98in}  \quad  \quad  \quad  \quad  \quad \qquad 
 Q \, \, = \,\, \,\, 
  1 \,\, \, \,  -(x + y + z \, \, + x^4 y^4 u^4 v^5  \, \, +x z v u^2),
\end{eqnarray}
the diagonal of $\, 1/Q$, where $\, Q$ is given by (\ref{x4y4u4v5}), reads:  
\begin{eqnarray}
  \label{diagx4y4u4v5}
&&  \hspace{-0.98in}   
 {\rm Diag}\Bigl( {{1} \over {Q}} \Bigr) \, \, = \, \,  \,  \, 
 \,  1 \,  \,\,  +15120 \,t^{6}\,  +3491888400  \,{t}^{12}  \, +1304290155168000 \,{t}^{18}
   \,  \,   \,  + \,  \cdots
\end{eqnarray}
The other polynomial
\begin{eqnarray}
\label{xyvu2}
&&  \hspace{-0.98in}  \quad  \quad  \quad  \quad \qquad
Q \,\, = \,\, \, \, \,
1  \, \,\,  \,  - \left( x y z u + x z u v + y z u v \, \, + x y v u^2+ x y z v^3 \right), 
\end{eqnarray}
yields for the diagonal of $\, 1/Q$, where $\, Q$ is given by (\ref{xyvu2}):  
\begin{eqnarray}
\label{diagxyvu2}
&&  \hspace{-0.98in}\, \, \,
{\rm Diag}\Bigl( {{1} \over {Q}} \Bigr) \, \, = \, \, 
  \,   1 \,  \,  +15120 \,t^9\, \,  +3491888400\,{t}^{18} \, \, +1304290155168000\,{t}^{27}
 \, \,   \,   + \, \cdots
\end{eqnarray}
The series (\ref{diagx4y4u4v5}) and (\ref{diagxyvu2}) are the series (\ref{CY2})
where $\, t$ is changed respectively into $\, t^6$ and $\, t^9$.

\vskip 0.1cm

\subsubsection{Calabi-Yau number 7 \\}
\label{trickcY2}

The series for Calabi-Yau number 7
\begin{eqnarray}
  \label{CY7}
&&  \hspace{-0.98in}  \quad  \quad \quad \quad
CY_7 \, \, = \,\, \, \,
\sum_{n=0}^{\infty} {\frac{(8n)!}{n!^4  \, (4n)!}} \cdot \, t^n \,\,   = \, \, \, 
{}_4F_3\left( [{1 \over 8}, {3 \over 8}, {5 \over 8}, {7 \over 8}],[1, 1, 1], \,  \, 2^{16} \,t \right) 
   \nonumber \\
&&  \hspace{-0.98in}  \quad  \quad \qquad \quad \quad
   = \, \, \, \,  \, \,
 1 \, \, \,  \,+1680\,t \, \, \,+32432400\,{t}^{2}  \,\,+ 999456057600\,{t}^{3}
 \,\, \,\,  \,  + \, \cdots
\end{eqnarray}
is actually the diagonal of $1/Q$ with  
\begin{eqnarray}
  \label{v1sur4}
&&  \hspace{-0.98in}  \quad  \quad \quad \quad \quad \quad \quad \quad
  Q\, \, = \,\, \, \,
  1 \,\,  \, \, -(x+y+z+u \, \, \, +v^{1/4}), 
\end{eqnarray}
the diagonal series being in $\, t = \, xyzuv$.
If one wants to get rid of the $\, N$-root $\, v^{1/4}$, one can alternatively consider
\begin{eqnarray}
  \label{v1sur44}
&&  \hspace{-0.98in}  \quad  \quad \quad \quad \quad \quad \quad \quad
Q\, \, = \,\, \, \,   1 \,\,  \, -(x^4+y^4+z^4+u^4 \, \,  \, +v), 
\end{eqnarray}
which gives, for the diagonal of $\, 1/Q$, the series (\ref{CY7}) where $\, t$ is changed into $\,t^4$.
Other polynomials, instead of the "polynomial" (\ref{v1sur4}), can be considered, such as 
\begin{eqnarray}
&&  \hspace{-0.98in}  \quad  \quad \quad \quad \quad \quad \quad \quad
Q \, \,=\, \, \, \,
 1 \, \, \,  \,- \left( x + y + z + u \,\, \,  +v^4 y^3 z^3 u^3 \right), 
\end{eqnarray}
yielding for the diagonal of $\, 1/Q$
\begin{eqnarray}
 &&  \hspace{-0.98in}  \quad  \quad \quad
 {\rm Diag}\Bigl( {{1} \over {Q}} \Bigr) \, \, = \, \, \,\,
 1 \, \, \, \, +1680 \,t^4\,  \,+32432400 \,{t}^{8}  \,\, + 999456057600\,{t}^{12}
  \,\, \,  + \, \cdots
\end{eqnarray}
or the polynomial 
\begin{eqnarray}
&&  \hspace{-0.98in}  \quad  \quad \quad \quad \quad \quad \quad
 Q \,\,  = \, \, \, \,
  1 \, \,  \,- \left( x y z u +y z u v+ z u v x+ u v x y \,  \, \,  + v x y z^4 \right), 
\end{eqnarray}
yielding
\begin{eqnarray}
&&  \hspace{-0.98in}  \quad  \quad \quad
   {\rm Diag}\Bigl( {{1} \over {Q}} \Bigr)
\, \, = \, \,\, \,
1 \, \,  \,+1680 \,t^7\,  \,+32432400 \,{t}^{14}  \,\, + 999456057600\,{t}^{21}
\, \,\,\,   + \, \cdots
\end{eqnarray}
which are the series (\ref{CY7}) where $\, t\, $ is respectively
changed into $\,t^4$ and $\,t^7$.

\vskip 0.1cm

{\bf Remark 5.1.}
Some of the Calabi-Yau equations in~\cite{TablesCalabi-2010}
are given in~\cite{batyrev-kreuzer-2010} from reflexive $4$-polytopes.
With the Laurent polynomial of
four variables\footnote[5]{Polytope v7.7, topology 67 in~\cite{lairez-2014, batyrev-kreuzer-2010}.}
giving the Calabi-Yau equation number 8, the rational function, depending on
five variables is $\, 1/Q$, with:
\begin{eqnarray}
  \label{4poly}
&&  \hspace{-0.98in}  \quad  \quad \quad \quad \quad \quad
Q \, \,= \, \,\, \,  \,
 1 \, \,\,  \, - v \cdot \, (1 \, \, +y+z) \cdot \, ( x^3+y^2  z^2) \, \,  \, -y z v u^2. 
\end{eqnarray}
The diagonal of $\, 1/Q$, with $\, Q$ given in (\ref{4poly}), reads:
\begin{eqnarray}
\label{diag4poly}
&&  \hspace{-0.98in}  \quad  \quad \quad \quad \, \, 
{\rm Diag}\Bigl( {{1} \over {Q}} \Bigr)
 \, \, = \, \, \,\,
1 \, \,  \,+360 \,t^6\,  \,+1247400 \,{t}^{12}  \,\,+6861254400\, t^{18}
\,\, \,   \,  + \, \cdots
\end{eqnarray}
Also, with the Laurent polynomial of
four variables\footnote[2]{Polytope v8.5, topology 61 in~\cite{lairez-2014,batyrev-kreuzer-2010}.}
giving the Calabi-Yau equation number 10, the rational function $\, 1/Q$ depends on
5 variables, where $\, Q$ reads:
\begin{eqnarray}
 \label{4poly2}
Q \,\,= \,\,\,\,  \,
 1 \,\, \, \, - v \cdot \, (x +u^2) \cdot \, \left( x^2 +y z\, +y z^2 +y^2 z \right). 
\end{eqnarray}
The diagonal of $\, 1/Q$, with $\, Q$ given in (\ref{4poly2}), reads:
\begin{eqnarray}
 \label{diag4poly2}
 &&  \hspace{-0.98in}  \quad  \quad \quad \quad \, \, 
{\rm Diag}\Bigl( {{1} \over {Q}} \Bigr) \, \, = \, \, \, \, \, 
 1 \, \,  \,+144 \,t^4 \,  \,+176400 \,{t}^{8}  \,\, +341510400\, t^{12}
 \,\, \, \,  \,    + \, \cdots
\end{eqnarray}
These expressions (\ref{4poly}), (\ref{4poly2}), different from the corresponding polynomials in Table 1,
{\em still have} five variables.

\vskip 0.1cm

\subsection{The rational function versus the pullbacked solution}

From the Calabi-Yau equations in~\cite{TablesCalabi-2010}, two (or more) may have the {\em same Yukawa
coupling}~\footnote[1]{See, e.g.~\cite{Lian-Yau-1996,Almk-Zud-2006,2011-calabi-yau-ising}
  for the definition of the Yukawa coupling, and \ref{sqrtR2CY}.},
which means~\cite{Almk-Zud-2006}  that one solution can be written in terms of the other one. 
Let us see {\em how this property appears} in the multivariate rational functions $\, 1/Q$.

\vskip 0.1cm

\subsubsection{Calabi-Yau number 76 \\}
The series for Calabi-Yau  number 76 reads:
\begin{eqnarray}
&&  \hspace{-0.98in}  \,   \quad \quad \quad \quad \quad \quad \quad \quad
CY_{76}\, \,= \,\,\,
    \sum_{n=0}^{\infty} \sum_{k=0}^{n} \,  {n \choose k} \cdot  \, {\frac{(4k)! \, (2k)!}{k!^6 }} \cdot \,t^n 
\nonumber \\
&&  \hspace{-0.98in}  \quad  \quad \quad \quad  \quad \,  \quad \quad \quad \quad
 \label{series76}
\,\, =  \,\,\,  \,  \, \, 
1 \, \,   \,\,  +49\,t\, \,   \,\,   +15217\,{t}^{2}\,\,  \,   +7437505\,{t}^{3}
\, \,\,\,  \,   + \,\cdots
\end{eqnarray}
The "ratio of factorials part" of the general coefficient of $\, CY_{76}$ is that of
$\, CY_6$ (see Table 1). Both equations
have the {\em same} Yukawa coupling. One actually has
\begin{eqnarray}
&&  \hspace{-0.98in}  \quad  \, \, \quad  \quad \quad  \quad \quad 
  CY_{76}\, \, = \, \,
  {\frac{1}{1 \,-t }} \cdot \, CY_6 \left( {\frac{t}{1 \, -t }} \right)
  \nonumber \\
&&  \hspace{-0.98in} \quad   \, \, \quad  \quad  \quad  \quad \quad \quad\quad 
  \, = \,\,\,
   {\frac{1}{1 \,-t }} \cdot \,
   {}_4F_3\left( [{1 \over 2}, {1 \over 2}, {1 \over 4}, {3 \over 4}],[1, 1, 1],
   \, \, 2^{10} \cdot \, { t \over 1-t}\right),     
\end{eqnarray}
and the rational function $ \, 1/Q_{76}$, corresponding to $ \, CY_{76}$,
reads (with $ \, t =  \,x y z u v$):
\begin{eqnarray}
  \label{Q76}
&&  \hspace{-0.98in}    \quad  \quad  \quad  \quad       
Q_{76} \, \, = \, \, \,\,
(1 \, -t) \cdot \,
\left( 1 \,\,  \, \, - \left( {\frac{x}{1 \,-t }} \,   \, \, + y  \,\, + (z+u) \cdot \, (1+v) \right) \right). 
\end{eqnarray}
 One has a rescale of one variable with respect to $ \, Q_6$ (see Table 1).
The diagonal of $\, 1/Q$, with $\, Q$ given in (\ref{Q76}), identifies with the series in (\ref{series76}).

\vskip 0.1cm

\subsubsection{Calabi-Yau number 79 \\}
Similarly the series for Calabi-Yau number 79 reads
\begin{eqnarray}
\label{Q79}
&&  \hspace{-0.98in}  \quad  \quad  \quad \quad  \quad \quad \quad 
CY_{79}\,  \, = \, \,\,
\sum_{n=0}^{\infty} \sum_{k=0}^{n} \, {n \choose k} \cdot \, {\frac{(5k)!}{k!^5 }} \cdot \, t^n
\nonumber \\
 &&  \hspace{-0.98in}  \quad  \quad  \quad \quad  \quad \quad  \quad  \quad \quad
\label{series79}
\, = \,\, \,\,\,
1 \,\,\,   +121\,t \,\,   \, +113641\,{t}^{2} \, \,\, +168508561\,{t}^{3}
 \, \, \, \, + \,  \cdots
\end{eqnarray}
and has the {\em same} Yukawa coupling~\cite{TablesCalabi-2010} as $\, CY_1$.  It reads
\begin{eqnarray}
  \label{C79}
&&  \hspace{-0.98in}  \quad     \quad   \quad   \quad   \quad        
  CY_{79} \,\, = \,\, \,   {\frac{1}{1 \, -t }} \cdot \, CY_1 \left( {\frac{t}{1-t }} \right)
  \nonumber \\
&&  \hspace{-0.98in}  \quad    \quad   \quad  \quad   \quad   \quad   \quad     \quad       
  = \, \, 
  {\frac{1}{1 \,-t }} \cdot \,
   {}_4F_3\left( [{1 \over 5}, {2 \over 5}, {3 \over 5}, {4 \over 5}],[1, 1, 1],
   \, \, 5^{5} \cdot \,  {  t \over 1-t}\right),     
\end{eqnarray}
which is the diagonal of $\, 1/Q_{79}$ where  $\, Q_{79}$ reads (with $\,t =\, x y z u v$):
\begin{eqnarray}
\label{Q79}
&&  \hspace{-0.98in}  \quad  \quad  \quad \quad  \quad   \quad 
  Q_{79}  \,\, = \,\, \, \,
  (1 \, -t) \cdot \, \left( 1 \,  \, \, \, - \left( {\frac{x}{1-t }} \, \, + y + z + u + v) \right) \right).
\end{eqnarray}
The diagonal of $\, 1/Q$, with $\, Q$ given in (\ref{Q79}), is the same series as in (\ref{series79}).

\vskip 0.1cm

\subsubsection{Calabi-Yau number 128 \\}
Another example is Calabi-Yau number 128 which series reads
\begin{eqnarray}
\label{seriesCY128}
&&  \hspace{-0.98in}  \quad   \, \quad  \quad \quad
CY_{128} \,\, = \, \, \, \,
   \sum_{n=0}^{\infty} \, \sum_{k=0}^{n} \,
   {2 n \choose n} \cdot {n \choose k} \cdot \, {\frac{(5k)!}{k!^3 \, (2k)! }} \cdot \, t^n
  \\
&&  \hspace{-0.98in}  \quad \quad  \, \quad \quad  \quad
=  \,\,\, \, \,
1\, \, \,+122\,t \,\, +114126\,{t}^{2}\, \,+169305620\,{t}^{3} \, +307902541870\, t^4
  \,  \,  \, \, \, + \, \cdots
   \nonumber
\end{eqnarray}
and has the {\em same} Yukawa coupling~\cite{TablesCalabi-2010} as $\, CY_1$,
in terms of which it writes:
\begin{eqnarray}
  \label{CY128} 
&&  \hspace{-0.98in}  \quad  \quad  \quad \quad \quad
\, \,  \,   CY_{128} \, \, = \, \,\, \,
   {\frac{1}{\sqrt{1 \,-4 t} }}  \cdot \, CY_1 \left( {\frac{t}{1 \, -4 t }} \right)
 \nonumber  \\
&&  \hspace{-0.98in}  \quad  \quad   \quad \quad \quad \quad \quad
 \, \,   \, = \,  \,  \,
   {\frac{1}{\sqrt{1 \,-4 t} }}   \cdot \,
   {}_4F_3\left( [{1 \over 5}, {2 \over 5}, {3 \over 5}, {4 \over 5}],[1, 1, 1],
   \, \, 5^{5} \cdot \, { t \over 1 \, -4\, t}\right).     
\end{eqnarray}
The Calabi-Yau series (\ref{seriesCY128}) or  (\ref{CY128}), is the diagonal of
 $\, 1/Q_{128}$, where the denominator $\, Q_{128}$
has an ``algebraic term''
{\em depending only on the variable} $\,t =\, x y z u v$: 
\begin{eqnarray}
  \label{Q128}
 &&  \hspace{-0.98in}  \quad  \quad  \quad  \quad  \quad  \quad \, \, 
  Q_{128} \,  \, = \, \, \,\,\,
      \sqrt{1\, -4 t} \cdot \, \Bigl( 1 \, \,  - (z+u+v) \Bigr) \, \, \,\,   - x - y.
\end{eqnarray}

\vskip .1cm

{\bf Remark 5.2.} Let us consider the rational function $\, 1/Q_{\alpha} \,$ where,
instead of (\ref{Q128}),  the polynomial denominator $\,Q_{\alpha} \,$ reads:
\begin{eqnarray}
  \label{Qalpha}
 &&  \hspace{-0.98in}  \quad  \quad  \quad  \quad \quad \quad    \quad  \quad \, \, 
  Q_{\alpha} \,  \, = \, \, \,\,\,
      \alpha \cdot \, \Bigl( 1 \, \,  - (z+u+v) \Bigr) \, \, \,\,   - x\,  - y.
\end{eqnarray}
The diagonal of  $\, 1/Q_{\alpha} \,$  is the hypergeometric function;
\begin{eqnarray}
  \label{diagQalpha}
 &&  \hspace{-0.98in}  \quad  \quad  \quad  \quad   \, \, 
{\rm Diag} \Bigl( {{1} \over { Q_{\alpha}}} \Bigr)  \,  \, = \, \, \,\,\,
    {{1} \over { \alpha}} \,  \cdot \,
   {}_4F_3\left( [{1 \over 5}, {2 \over 5}, {3 \over 5}, {4 \over 5}],[1, 1, 1],
    \, \, 5^{5} \cdot \, { t \over \alpha^2}\right)
  \\
 &&  \hspace{-0.98in}  \quad  \quad \quad  \quad     \quad   
 \, \, = \, \, \,
 {1 \over \alpha }\cdot  \, \Bigl(
 1 \,   \,   + 120 \,{t \over \alpha^2}   \,   \, + 113400 \,\left({t \over \alpha^2}\right)^{2}  \,  \, 
    + 168168000 \, \left(t \over \alpha^2\right)^{3} \,  \,   + \, \cdots \Bigr).
 \nonumber 
\end{eqnarray}
As remarked\footnote[1]{See, for instance, section 4.3 or Appendix G in~\cite{Diag2F1}.}
in previous papers~\cite{Diag2F1}, 
this series expansion result (\ref{diagQalpha}) is valid for $\, \alpha$
being {\em any arbitrary constant}, but {\em also any function} $\, \, f(t)$
of the product variable $\,t =\, x y z u v$ having a globally bounded series expansion\footnote[2]{For instance
algebraic series.}.
For $ \,\alpha = \, 1$,  the series corresponds to $ \,CY_1$ (see Table \ref{Ta:1}). 
The Calabi-Yau series (\ref{seriesCY128}), or  (\ref{CY128}), correspond to $\, \alpha \, = \,\sqrt{1\, -4 t}$.
Note that  the diagonal of $\, 1/Q_{128} \,$ can be written as
\begin{eqnarray}
  \label{diagQalphaq}
 &&  \hspace{-0.98in}  \quad \quad  \quad   \quad   \quad  \quad   \quad  \quad   \, \, 
    {\rm Diag} \Bigl( {{1} \over { Q_{128}}} \Bigr)  \,  \, = \, \, \,\,\,
   {\frac{1}{\sqrt{1 \,-4 t} }}  \cdot \,  {\rm Diag} \Bigl( {{1} \over { q_{128}}} \Bigr),  
\end{eqnarray}
where:
\begin{eqnarray}
  \label{q128}
 &&  \hspace{-0.98in}  \quad  \quad \quad \quad    \quad  \quad  \quad  \quad \, \, 
  q_{128} \,  \, = \, \, \,\,\,
     1 \, \,  \, \,   - (z+u+v)  \, \, \,\, \,   - x   \,  \,  \,   \, - {{y} \over { 1 \, -4 t}}.
\end{eqnarray}
Thus we see that the Calabi-Yau series (\ref{seriesCY128}), or  (\ref{CY128}), are, {\em up to a simple
  algebraic overall} $\, \sqrt{1 \,-4 t}$,  the diagonal of a {\em rational function} $\, 1/q_{128}\, $ of
{\em five} variables.  

\vskip .1cm

\section{Factorizable denominator $ \, Q$: examples with two factors}
\label{facto}

\subsection{The linear differential operators occurring in the $\chi^{(n)}$'s of the Ising model}
\label{factosub}

The rational functions, considered in the previous sections, are in the form $ \, 1/Q$,
where the denominator $ \, Q$ is a {\em non factorizable} multivariate polynomial,
the (minimal order) linear differential operator annihilating these diagonals, being irreducible,
contrary to the linear differential operators annihilating integrals corresponding to the
$\, n$-fold integrals $\, \chi^{(n)}$'s
of the susceptibility of the Ising
model~\cite{2004-chi3,2005-chi4,2008-experimental-mathematics-chi,2009-chi5,2010-chi5-exact,2010-chi6}.  These
integrals are very convoluted forms of algebraic fractions,
which have been shown to be diagonal of rational
functions~\cite{2013-rationality-integrality-ising},
but are far from being in the form $\, 1/Q$.

All the factors, occurring in the linear  differential operators for the $\chi^{(n)}$'s,
have been shown
to have  differential Galois groups
either  symplectic  or  orthogonal  (see~\cite{2014-Ising-SG} and references therein).
Furthermore, focusing on the blocks of factors (occurring in the
linear differential operators corresponding to the $ \, \chi^{(n)}$'s) which have 
a {\em unique factorization}, i.e. that write as, (e.g. for three factors)  $ \, L_n \, L_p \, L_q$,
it appears that if $ \, L_q$ is in the orthogonal group 
(resp. symplectic group), the left factor $ \, L_p$ is in the symplectic group (resp. orthogonal group) and so on,
in an alternating way.
\ref{chin} gives the situation for all the factors occurring in the linear differential operators corresponding 
to $\,\chi^{(3)}$, $ \, \cdots$, $\,\chi^{(6)}$. 

\vskip 0.1cm

This occurence of products, as well as {\em direct sums} of products, for factors of
the linear differential operators annihilating these diagonals, 
seems to be related to the fact that the denominator $ \, Q$ is {\em not} irreducible. In
the sequel, we address this case, {\em restricting for pedagogical reason, to only two factors}.
We will see, in these examples with {\em two} factors, that conjecture (\ref{conjmain}) is still valid.
The fact that the denominator $ \, Q$ is not irreducible
yields linear differential operators that are not irreducible:
thus one cannot simply introduce
a differential Galois group for these linear differential  operators. One has to consider the
differential Galois group
of {\em each factor} of these reducible linear differential operators. We will see that an
alternating symplectic/orthogonal structure seems to systematically occur (as \ref{chin}
shows for the $ \, \chi^{(n)}$'s). 

\vskip 0.1cm

\subsection{Foreword: two factors}
\label{forewo}

In all the examples of the form $\, 1/Q \, = \, 1/Q_1/Q_2\, $ displayed below, and many more not given
in this paper, we have
systematically obtained the following results.  If the number of variables,
and the variables for $\, Q_1$ and $\, Q_2$ 
are the {\em same}, the linear differential operator, annihilating the diagonal
of $\, 1/Q \, = \, 1/Q_1/Q_2$, is of the form
\begin{eqnarray}
  \label{foreword1}
  &&  \hspace{-0.99in} \quad \quad \quad \quad \quad \quad  \quad \quad \quad \quad \quad \quad \quad 
         \Bigl( {\cal L}^{(1)} \, \oplus \,  {\cal L}^{(2)} \Bigr) \cdot \, N, 
\end{eqnarray}
where the linear differential operators $\, {\cal L}^{(1)}\, $ and $\, {\cal L}^{(2)}\, $
are {\em homomorphic to the  linear differential operators annilating, respectively, the diagonal of}
$\, 1/Q_1\, $ and $\, 1/Q_2$,
and where the linear differential operator  $\, N$, at the right, is
some ``dressing''\footnote[5]{For $\, Q_1$ and $\, Q_2$  polynomial of three variables, the ``dressing''
  operator  $\, N$ seems to always have algebraic solutions. }. 
We have not found any simple interpretation of this ``dressing''.

In contrast, if the number of variables for $\, Q_1\, $ and $\, Q_2\, $ is different, the variables for (for instance)
$\, Q_2$ being a subset of the set of variables for $\, Q_1$, 
the linear differential operator, annihilating the diagonal of $\, 1/Q \, = \, 1/Q_1/Q_2$,
is of the form
\begin{eqnarray}
  \label{foreword2}
  &&  \hspace{-0.99in} \quad \quad \quad \quad  \quad \quad \quad \quad \quad \quad \quad \quad  \quad \quad 
     {\cal L}^{(1)} \, \cdot \, N, 
\end{eqnarray}
where the linear differential operator $\, {\cal L}^{(1)}\, $ 
is {\em homomorphic to the  linear differential operator annilating the diagonal of} $\, 1/Q_1$, 
and where, again, we have not found any
simple interpretation\footnote[1]{In particular,
  in the case where both operators have the same order and  the same singularities,
  the "dressing" operator $\,N$ in
  (\ref{foreword2}) is not homomorphic to the linear differential operator annihilating
  the diagonal of $ \, 1/Q_2$.}
 of this ``dressing'' linear differential  operator $\, N$.

\vskip 0.1cm

\subsection{First example}
\label{firstexa}

As a first example, consider the diagonal of $ \, 1/Q$ 
\begin{eqnarray}
  \label{Q68}
&&  \hspace{-0.98in}  \quad  \quad  \quad  \quad 
   Q \, \,\, = \, \, \,  \,\,
   1 \, \,\,\, -x-y -z \,\,\, \,+{y}^{2}-{z}^{2}+xy+yz \, \, \, +y{z}^{2}+x{z}^{2} +{z}^{3}, 
\end{eqnarray}
which is annihilated by an  {\em irreducible} order-ten linear differential operator $ \,  L_{10}$
which has a differential Galois group (included) in $ \, Sp(10, \, \mathbb{C} )$.
The  formal solution with the highest log-power at the origin is in $ \, \ln(t)^1$,
in agreement with $\, 3 \, = 1\, +2$ variables.

\vskip 0.1cm

Changing the monomial $ \, -y \,$ into $\,-2\,y \,$ in (\ref{Q68}),
one obtains a polynomial $ \, {\tilde Q}\, $ that, now, factorizes:
\begin{eqnarray}
  \label{Q69}
&&  \hspace{-0.98in}  \quad \quad \quad \quad \quad  \quad \quad \quad  \quad 
  {\tilde Q} \,\, \,= \, \,\, 
  \Bigl(1 \, -(x+y+z) \Bigr) \cdot \, (1 \, -y - z^2). 
\end{eqnarray}
The diagonal of $\,1/{\tilde Q}$ is, now, annihilated by an order-six linear
differential operator with the unique\footnote[5]{No direct sum factorization.} factorization
$ \, L_6 \,  = \, M_2 \cdot \, M_4$, where the  order-two linear differential operator $ \, M_2$
has a differential Galois group (included) in $\, Sp(2, \, \mathbb{C} )$, and where the order-four
linear differential operator
$ \, M_4$ has a differential Galois group (included)  in\footnote[1]{Its symmetric square has
  the rational solution $\,(175\, t \, +48)/t/(3125\, t^2 \, +1644\, t \, +128) $. Its symmetric cube
  (of order 20)  has a rational solution, $\, (3125\, t^2 \, +1644\, t \, +128)/t$. Its
  symmmetric fourth power (of order 35),
  has a rational solution, $\, (3125\, t^2 \, +1644\, t \, +128)/t^2$, and
  its symmmetric fifth power (of order 56),
   again, has a rational solution, $\, (3125\, t^2 \, +1644\, t \, +128)/t^2$, 
  suggesting a differential Galois group smaller than $ \, SO(4, \, \mathbb{C})$.}
$ \, SO(4, \, \mathbb{C})$. 
While the (left) linear differential differential operator $ \, M_4$ is {\em homomorphic}
to the order-two linear differential operator annihilating the diagonal 
of $ \, 1/\left( 1 \, -(x +y +z) \right)$,
we have {\em no interpretation}
for the "dressing" right factor $\, L_2$ {\em with respect to} the right factor of $\, {\tilde Q}$.

\vskip 0.1cm

This example shows that changing the coefficients in front of the monomials
to a value that makes the polynomial $\, Q \,$ 
factorizable leads to a {\em reducible} linear differential operator. A more subtle situation
is addressed in \ref{split}.

\vskip 0.1cm

The order-four linear differential operator $ \, M_4\, $ reads:
\begin{eqnarray}
\label{M4reads}
&&  \hspace{-0.99in}  \quad   \quad   \quad  \, \, 
M_4 \,  = \, \, 
(17\, t+104) \cdot \, (3125\, t^2+1644\, t+128) \cdot \, t^2 \cdot\, D_t^4
\nonumber \\
&&  \hspace{-0.99in}    \quad   \quad    \quad   \quad \, \, 
\, \, 
+5 \cdot \, (116875\, t^3+821922\, t^2+292400\, t+13312) \cdot \, t \cdot \, D_t^3
 \\
&&  \hspace{-0.99in}    \quad   \quad    \quad   \quad \, \, 
\, \,  \, \, 
+20 \cdot \, (82875\, t^3+633474\, t^2+140908\, t+2496)\cdot \, D_t^2
\nonumber \\
&&  \hspace{-0.99in}    \quad   \quad    \quad   \quad \, \, 
\, \,  \, \, 
+60 \cdot \, (19975\, t^2+173574\, t+16956)\cdot \, D_t \, \,  \, \, 
+6720\cdot \, (17\, t+186).
\nonumber 
\end{eqnarray}
The  formal solution of $ \, L_6$ (and of $ \, M_2$)  with the highest log-power
at the origin is in $\, \ln(t)^1$,
in agreement with the fact that $ \, {\tilde Q}$ (or the first factor in (\ref{Q69}))
depends on $\, 3 \, = \, 1 \, +2 \,$
variables. The ``dressing'' linear differential operator $ \, M_4$  has {\em no logarithmic} formal
solution around {\em all} the singularities, and all these formal 
solutions are globally bounded. According to Christol's conjecture~\cite{Chrisconj} 
these series should be diagonal of rational functions. According to our conjecture (\ref{conjmain})
the absence of logarithm in the formal solutions suggests that these globally bounded
series are diagonal of rational functions of only {\em two} variables, and, therefore,
should be {\em algebraic series}. In fact it is quite hard to see that these
globally bounded series are algebraic series. The calculation of the $p$-curvature of the 
linear differential operator $ \, M_4$ gives zero\footnote[9]{We thank A. Bostan for
  calculating the first $\, p$-curvatures, thus showing that these
$\, p$-curvatures are zero for $\, 7 \le p \le \, 73$.} for every prime $\, p$, 
in agreement with  algebraic series (according to Grothendieck-Katz
conjecture~\cite{Grothendieck-Katz}).
One of these four formal series is the globally bounded series
\begin{eqnarray}
\label{oneofthese}
&&  \hspace{-0.99in}  
{\cal S}_0 \,  = \, \, 
      -1 \,  +{{4239} \over {416}} \cdot \, t    \, -{{15802593} \over {173056}} \cdot \, t^2
   \, +{{11080683} \over {13312}} \cdot \, t^3  \, -{{86282904895} \over {11075584}} \cdot \, t^4
  \, +  \, \cdots 
\end{eqnarray}
which is an algebraic series\footnote[2]{One needs more than 1200 terms in the
  series (\ref{oneofthese}) in order to find this polynomial. It has been checked with 3010 terms series.
  More directly one can use the gfun command ``algeqtodiffeq'' on polynomial (\ref{POLoneofthese})
and {\em one actually recovers} the order-four linear differential operator $\, M_4$.}
solution of a polynomial
  \begin{eqnarray}
\label{POLoneofthese}
&&  \hspace{-0.98in}  \quad  \quad  \quad  \quad  \quad  \quad \quad  \quad  \quad  \quad \quad  \quad  \quad
P(t, \, \, y) \, \, = \, \, \, 0,
\end{eqnarray}
where the polynomial $\, P$ is of degree 38 in $\, t$ and of degree 30 in $\, y$, and is the sum of $\, 586$
monomials. The coefficient in $\, y^{30}$ reads:
 \begin{eqnarray}
\label{POLoneofthesey30}
&&  \hspace{-0.98in}  \quad \quad \quad   \quad   \quad  \quad  \quad
  2^{120} \cdot \, 13^{60}   \cdot \, t^{14} \cdot \, (3125\, t^2 \, +1644\, t \, +128)^{12}. 
\end{eqnarray}
 The discriminant  $\, \delta \, $ of this polynomial, in $\, y$, reads (the $\,P_n$'s
 are polynomial of degree $\, n$):
\begin{eqnarray}
  \label{Discrimoneofthese}
  &&  \hspace{-0.98in}  \quad  \quad \quad  \quad  \quad\quad 
\delta \,\, = \, \, \, P_{42}(t)^2 \cdot \, P_{36}(t)^2 \cdot \, P_{36}(t)^4 \cdot \, P_7(t)^2 \cdot \, Q_7(t)^{24}
 \nonumber \\
&&  \hspace{-0.98in}  \quad \quad \quad   \quad  \quad \quad \quad \quad \quad \quad \quad
\times \,  (3125\, t^2 \, +1644\, t \, +128)^{318} \cdot \, t^{378},
\end{eqnarray}
where $\, 3125\, t^2 \, +1644\, t \, +128\, = \, \, 0\, $ corresponds to the singularities\footnote[5]{
  The order-four linear differential operator $\, M_4\, $ also has $\, 17\, t \, + \, 104 \, = \, 0\, $
  as a singularity, but it is an {\em apparent singularity}. } of $\, M_4$.
Another solution of $\, M_4$ reads:
\begin{eqnarray}
  \label{Z2}
&&  \hspace{-0.98in}  \quad  \quad  \quad   
{\cal S}_1 \, \, = \, \, \, \,
 t \,\,  \, -{{4239} \over {416}}\, t^2 \, \, \, +{{3045} \over {32}} \, t^3
 \,\, \, -{{23833425} \over {26624}} \,  t^4 \,\, +{{17539767} \over {2048}}\, t^5
 \, \, \, \,  + \, \, \,\cdots 
\end{eqnarray}
The polynomial (\ref{POLoneofthese}) has  not only solution $\, y\, = \,  {\cal S}_0$, 
but also the solution:
\begin{eqnarray}
  \label{butalso}
&&  \hspace{-0.98in}  \quad  \quad   
  -{{21719} \over {10816}} \cdot \,  {\cal S}_0 \, \, +{{39414345} \over {4499456}}  \cdot \,  {\cal S}_1
  \nonumber \\
&&  \hspace{-0.98in}  \quad  \quad  \quad \quad  \quad 
\, \, = \, \, \, \,
   {{21719} \over {10816}} \, \, \, -{{253137} \over {21632}} \, t \, \,\, 
   +{{65140203} \over {692224}} \, t^2 \, \, \, -{{44617113} \over {53248}} \, t^3
   \, \,  \, \, + \, \, \, \cdots 
\end{eqnarray}
Note that $\, {\cal S}_1$ is solution of {\em another}
polynomial\footnote[1]{Again, one can use the gfun command
  ``algeqtodiffeq'' on polynomial $\, Q(t, \, \, y)$
  {\em to actually recover} the order-four linear differential operator $\, M_4$,
  but with different initial conditions.}
equation $\, Q(t, \, \, y) \, \, = \, \, \, 0$,
where $\, Q$ is also of degree 38 in $\, t$, and of degree 30 in $\, y$, and is sum of $\, 585$
monomials. 
The coefficient in $\, y^{30}$ reads:
 \begin{eqnarray}
\label{POLoneofthesey30}
&&  \hspace{-0.98in}  \quad  \quad  \quad  \quad \quad  \quad \quad  \quad
13^{30}   \cdot \, t^{14} \cdot \, (3125\, t^2 \, +1644\, t \, +128)^{12}. 
\end{eqnarray}
 The discriminant of $\, Q(t, \, \, y)$, in $\, y$, is
 similar to (\ref{Discrimoneofthese}): it has the same last three factors.
 
 \vskip .1cm
 
 {\bf Remark 6.1.} Other examples generalizing (\ref{Q69}), or
 similar to (\ref{Q69}), are displayed in \ref{otherfactofamillies}.
 These examples illustrate the fact that in the case of factorization
 of the denominator in a polynomial of {\em three} variables and another one
 of just  {\em two} variables,  the ``dressing'' right factor in the product (\ref{foreword2})
 corresponds to {\em algebraic solutions} which are so involved that they rule out
 any simple interpretation of the ``dressing'' right factor. 
 
 \vskip .1cm
 
\subsection{Second example}
\label{firstex}

Consider, now, the denominator $\, Q$ 
\begin{eqnarray}
\label{exempFORappD}
 Q\, \, = \,\,\, 
    \Bigl( 1 \,\,  -(x+y+z+u) \Bigr) \cdot \, \left( 1 \,\, -x y\, - z u \right), 
\end{eqnarray}
and the diagonal of the rational function  $ \, 1/Q$:
\begin{eqnarray}
\label{exempFORappD}
&&  \hspace{-0.98in} \quad  \,\,\,  
 {\rm Diag}\Bigl({{1} \over {Q}}\Bigr)
 \, \, = \,\,\,  \,  
 1 \,\,\,   \, +30\,t \,\, \,  +2958\,t^2 \, \,   +428652\,t^3 \, \,   +72819090\,t^4
 \, \,  \, \,\, + \, \, \, \cdots
\end{eqnarray}
This diagonal is annihilated by an order-five linear differential operator that
factorizes as $\, L_{5} \, = \, L_3 \cdot \, L_2$,
where $ \, L_3 \,$ has a differential Galois group (included) in $\, SO(3, \, \mathbb{C})$,
and $\, L_2$ has a differential Galois group (included) in $\, Sp(2, \, \mathbb{C} )$.
The linear differential operator $ \, L_3$ is homomorphic with the order-three linear
differential operator annihilating the diagonal 
of $ \, 1/\left( 1 \, \,  -(x +y +z +u) \right)$.
Note that the  formal solution of $ \, L_5$
with the highest log-power at the origin, is in $ \, \ln(t)^2$,
indicating that we should deal with a minimum number of $\, 4 \, =  \, 2 \, +2 \, $
variables, in agreement with (\ref{conjmain}). 

\vskip 0.1cm

Here again, while the left factor of the  linear differential operator $\, L_5$, and the linear  differential operator
annihilating the diagonal of the reciprocal of the first factor of $\, Q$,
are actually related by operator homomorphism, we have {\em no interpretation}
on the right factor $\, L_2$ with respect
to the right factor of $\, Q$. The linear differential operator $\, L_2$ has a differential Galois group (included)
in $\, Sp(2, \, \mathbb{C})$, and carries $\, \ln(t)^1$ as the formal solution with the highest log-power, meaning 
that it is, {\it per se}, given by the diagonal of a rational function with $\,\, 3 \, = \,\, 1 \, +2 \, $ variables:
\begin{eqnarray}
&&  \hspace{-0.98in}   \quad 
  {\rm sol} (L_2) \, \,  = \, \, \,
  {\frac{1}{\sqrt{1 \, -4 t}}} \cdot \, 
{}_2 F_1 \left( [{1 \over 4}, {3 \over 4}], [1], {{ 64} \over {9}}  \,t \right)
\\
&&  \hspace{-0.98in}   \, \,  \quad   \quad \,
\, \,  =  \, \,\,  \, 1 \,\,   \, +{{10} \over {3}} \, t \, \, +{{374} \over {27}} \, t^2 \,\, 
+{{15484} \over {243}} \, t^3 \,\,  +{{230210} \over {729}} \, t^4\,  \, +{{10919020} \over {6561}} \, t^5
\, \,  \, + \, \, \, \cdots 
\nonumber \\
&&  \hspace{-0.98in}   \, \,  \quad   \quad \,
 \, \,  =  \, \,
   {\frac{1}{\sqrt{1-4 t}}} \cdot \,
 {\rm Diag}\left( {\frac{1}{1 \, \, \, -c_1 \, x - c_2 \, y \,  -c_3/3 \cdot \,  z^{1/2} }}\right)
 \quad \hbox{where:} \,    \,\, \,\, \,\, \, \, \,
  \,  c_1 c_2 c_3^2  \, = \,   \, 1.
\nonumber 
\end{eqnarray}

\vskip 0.1cm

{\bf Remark 6.2.}
One should note that the second factor in the polynomial (\ref{exempFORappD}), {\em in itself}, depends
on {\em only two} variables $ \, x\, y \, $ and $ \, z\,u$. However, in the product of polynomials given in (\ref{exempFORappD}),
all the four variables contribute to the diagonal. This is not the kind of situation of section~\ref{forewo} where the two
polynomials should depend {\em separetly} on the same  number of the same variables for which the diagonal is not 
blind (see section~\ref{someRq}). The resulting linear differential operator $\, L_{5} \, = \, L_3 \cdot \, L_2$,
annihilating the diagonal of $\, 1/Q$  with the differential operator $\, L_2$ shows that {\em there is}
a product of polynomials $\, Q_1 \,Q_2$ where $\, Q_1$ and $Q_2$ depend
respectively on four and three variables and such that the diagonals of $\, 1/Q$ and of  $\, 1/Q_1/Q_2$ identify.

\vskip 0.1cm

{\bf Remark 6.3.}
The change of values of the coefficients in front of the monomials are irrelevant with respect  to the conclusions,
{\em as far as} this change of values does not change the factorization character of the denominator of the rational function. 
Changing in the polynomial $\,Q$ the coefficients in front of the
monomials $ \, x$, $ \, y$ and $ \, u$, to obtain the new polynomial
\begin{eqnarray}
  \label{Q67}
  {\tilde Q} \, \,=  \, \, \,
  \Bigl( 1 \, \, \, -(2x+7y+z-u) \Bigr) \cdot \, \left( 1 \, \, -x y \, -z u  \right), 
\end{eqnarray}
gives an order-seven linear differential operator that factorizes as $ \, L_{7} = \, N_3 \cdot \, N_4 \, $
where the differential Galois groups of $\, N_3$,  and $ \, N_4$, are respectively included
in $ \, SO(3, \, \mathbb{C}) \, $ and in $ \, Sp(4, \, \mathbb{C} )$.
Instead of an order-five  linear differential operator, we obtain,
for (\ref{Q67}), an order-seven  linear differential operator. 
The factorization scheme is still present and the linear differential  operators $ \, L_3$,
and $ \, N_3$, are homomorph once the variable $ \, t$ is
scaled: $ \, \, \, t \, = \, x y z u \,\, \longrightarrow \,\,  \, \,t \, = \, -14 \, x y z u$.
The formal solution of $ \, L_7$, with the highest log-power at the origin, is in $ \, \ln(t)^2$,
in agreement with $\, 4 \, = \, 2 \, +2$ variables. 

\vskip 0.1cm

In the previous examples, the denominator polynomial $\, Q$ factorizes
into two polynomials $ \, Q = \, Q_1\,Q_2$,
where the number of variables  occurring in each  polynomial is different. 
\ref{directsumtounique} addresses the case when the product $\, Q_1\,Q_2 \, $
switches from the situation
where both polynomials $\, Q_j$ carry the {\em same number}
of variables, to the situation where one of them has a smaller number of variables.
\ref{rat4to3} considers a denominator $ \, Q = \, Q_1\,Q_2$, depending on one parameter $\,b$,
and where the polynomials $\, Q_1$ and $ \, Q_2$ 
depend, respectively, on 4 and 3 variables. We address in \ref{rat4to3} the situation where,
for one particular value of the parameter $\,b$, the polynomial $\, Q\, $ reduces to a polynomial
depending on {\em only} three variables instead of the four variables we started with.

\vskip 0.1cm

\section{Rational versus algebraic functions and powers of rational functions}
\label{powers}

In this section we consider some examples of diagonals of {\em algebraic} functions,
square root of  rational functions, and  powers of rational functions.

\vskip 0.1cm

\subsection{A modular form example}

Let us consider the order-two linear differential operator $\, L_2$
annihilating the hypergeometric function
$ \, _{2}F_1 \left( [1/12, 5/12], \,  [1],\,  1728\, t \right)$:
\begin{eqnarray}
&&  \hspace{-0.98in}  \quad \, \, \, \quad \quad  \quad  \quad 
  L_2  \,  \, = \, \, \,\,
  (1 -1728\, t) \cdot \, t  \cdot \, D_t^2 \,\,\, +(1 -2592\, t) \cdot \, D_t \,\, \, -60.
\end{eqnarray}
This order-two linear differential operator $\, L_2$ is homomorphic to its adjoint
provided\footnote[2]{Alternatively one can use the ``Homomorphisms'' command
  of DEtools in Maple on the {\em symmetric square} of $\, adjoint(L_2)$ and $\, L_2$.}
one considers a simple square-root algebraic extension:
\begin{eqnarray}
  &&  \hspace{-0.98in}  \quad \quad   \quad \quad \quad  \quad
  (1 -1728\, t)^{1/2} \cdot \, adjoint(L_2)
  \,  \, \, = \, \, \,  \,
  L_2 \cdot \, (1 -1728\, t)^{1/2}.
\end{eqnarray}
This  hypergeometric function corresponds to the diagonal 
\begin{eqnarray}
  \label{2F2diag}
&&  \hspace{-0.98in}  \quad \quad  \quad \quad \quad \quad  \quad  \quad 
  _{2}F_1 \left( [{1 \over 12}, {5 \over 12}], [1], \, \, 1728\, t \right)
  \,  \, =\, \, \,  {\rm Diag} \left(  R \right), 
\end{eqnarray}
of the {\em algebraic function} $\, R$
(depending on {\em two} variables with $\, t \, = \, x y$) given by:
\begin{eqnarray}
&&  \hspace{-0.98in}  \quad \quad  \quad \quad \quad \quad  \quad  \quad 
R \, \,= \, \, \,
\Bigl( \sqrt{1 \, -1728 \, x y}\, \,\, \,  -\sqrt{432} \cdot \, (x-y) \Bigr)^{-1/6}. 
\end{eqnarray}
According to conjecture (\ref{conjmain}), we would like to be able to write (\ref{2F2diag}),
not as a diagonal of an {\em algebraic} function of {\em two} variables, but as a
{\em rational} function of {\em three} variables.  

\vskip 0.1cm

The procedure in~\cite{Denef-lipschitz-1987} by Denef and Lipschitz gives,
on this example, a {\em rational} function $ \, {\tilde R}$
depending on {\em four} variables $\, (x,\, y,\, u, \, v)\, $ such that (\ref{2F2diag}), the diagonal
of $\, {\tilde R}$ on the four variables $\, (x,\, y,\, u, \, v)$, identifies
with the diagonal of the  {\em algebraic} function $\, R$ on the variables $\, (x,\, y)$. The
rational function $\, {\tilde R}$ has the expression:
\begin{eqnarray}
 &&  \hspace{-0.98in}  \quad \quad \quad \quad \quad \quad \quad \quad  \quad 
  {\tilde R} \,  \, = \,\,
  {\frac{P(x, \,y, \,u,\, v)}{{\tilde Q}(x,\, y, \, u) \cdot \, {\tilde Q}(x, \, y, \, v)}}. 
\end{eqnarray}
The polynomial $\, P(x,\, y,\, u,\, v)$ contains 2848 terms, and $\, {\tilde Q}$ reads:
\begin{eqnarray}
 &&  \hspace{-0.98in}  \quad \, 
{\tilde Q}(x,y,u)\,  \,=\,\,   \,
  432 \cdot \, u \cdot \, ( 1\,  \,+u))^{12} \cdot \, (x+y)^{2}
  \, \, +24 \cdot \,\sqrt {3} \cdot \, (1+u)^{6} \cdot \, (x-y)
  \\
 &&  \hspace{-0.98in}  \quad \, 
  - \left( 2+u \right) \,  \left( 1+u+{u}^{2} \right) \, \left( 2+2\,u+{u}^{2} \right)
  \,\left( 3+3\,u+{u}^{2} \right)
  \nonumber
 \left( 1+2\,u+5\,{u}^{2}+4\,{u}^{3}+{u}^{4} \right).   
\end{eqnarray}
One remarks that the denominator of $ \,{\tilde R}\, $ is in a factorized form by {\em construction},
and that each factor {\em does not contain} all the variables. 
For an algebraic function with $n$ variables $ \,(x_1, \, x_2, \, \cdots, \, x_n)$,
the Denef and Lipschitz procedure~\cite{Denef-lipschitz-1987} will give, as denominator, the product
$ \, Q(x_1, \, x_2,  \, \,\cdots,  \, \,x_n, \,  u_1) \cdot \, Q(x_1,x_2,
\, \cdots, \, x_n, u_2) \, \cdots \, \, Q(x_1, x_2,  \,\cdots,  \, x_n, u_n)$.

\vskip 0.1cm

For the hypergeometric function $\, _{2}F_1 \left( [1/12, 5/12], [1],\,  1728\, t \right)$,
the procedure sketched in \ref{binomialVSfactorial}, is not applicable. 
However, one may imagine that, introducing a pullback, the general term may be cast
in the appropriate form of {\em ratio of factorials}. 
One actually has (with a double expansion, index summation change,
and summation of the inner sum):
\begin{eqnarray}
&&  \hspace{-0.98in}    \quad  \quad  \quad \,  \quad \quad 
_{2}F_1 \left( [{1 \over 12}, {5 \over 12}], [1], \, \, 1728\, t  \cdot \,(1 \, -432\,t) \right)
\nonumber \\ 
&&  \hspace{-0.98in}   \quad  \quad  \quad   \, \,  \, \,  \, \quad  \quad \,\,\,
\,\, = \,  \,\, \, \,
\sum_{n=0}^{\infty} \,   \sum_{k=0}^{n} \,  \, \, 
(-4)^k  \cdot \, {k \choose n-k}  \cdot \, 
   {\frac{\left( 1/12 \right)_k \left( 5/12 \right)_k }{\left( 1 \right)_k \,  k!}}   \cdot \,
   \left( -432\, t \right)^n
  \nonumber    \\
 &&  \hspace{-0.98in}   \quad  \quad  \quad  \,  \quad \quad \quad \quad  \quad
\,   \, =  \, \, \, \,  \, 
\sum_{n} \,
{\frac { \left( 6\,n \right) !}{  n!\, \left( 2\,n \right) !\,\left(3\,n \right) !}} \cdot \, {t}^{n}.
\end{eqnarray}
The last multinomial form  leads to (with $\, t\, = \, x y z$):
\begin{eqnarray}
  \label{agreement}
&&  \hspace{-0.98in}   \, \, \, 
  _{2}F_1 \left( [{1 \over 12}, {5 \over 12}], [1], \,  \, 1728\, t \cdot  \, (1-432\,t) \right)
  \,  \, =\, \,  \,  \,   \, 
{\rm Diag} \left( {\frac{1}{1 \,  \,  \,  -x \, \, -y^{1/2} \, \, - z^{1/3} }} \right).   
\end{eqnarray}
The order-two linear differential operator, annihilating the diagonal (\ref{agreement}),
is not only {\em homomorphic} with its adjoint, it is self-adjoint. 
Its differential Galois group  is (included) in $ \, Sp(2, \, \mathbb{C} )$,
and the formal solution of the linear differential operator
with the highest log-power carries a $ \, \ln(t)^1$,
in agreement with the $\, 3 \, = \, 1 \, + 2 \, $  variables of the diagonal
in (\ref{agreement}), and conjecture (\ref{conjmain}). 

\vskip 0.1cm

{\bf Remark 7.1.} From (\ref{agreement}) we easily get that
\begin{eqnarray}
  \label{agreement2}
&&  \hspace{-0.98in}   \, \quad \quad  \quad \, \, 
  _{2}F_1 \left( [{1 \over 12}, {5 \over 12}], [1], \,  \, 1728\, t  \right)
  \,  \, = \, \,  \,  \,   \, 
{\rm Diag} \left( {\frac{1}{1 \,  \,  \,  -\alpha \cdot \, x \, \, -y^{1/2} \, \, - z^{1/3} }} \right),    
\end{eqnarray}
where:
\begin{eqnarray}
\label{wherealpha}
&&  \hspace{-0.98in}  \quad    \quad   \, \, \,
\alpha \, \, = \, \, \,  {{1 \, -(1\, -1728 \, t)^{1/2} } \over {864 \, t }}
\, \, = \, \, \, \,
1 \, \, +432\, t \,\, +373248\, t^2 \,\, +403107840\, t^3
\nonumber \\
&&  \hspace{-0.98in}  \quad \quad   \quad   \quad  \, \, \,
\, +487599243264\, t^4 \, \, \, + \, \, \, \cdots
\quad  \quad \quad \hbox{where:} \quad \quad  \quad  
t \, = \, \, x \, y \, z.
\end{eqnarray}
{\em Up to an algebraic function of the product} $\, t \, = \, \, x \, y \, z$,
we thus have a representation of (\ref{2F2diag}) as a diagonal of a rational function
of {\em three} variables, or $\, N$-th root of variables,  $\, x$, $\, y$ and $\, z$.

\vskip 0.1cm

{\bf Remark 7.2.} Note that the pullback in   (\ref{agreement}) is precisely the one that matches
 (\ref{agreement}) with one of a {\em modular form} of Appendix B in~\cite{Heun-jmm-2020}:
\begin{eqnarray}
&&  \hspace{-0.98in}  \quad \quad\,  
_{2}F_1 \left( [{1 \over 12}, {5 \over 12}], [1], \, 1728 \cdot \, t  \cdot \,(1 \, -432\,t) \right)
   \, \, \,  = \, \, \,
_{2}F_1 \left( [{1 \over 6}, {5 \over 6}], [1], \, 432 \, t \right)
\nonumber \\
 &&  \hspace{-0.98in}  \quad \quad \quad  \quad  \,  \,  \, 
 \, \, = \, \, \,\,\,
1 \,\, \,\, +60 \, t \,\,  +13860 \, t^2 \,\,  +4084080\, t^3 \,\,  +1338557220\, t^4
\, \,  \, \,\,  + \, \cdots 
\end{eqnarray}

\vskip .1cm

\subsection{From the LGF of 3-D s.c to Calabi-Yau number 69}
\label{algCONNrat1}

Recall the rational function $\, 1/Q$,  depending on four variables,
and corresponding\footnote[5]{This Heun function (\ref{Heun})
  can be written as pullbacked $\, _2F_1$ hypergeometric function with algebraic pullbacks (see equations
  (12), (13), (14) in~\cite{Heun-jmm-2020}).} to  
the lattice Green function of the 3-dimensional simple cubic (\ref{LGFsc}), 
\begin{eqnarray}
  \label{QHeun} 
&&  \hspace{-0.98in}  \quad \quad\,   \quad \quad \quad
Q  \, \,= \,  \, \,\, \, 
1 \, \, \, \,  - \Bigl(x + y + z \, \,  \,  \,  + u \cdot \, (x y + x z + y z) \Bigr),
  \\
\label{Heun}  
&&  \hspace{-0.98in} \quad \, \quad   \,  \quad \quad  \quad
{\rm Diag} \Bigl( {{1} \over { Q}} \Bigr) \,  \, = \,  \,  \, \, 
{\rm HeunG} \left( {1 \over 9}, {1 \over 12}, {1 \over 4}, {3 \over 4}, 1, {1 \over 2}, \,\, 4 \cdot \,t \right)^2
 \\
&&  \hspace{-0.98in}  \quad  \quad  \quad    \quad \quad \quad \quad \,  \, 
 \, = \,  \,  \, \, \,  \,
 1 \,  \, \, +6\,t \,\,  +90\, t^2 \,  \, +1860\, t^3 \, \, +44730\, t^4 \,\,   +1172556\, t^5
 \, \,\, \, + \, \cdots
   \nonumber 
\end{eqnarray}
Considering, now, the diagonal of the reciprocal of the {\em square root}
of the polynomial (\ref{QHeun}) 
\begin{eqnarray}
\label{Qhalf3dimsc}
&&  \hspace{-0.98in}  \quad \quad\,  \quad \quad  \quad \quad \quad
  Q \,  \,\, = \,  \, \,
  \Bigl( 1 \,  \,  \,\,  -  \Bigl(x + y + z  \, \, \,\, 
   + u  \cdot\, ( x y + x z + y z) \Bigr) \Bigr)^{1/2}, 
\\
&&  \hspace{-0.98in} \quad \quad \quad \quad  \quad \, \,  \quad \quad
{\rm Diag} \Bigl( {{1} \over { Q}} \Bigr) \,  \, = \,  \,  \, \, \, 
1 \, \,\,\,  +{9 \over 4} \,t \,\,\,  +{1575 \over 64} \, t^2 \, \, \, +{107415 \over 256} \, t^3
   \,    \, \,\, \, + \, \cdots
   \nonumber 
\end{eqnarray}
one obtains an annihilating irreducible order-four linear differential operator, with
a differential Galois group included in the symplectic group $\, Sp(4, \, \mathbb{C} )$,
and with $ \, \ln(t)^3 \,$ highest log-power formal solution. According to
conjecture (\ref{conjmain}), a rational function $ \, 1/Q$, 
{\em depending on $\, 5= \, 3 \, +2\, $ variables} should exist. It actually reads:
\begin{eqnarray}
  \label{toCY69}
&&  \hspace{-0.98in}  \quad  \quad   \quad  \quad  \quad  \quad  
  Q  \, \, = \,  \,\,\,
   1 \, \,\,\,
   - \left(x + y + z \, \,\,  + u \cdot  \, ( x y + x z + y z) \,  \,  \,  + {v^{1/2} \over 4} \right).
\end{eqnarray}
The diagonal of $\, 1/Q$, with $\, Q\, $ given in (\ref{toCY69}), identifies with the diagonal
of $\, 1/Q$, with $\, Q \, $ given in (\ref{Qhalf3dimsc}).
With the rescale $\, v^{1/2}/4 \, \rightarrow \, v^{1/2} \, $
(i.e. $\, t \, \rightarrow \,  16\, t$) 
it corresponds to the Calabi-Yau series (number 69  in~\cite{TablesCalabi-2010}):
\begin{eqnarray}
 \label{CY69}
&&  \hspace{-0.98in}  \quad  \,   \,  \quad \quad \quad \quad\quad
CY_{69}  \, \,= \,  \,\,\,
   \sum_{n=0}^{\infty } \, {\frac{(4n)!}  {n!^2 \, (2n)!}}
   \cdot \, \sum_k^{n} {n \choose k}^2 {2k \choose k} \cdot \, t^n
  \\
&&  \hspace{-0.98in}  \quad  \,   \,  \quad  \quad \quad \quad \quad \quad \quad \quad
 \, \,= \, \, \, \,\, 
1 \, \,\,\, + 36\, t  \, \,\, + 6300\,t^2 \,\,  +1718640\, t^3
   \,  \,  \, \, \, + \, \cdots
\nonumber
\end{eqnarray}
 
\vskip 0.1cm

\subsection{From the LGF of 3-D f.c.c to Calabi-Yau number 14}
\label{algCONNrat2}

The diagonal of the rational function $\, 1/Q$ with denominator
\begin{eqnarray}
&&  \hspace{-0.98in}  \quad  \quad \, \quad \quad  \quad\quad  \quad \quad  \,  \, 
 Q \, \, = \, \,\,\,
   1 \, \,\, \,\, - \left(x + y + z \, \, \, + u \cdot \, (y +z) \right),
\end{eqnarray}
reads:
\begin{eqnarray}
\label{sol3fccSans pullback}
&&  \hspace{-0.98in}  \,  \, \quad  \quad \quad  \quad \quad  \quad \quad  \, \, 
{\rm Diag}\Bigl( {{1} \over { Q}} \Bigr) \, \, = \, \, \,\,
_{3}F_2 \left( [{1 \over 2}, {1 \over 3}, {2 \over 3}], [1, 1], \, 108\, t \right). 
\end{eqnarray}
This hypergeometric series (\ref{sol3fccSans pullback})  actually
occurs\footnote[2]{The solution of the LGF of 3-D f.c.c is given in (\ref{On4vars}),
  a pullbacked form of (\ref{sol3fccSans pullback}).}  
for the lattice Green function of the 3-dimensional {\em face-centred cubic} lattice.
Let us, now, consider a {\em rational number} $\, \alpha$,
and let us introduce the {\em algebraic function} $\, 1/Q(\alpha)$: 
\begin{eqnarray}
\label{Qalpha13}
&&  \hspace{-0.98in}  \quad  \quad  \quad \,  \quad  \quad \quad  \quad  \,   \,   \,   \, 
  Q(\alpha) \, \, = \,\, \, 
  \Bigl( 1 \, \,  \, \, -  \left(x + y + z \, \, \, + u \cdot  \, (y+z) \right) \Bigr)^{\alpha}.
\end{eqnarray}
One actually obtains for the diagonal of $\, 1/ Q(\alpha)$:
\begin{eqnarray}
  \label{4F3homo}
 &&  \hspace{-0.98in} \quad \quad \, \quad \, \,   \, \quad  \,   \, 
    {\rm Diag}\Bigl( {{1} \over {Q(\alpha)}} \Bigr)
    \,  \, = \, \, \,  \, 
_{4}F_3 \left( [{1 \over 2}, {\alpha \over 3}, {\alpha \over 3} +{1 \over 3}, {\alpha \over 3} +{2 \over 3}], \, 
[1, 1,1],\,\,  108\, t \right).  
\end{eqnarray}
For $\, \alpha= \, 1/2$, the diagonal given in (\ref{4F3homo}) 
\begin{eqnarray}
\label{seriesQhalf}
&&  \hspace{-0.98in} \quad \quad   \quad  \, \quad \, 
  {\rm Diag}\Bigl( {{1} \over {Q(1/2) }} \Bigr) \,  \, = \,  \,  \, \,
  \,  1 \, \, \,  +{\frac {15}{4}} \, t \, \,  \,\, +{\frac {31185}{256}} \, {t}^{2}
  \,\,  \, +{\frac {6381375}{1024}} \, {t}^{3}
  \, \,  \, \, + \, \, \,  \cdots
\end{eqnarray}
is annihilated by an irreducible order-four linear differential operator, homomorphic to its adjoint,
with $ \, \ln(t)^3 \, $ as the maximum power 
in its formal solutions around the origin. This globally bounded power series (\ref{seriesQhalf})
is actually the diagonal of a rational function $ \, 1/{\tilde Q}(1/2)\, $ with $\, 5 \, = \,3 \, +2 \, $
variables where the polynomial denominator reads:
\begin{eqnarray}
  \label{Qvthird}
&&  \hspace{-0.98in} \quad \quad \quad \, \, \,   \quad \quad \quad  \quad 
  {\tilde Q} (1/2) \, \, = \,\, \, \,  \,
 1 \, \, \,  \, -  \left(x + y + z \, \,\,  + u \cdot  \, (y+z) \, \,  \, +{v^{1/3} \over 4} \right). 
\end{eqnarray}
The diagonal of $\, 1/{\tilde Q} (1/2) \,$ identifies with the series given in (\ref{seriesQhalf}).
The  polynomial (\ref{Qvthird}) (with the change $ \, v^{1/3}/4 \,\rightarrow \,v^{1/3}$) 
is the same polynomial as the one given in Table 1, which corresponds to Calabi-Yau equation number 14.

\vskip 0.1cm

{\bf Remark 7.3.}
The relations between the square root of the rational function of the LGF of simple cubic
(resp. face centred cubic) and the rational function corresponding to Calabi-Yau number 69
(resp. Calabi-Yau number 14) are similar to the 
connection~\cite{guttmann-2010,guttmann-prellberg-1993,glasser-montaldi-1993}
between the LGF of the diamond and the LGF of the simple cubic lattices (see \ref{conndiasc}).
Other examples are given in \ref{sqrtR2CY}.

\vskip 0.1cm

Let us return to  (\ref{4F3homo}), and let us denote
$ \, L_4(\alpha)$ the order-four linear differential operator
annihilating the diagonal (\ref{4F3homo}).
For generic ({\em non integers}) rational values of $\, \alpha$,
the linear differential operator $\, L_4(\alpha)$
is not\footnote[1]{Using the Homomorphisms command in Maple.}
homomorphic to its adjoint.
However, for half integer values of $\, \alpha = \,  m \, +1/2$, $ \, L_4(\alpha) \, $
is irreducible, and is homomorphic to its adjoint. 
The diagonal of $ \, 1/Q(\alpha)$ reads:
\begin{eqnarray}
  \label{82}
&&  \hspace{-0.98in}  \quad  \,  \quad \quad   \quad  \quad  \quad  \quad   \quad  
  {\rm Diag}\Bigl( {{1 } \over { Q(\alpha) }} \Bigr)
  \,  \,= \, \, \, \, 
 \sum_{n=0}^{\infty} \, \left(\alpha\right)_{3n} \cdot \, {\frac{(2n)!}{ n!^5}}  \cdot \,  t^n. 
\end{eqnarray}
For  half integer values of  $\, \alpha$, $\, \alpha = \, m\, +1/2$, the Pochhammer symbol
$\left(\alpha\right)_{3n}$
can be written as multiple argument of factorials and, thus, (\ref{82}) reads:
\begin{eqnarray}
  \label{Qm1sur2}
&&  \hspace{-0.98in}  \,  \quad  \quad \quad \quad   \quad  
{\rm Diag} \Bigl( {{1 } \over { Q(m +1/2) }} \Bigr) \,  \, = \, \, \, 
   \sum_{n=0}^{\infty} \, {\frac{(6n+2m)!}{(3n+m)!}}\, {\frac{m!}{(2m)!}}\,  {\frac{(2n)!}{ n!^5}}
   \cdot \,   {\frac{t^n}{64^n}}. 
\end{eqnarray}
For $ \, \alpha = \, 3/2$, one has:
\begin{eqnarray}
&&  \hspace{-0.98in}  \quad \quad  \quad  \quad  \quad     \quad  \quad    \quad  
{\rm Diag}  \Bigl( {{1 } \over { Q(1 +1/2) }} \Bigr) 
\, = \, \,  \,  \, 
   {\frac{1}{2!}} \cdot \, \sum_{n=0}^{\infty} \, {\frac{(6n+2)!}{(3n+1)!}}\,
   {\frac{(2n)!}{ n!^5}}  \cdot \,   {\frac{t^n}{64^n}}
\nonumber   \\
&&  \hspace{-0.98in}  \quad  \quad  \quad  \quad   \quad   \quad \quad  \qquad \quad \,\,
\,=\, \,  \,  \,
\sum_{n=0}^{\infty} \,  \, (6n+1) \cdot \,
    {\frac{(6n)!}{(3n)!}}\,   {\frac{(2n)!}{ n!^5}}  \cdot \,   {\frac{t^n}{64^n}}.
\end{eqnarray}
With the homogeneous derivative $\, \theta \,= \,  t \, D_t$, one obtains
\begin{eqnarray}
 &&  \hspace{-0.98in}   \,\,  \quad \quad  \quad  \quad   \quad  \quad  \quad 
  {\rm Diag}\left( {1 \over Q(3/2)} \right ) \,  \,= \, \, \,\,
  \left( 6 \cdot \,t \, D_t \, \, +1 \right) \cdot \,{\rm Diag}\left( {1 \over Q(1/2)} \right ), 
\end{eqnarray}
meaning that the order-four linear differential operators
$\, L_4(\alpha=3/2) \,$ and $\, L_4(\alpha=1/2)\, $
are homomorphic with an {\em order-one intertwinner}. 
The linear differential operators annihilating the diagonal of
$ \, 1/Q(1/2)$, or  $ \, 1/Q(3/2)$, have $ \, \ln(t)^3$
as maximum power in the formal solutions at the origin. In terms of rational functions, this gives
the relation: 
\begin{eqnarray}
 &&  \hspace{-0.98in}   \,\,    \quad \quad \quad   \quad  \quad  \quad  \quad  \quad  \quad 
  {\rm Diag}\left( {1 \over Q(3/2)} \right ) \,  \, = \, \, \,\,\,
  {\rm Diag}\left( {{\tilde Q}(1/2) \, \, +6 \, x \over {\tilde Q}(1/2)^2} \right ).
\end{eqnarray}

\vskip 0.1cm

The simple order-one intertwinner in the homomorphisms of $\, L_4(\alpha)\, $ and $\, L_4(\alpha+1)\, $
also occurs for integer values of $\, \alpha$.
In the sequel, we present this situation.

\subsection{Power of rational functions and homomorphisms}
\label{power}

Consider the polynomial $\, Q$ dependent on the parameter $\mu$:
\begin{eqnarray}
 &&  \hspace{-0.98in}   \quad   \quad  \quad  \quad  \quad   \quad \,\,  \,   \quad     \quad   \quad  
  Q \,\,= \,\,\,\,  \,  
     1\,\, \,  \,  \, -\left(x+y \, \, +z^2 \,\,\,  +\mu \cdot \, x z \right). 
\end{eqnarray}
The diagonal of $\, 1/Q$ is annihilated by an irreducible order-four
linear differential operator $\,L_4$, and the diagonal of the square $\, 1/Q^2 \, $
is annihilated by an irreducible order-four linear differential operator $\,N_4$. These two
linear differential operators are actually {\em homomorphic}~\cite{Plea}, giving
\begin{eqnarray}
 \label{intwQnQW}
 &&  \hspace{-0.98in}   \quad   \quad  \quad  \quad  \quad  \quad   \quad  \,  \, \, \, \, \,   \,  \, \, \,   \,  \, 
  {\rm Diag}\Bigl( {{1} \over {Q^2}}\Bigr) \,  \,= \,  \, \,
   W_3(\mu) \cdot \, {\rm Diag}\Bigl( {{1} \over {Q}}  \Bigr) , 
\end{eqnarray}
where $ \, W_3(\mu)$ is an intertwiner of order three.
For $ \mu=\, 0$, the order-three linear differential operator $ \,W_3$ reduces to an order-one  intertwiner:
\begin{eqnarray}
&&  \hspace{-0.98in}   \quad   \quad  \quad \quad   \quad  \quad  \quad  \quad \quad \, \,   \,  \, 
  W_3(\mu=0) \, \, = \,\,\,  \, \,
      {5 \over 2}   \cdot \, t  \cdot \, D_t\, \, \,  + 1.   
\end{eqnarray}
Let us take $\mu=\,0$, and generalize the polynomial $\, Q\, $ to  
(with $r$ a positive integer):
\begin{eqnarray}
 &&  \hspace{-0.98in}   \quad   \, \quad \quad   \quad   \quad  \quad  \quad  \quad  \quad   \quad 
  {\tilde Q} \,\,=\,\, \,\, \, 
  1\, \, \, \,  -\left(x+y\,\,\,   +z^r \right). 
\end{eqnarray}
The diagonal of $\, 1/{\tilde Q} \, $ is annihilated by an order $\, 2\,r$
linear  differential operator $\, L_{2r}^{(1)}$.
The linear  differential operator, annihilating the diagonal of $\,1/{\tilde Q}^n$,
is also of order $\, 2\,r$. Let us call it $\,L_{2r}^{(n)}$.
The linear  differential operator $\, L_{2r}^{(n)}$ is homomorph with the
linear differential operator $\, L_{2r}^{(n+1)}$, giving:
\begin{eqnarray}
  \label{intwQnQ}
 &&  \hspace{-0.98in}     \quad   \quad  \quad     \quad  \quad  \quad 
  {\rm Diag} \Bigl({{1} \over {{\tilde Q}^{n+1}}}  \Bigr)  \, \, = \,\, \,\,
    \left( {2 r +1 \over r \,n} \cdot \, t\,D_t \,\, +1 \right)
    \cdot \, {\rm Diag}\Bigl({{1} \over { {\tilde Q}^{n}}}  \Bigr) , 
\end{eqnarray}
i.e.
\begin{eqnarray}
&&  \hspace{-0.98in}     \quad   \quad   \quad   \quad  \quad  \quad 
{\rm Diag}\Bigl({{1} \over {{\tilde Q}^{n}}}  \Bigr) \, \, = \, \,\,\,\,
   \prod_{k=1}^{n-1} \left( {2 r +1 \over r \,k} \cdot \, t\,D_t \,\, +1 \right)
   \cdot \, {\rm Diag} \Bigl({{1} \over { {\tilde Q}}}  \Bigr). 
\end{eqnarray}
\ref{classratfun} shows the conditions for this property to happen.

\vskip 0.1cm

As far as conjecture (\ref{conjmain}) is concerned,  the rational functions $\,1/{\tilde Q}$ and $\,1/{\tilde Q}^n$,
have the same minimum number of variables, therefore the formal solutions
of the corresponding linear differential operators {\em have  the same maximum log-power}. 

\vskip 0.1cm

{\bf Remark 7.4.} 
For an order-$q$ linear differential operator $\, L_q$, annihilating $\, {\rm Diag}(1/Q)$,
one can easily show, from (\ref{intwQnQ}), that 
\begin{eqnarray}
&&  \hspace{-0.98in}   \quad   \quad  \quad  \quad   \quad   \quad  \quad   \quad  \quad  \quad 
   \sum_{n=1}^{q+1} \, p_n(t) \cdot \, {\rm Diag}\left( {1 \over Q^n} \right)
   \,\, =  \,\,\,\, 0, 
\end{eqnarray}
where the $ \, p_n(t)$'s are polynomials in $ \, t$.
In fact, such relation occurs also for the diagonal of $ \, 1/Q \, $ satisfying
the more general homomorphisms (\ref{intwQnQW}).

\vskip 0.1cm

\subsection{From reciprocal of square root to rational functions}
\label{Fromto}

Let us recall\footnote{See also examples in \ref{sqrtR2CY}.}  the diagonals
$\, {\rm Diag}(1/Q^{1/2})$ given in (\ref{Qhalf3dimsc})
and  (\ref{seriesQhalf}), where $Q$ depends on four variables, and which identify
with the diagonals of $\, {\rm Diag}(1/Q_{eq})$ given in (\ref{toCY69}) and (\ref{Qvthird})
\begin{eqnarray}
\label{sqQisQeq}
&&  \hspace{-0.98in}  \quad  \quad  \quad  \quad  \quad  \quad  \quad  \quad  \quad 
      {\rm Diag}\left( { 1 \over Q^{1/2} } \right)
      \,  \,=\, \,\,
      {\rm Diag} \left( { 1 \over Q_{eq} } \right), 
\end{eqnarray}
where 
\begin{eqnarray}
  \label{Qeq}
&&  \hspace{-0.98in}  \quad  \quad  \quad  \quad   \quad  \quad   \quad  \quad    \quad  \quad  
Q_{eq}\, \,  =  \,\,\,\,   Q \,\,\, -{1 \over 4}\, u^\alpha, 
\end{eqnarray}
 with an additional fifth variable $\,u$, and where $\alpha$ is a rational (not integer) number.  
 Let us show how this occurs with the procedure of \ref{binomialVSfactorial}, and show
 whether the form (\ref{Qeq}) is general.

With the multivariate polynomial  $\, Q\, $
\begin{eqnarray}
  \quad      \quad   
  Q  \, \,\,  = \, \, \, \, \,
  1 \, \, \,  \, - \left( T_1 \, + T_2 \, \, + \cdots \, +T_n \right), 
\end{eqnarray}
where the $\, T_j$'s are monomials, the expansion, around the origin of the rational function $\, 1/Q\, $ reads:
\begin{eqnarray}
\label{expQsec73}
&&  \hspace{-0.98in}  \quad  \quad  \quad \quad  \quad  \quad 
{{1} \over {Q}} \,  \, = \, \,\,\,
\sum \,\,   {\frac{(k_1 +k_2 + \, \cdots \, + k_n)!}{k_1! \,  k_2! \, \cdots \, k_n!}}
\cdot  \, T_1^{k_1} \,T_2^{k_2} \,\cdots \, T_n^{k_n}. 
\end{eqnarray}
With the square root of the polynomial $\, Q$, the expansion of $\, 1/Q^{1/2}\, $ is
\begin{eqnarray}
&&  \hspace{-0.98in}   \quad  \quad 
{{1} \over {Q^{1/2}}} \,  \, = \, \,\sum \,\,   {1 \over 4^{k_0} } {(2\,k_0)! \over k_0!^2 }\,
\left( T_1 \, + T_2 \, \, + \cdots \, +T_n \right)^{k_0}
\nonumber \\
\label{expQeqsec73}
&&  \hspace{-0.98in}   \quad  \quad    \quad  \quad 
 \,  \, = \, \,  \,   \sum \,\,   {1 \over 4^{k_0} } {(2\,k_0)! \over k_0!^2 }\,
 {\frac{(k_1 +k_2 + \, \cdots \, + k_n)!}{k_1! \,  k_2! \, \cdots \, k_n!}}
\cdot  \, T_1^{k_1} \,T_2^{k_2} \,\cdots \, T_n^{k_n}. 
\end{eqnarray}
With the constraint $\,k_0 =\,  k_1 + k_2 \,  + \cdots \,  + k_n\,$,
there will be one more factorial (i.e. $\,k_0!$)
in the denominator with respect to the
expansion given in (\ref{expQsec73}).  One has then one additional monomial in the polynomial $\, Q_{eq}$
if we consider the right hand side of (\ref{expQeqsec73}) as an expansion of $1/Q_{eq}$.

By performing the diagonal on (\ref{expQeqsec73}), one factorial in the denominator $\, k_j!$
takes a value like $\, k_j! = \,  (r\, p+\cdots)!$,
where $p$ is the running index of the diagonal and $\, r$ an integer. The additional monomial
in $\, Q_{eq}$ will be $\, v^{1/r}$. Now depending on the monomials $\, T_j$, there is no reason
why the additional variable will not occur elsewhere.

Hereafter, let us display some examples of polynomials $\, Q$ and $\, Q_{eq}\,$
(with $\mu$ a parameter)
where the relation (\ref{sqQisQeq}) holds but 
with an expression for $\, Q_{eq}$, different from the simple expression given in (\ref{Qeq}):
\begin{eqnarray}
&&  \hspace{-0.98in}   \quad  \quad   \quad 
Q \, = \,\, \,
1 \, -\left( x+y+z+\mu \, x^5\,y \right),
\\
&&  \hspace{-0.98in}   \quad  \quad    \quad
Q_{eq}\, =\,\,
1\,- \left( x+y+z+\mu \, x^5\,y\,u^{5/3} + {1 \over 4} u^{1/3} \right),
\end{eqnarray}
\begin{eqnarray}
&&  \hspace{-0.98in}   \quad  \quad   \quad 
  Q \,=\,\, 1\,-\left( x+y+z+\mu \, x^4\,(y + z) \right),
  \\
&&  \hspace{-0.98in}   \quad  \quad  \quad
Q_{eq}\, = \,\,
1\,- \left( x+y+z+\mu \,x^4\,(y+z)\,u^{4/3} + {1 \over 4} u^{1/3} \right),
\end{eqnarray}
\begin{eqnarray}
&&  \hspace{-0.98in}   \quad  \quad  \quad 
Q \,=\,\,
1 \,-\left( x+y+z+x y + y z + \mu \, x^3\,y \, z \right),
\\
&&  \hspace{-0.98in}   \quad  \quad  \quad
Q_{eq}\, =\,\,
1\,- \left( x+y+z\, + (x y + y z) \cdot \,u^{1/3} \,\,
+\mu \,x^3\,y\, z \, u^{4/3}\, +{1 \over 4} u^{1/3} \right).
\end{eqnarray}

Such relations and relation (\ref{Qeq}) single out {\em square roots},
and {\em cannot} be simply generalised to $\, N$-th roots of polynomials. 
This special role played by reciprocal of {\em square roots} comes, in fact, from
the emergence of ratio of factorials of multiple argument in the relation:
\begin{eqnarray}
&&  \hspace{-0.98in}   \quad \quad  \quad  \quad  \quad    \quad  \quad  \quad    \quad    \quad    \quad  
  {\frac{\left( 1/2 \right)_n }{n!}}
  \, \, =\,\, {1 \over 4^n} \, {\frac{\left(2\,n \right) }{n!^2}}.
\end{eqnarray}

\vskip 0.1cm

In view of the expressions of the polynomials $\,Q_{eq}$ a natural question arises for the case
where the polynomial $\,Q$ is factorizable (in two factors, for instance) $\, Q\,  =\,  Q_1 \,Q_2$. 
Does the additional variable $\, u$ occurs in one factor, or in both, or as an additive monomial ?
In the last situation, the equivalent polynomial $\,Q_{eq}$ may come out {\em non factorizable},
and the equivalent linear differential operator will be {\em irreducible}.

\vskip 0.1cm

Let us consider the polynomial ${\tilde Q}$ given in   (\ref{Q69}):
\begin{eqnarray}
  \label{Q1yz2}
&&  \hspace{-0.98in}  \quad \quad \quad \quad \quad  \quad \quad \quad  \quad 
  {\tilde Q} \,\, \, = \, \,\, 
  \Bigl(1 \, -(x+y+z) \Bigr) \cdot \, (1 \, -y - z^2).
\end{eqnarray}
Here the variables in the second factor of  of ${\tilde Q}$ are
a subset of the variables in the first factor.
The diagonal of $\,1/{\tilde Q}\, $ is annihilated by an order-six linear differential operator
with the unique factorization $ \, L_6 \,  = \, M_2 \cdot \, M_4$.
The diagonal of $\, 1/{\tilde Q}^{1/2}$ has the following expansion
\begin{eqnarray}
\label{sqQtoQeq}
&&  \hspace{-0.98in}   \quad \quad \quad  \quad    \quad    \quad
 {\rm Diag} \left( {1 \over {\tilde Q}^{1/2} } \right) \, \, = \,\,\,\,
 \sum_p \, (-1)^{s_0} \cdot \, {\frac{ 2^{c_0} N_0 } {D_0 }} \cdot \, (x y z)^p, 
\end{eqnarray}
with
\begin{eqnarray}
&&  \hspace{-0.98in}   \quad  \quad  \quad  \quad
  s_0 \,= \,\, k_1+k_3+k_4+k_5+k_6+k_7,
\nonumber
\end{eqnarray}
\begin{eqnarray}
&&  \hspace{-0.98in}   \quad
c_0 \,= \, \, -5 p+4 k_1+2 k_2+k_3+k_4+4 k_5+3 k_7,
\nonumber  \\
&&  \hspace{-0.98in}   \quad
N_0 \,= \,\, \left( 6p-4 k_1 -2 k_2 -2 k_3 -2 k_4 -4 k_5 -2 k_6 -4 k_7 \right)!,
\nonumber  \\
&&  \hspace{-0.98in}   \quad
D_0 \,=\,\,
k_1! \, k_2! \, k_3! \, k_4!\, k_5! \,k_6!\, k_7! \, \left( p -k_4-k_5 \right)! \,  \left( p-k_3-k_4-2 k_6-k_7 \right)!
\nonumber \\
&&  \hspace{-0.98in}   \quad  \,  \, 
\times \, \left( p -3 k_1 -2 k_2 -k_3 -2 k_5-2 k_7 \right)! \, \left( 3 p -2 k_1 -k_2 -k_3 -k_4-2 k_5 -k_6-2 k_7 \right)!
\nonumber
\end{eqnarray}
where the first terms are ($t=\, x y z$):
\begin{eqnarray}
\label{seriesQeqsec75}
&&  \hspace{-0.98in}   \quad  \, \, 
{\rm Diag} \left( {1 \over {\tilde Q}^{1/2} } \right)
 \, \,= \, \, \, 
1 \, \, +{9 \over 4}\,t \, \, +{\frac {1695}{64}}\,{t}^{2} \, \, +{\frac {26215}{64}}\,{t}^{3} \,\,  
+{\frac {120986775}{16384}}\,{t}^{4} \, \, + \,  \cdots
\end{eqnarray}
This globally bounded power series is annihilated by an order-eight linear differential operator $\, L_8$.
The linear differential operator $L_8$ is {\em irreducible} with a differential Galois group
included in  $\, SO(8, \,\mathbb{C})$.
Among the formal solutions, at the origin,  of the differential operator $L_8$, two series behave as $\ln(t)^2$,
(i.e. the maximum log power).
One of these solutions is the series given in (\ref{seriesQeqsec75}) which, according to our conjectures,
should be diagonal of a rational function with four variables ($N_v = \, 2 \, +2$).

\vskip 0.1cm

In order to find such a rational function $\, 1/Q_{eq}$, we follow the procedure in \ref{binomialVSfactorial}
with the general term given in (\ref{sqQtoQeq}), to obtain:
\begin{eqnarray}
&&  \hspace{-0.98in}   \quad  \quad   \quad  \quad    \quad  \, 
Q_{eq} \, = \, \, \, \, 
1 \,\,\,  -x-y-z \,\,  \, +{1 \over 4} \cdot \,  \left(2\,xy +{y}^{2} \, \,  + 2\,yz \, \,  -4\,{z}^{2} \right) \cdot  \, u^{1/3}
\nonumber \\
&&  \hspace{-0.98in}   \quad \,  \quad    \quad \quad    \quad   \qquad
+{1 \over 2} \, {z}^{2}\cdot \,  \left(2\,x\,+y\, + 2\,z \right) \cdot \, {u}^{2/3} \,\,\,  -{1 \over 4}\, u^{1/3}.
\end{eqnarray}
Consider, now, the polynomial $\,Q$
\begin{eqnarray}
&&  \hspace{-0.98in}  \quad  \quad \quad  \quad \quad \quad  \quad 
   Q \,\, \,= \, \,\, 
  \Bigl(1 \,\, -(x+y+z) \Bigr) \cdot \, (1 \,\, - x -y \, - z^2), 
\end{eqnarray}
where both factors in $\,Q$ carry the same number of variables.
The diagonal of $\,1/Q$ is annihilated by an order-seven linear differential operator with the direct sum
factorization $ \, L_7 \,  = \, \, \left( L_2 \oplus \, L_4 \right) \cdot \, N_1$.
The diagonal of $ \,1/\,Q^{1/2}$ 
\begin{eqnarray}
\label{series2Qeqsec75}
&&  \hspace{-0.98in}   \quad \quad \quad  
{\rm Diag} \left( {1 \over  Q^{1/2} } \right)
 \, \, = \, \, \, \,
 1 \, \,\, +3\,t \, \, \,\, +{\frac {195}{4}}\,{t}^{2} \,\,
  +665\,{t}^{3} \,\,\, +{\frac {820575}{64}}\,{t}^{4}
\,\,\,\,\,\,  +\, \, \cdots
\end{eqnarray}
is annihilated by an {\em irreducible} order-eight linear differential operator 
with differential Galois group included in  $ \, SO(8, \,\mathbb{C})$.
One of the formal solutions, at the origin, of the  linear differential operator,
behaves as $\, \ln(t)^2$, (i.e. the maximum log power).
The power series in front of $\, \ln(t)^2\,$
identifies with the series given in (\ref{series2Qeqsec75}),
and should be the diagonal of a rational function $\, Q_{eq}$
with four variables. The polynomial $\, Q_{eq}$ reads:
\begin{eqnarray}
&&  \hspace{-0.98in}  \quad \quad   \quad  \quad  \quad 
Q_{eq} \,=\, \,  \, 
1 \,\, \,   -\left(x+y \right) \cdot \, \left(z \, -1 \right) \cdot \, \left(z+2 \right)
\,\, \, \, \,   +\left(x + y \right)^2 \cdot \, u^{1/2}
\nonumber \\
&&  \hspace{-0.98in}  \, \,   \quad \quad \quad \quad \quad \quad \quad  \quad 
+z \cdot \, \left(z^2-z-1 \right) \cdot \, u^{-1/2} \, \, \,  -{1 \over 4} \cdot \,  u^{1/2}.
\end{eqnarray}

\vskip 0.1cm

\section{On the homomorphism to the adjoint assumption}
\label{assumption}

All the examples, displayed in this paper (and many others not given here),
confirm the conjecture (\ref{conjmain}). 
One assumption for this conjecture to hold, is that the (minimal order) linear differential operator
is {\em homomorphic  to its adjoint}, thus yielding symplectic, or orthogonal, differential Galois 
groups~\cite{2014-Ising-SG,2014-DiffAlg-LGFCY,selectedGalois}. 
This assumption may look as an innocent caveat since, as we underlined
in several papers~\cite{selectedGalois}, the (minimal order) linear differential operators
annihilating diagonals of rational functions are, almost systematically, homomorphic
to their adjoint\footnote[1]{For reducible differential operator, each factor is
homomorphic to its adjoint.}.

However, some examples of diagonals, whose corresponding (minimal order)
linear differential operators are {\em not}
homomorphic to their adjoint,  have been seen to correspond to $\, _3F_2$
candidates to be counterexamples
to Christol's conjecture~\cite{Chrisconj,christol-EMS}. Such a candidate, for instance, reads:
\begin{eqnarray}
\label{Christolconj}
&&  \hspace{-0.98in} \, \, \quad  \quad \quad \quad \quad \,\,
_3F_2\Bigl([{{2} \over {9}},{{5} \over {9}}, {{8} \over {9}}], \, [1, \, {{2} \over {3}}], \, \,   27\, t \Bigr)
  \, \, = \, \, \, \,\,
 {\rm Diag}\Bigl( \, {{ (1\, -x \, -y)^{1/3} } \over { 1 \,\, -x \, -y \, -z}} \Bigr).  
\end{eqnarray}
This hypergeometric function is the diagonal of a quite simple {\em algebraic function}
of three variables.
The order-three linear differential operator
annihilating (\ref{Christolconj}) is {\em not} homomorphic to its adjoint. 
Its differential Galois group  is $\, SL(3, \, \mathbb{C})$.
One of the formal solutions (at the origin) of the order-three linear differential operator
has the highest logarithmic power $\, \ln(t)^{1}$.

Note that a representation of this hypergeometric function (\ref{Christolconj})
as diagonal of {\em rational} function of more than
{\em three  variables}, is possible~\cite{Chrisconj} using Denef-Lipschitz
formulation~\cite{Denef-lipschitz-1987}, but a representation as a diagonal
of a {\em rational} function of $\, 1 \, +2 \, = \, 3 \, $ variables does not seem possible. 

Thus this example does not seem to satisfy relation (\ref{conjmain}). It is outside
the framework of this paper, for which  
the assumption to be homomorphic to its adjoint is not superfluous, {\em but necessary}.

\vskip 0.1cm

\section{Conclusion}
\label{Conclusion}

In this paper, we addressed some properties of the diagonal of rational functions.
We restricted the analysis to the {\em rational} functions of the form $\, 1/Q$. 

It seems that the minimal number of variables $ \, N_v \, $ required to represent a  globally bounded
D-finite series as a diagonal of rational function, is simply related to the highest power $\,n$
of the logarithmic formal solutions
of the  (minimal order) linear differential operator annihilating the diagonals,
by $\, N_v = \, n\,+2$, provided one
 {\em assumes}  that this linear differential operator is  {\em homomorphic to its adjoint}.

 Furthermore it is observed that the  symplectic, or orthogonal, character of the differential Galois group
 seems to be related to the {\em parity} of this highest power $\,n$ of the logarithmic formal solution.

 In the situation where the polynomial $\, Q$ factorizes in two polynomial factors
 as $\, Q \, = \, Q_1 \cdot \, Q_2$,
 the linear differential operator annihilating 
 the diagonal of $\, 1/Q$ has either a direct sum or a unique factorization, 
 depending on whether
 both polynomials $\, Q_j$ carry all (the same) 
variables, or not. Furthermore, in the case of a unique factorization, the successive factors
in the linear differential operator are included 
in the symplectic and orthogonal differential Galois groups, in alternance.
Even in these factorized cases conjecture (\ref{conjmain})
remains valid.

All the results, and educated guess statements, of this paper are just conjectures.
One would like to have a demonstration of these various conjectures. At first sight,
 one can imagine that the number of variables $ \, N_v$ is
related to some ``complexity measurement'' of the (minimal order) linear differential
operator annihilating the diagonal. A very naive measure of the complexity is
the order of the operator, and the MUM examples, for which we have
a simple relation between the highest power $\,n$  and the order, seem to confirm
such a naive view-point.  We show, in this paper, that {\em this is not the case}.
The relation  the number of variables $ \, N_v$ is {\em not related with the order}
but with highest log-power $\,n$. Along this line, and as far as a demonstration
of the main conjecture (\ref{conjmain}) is concerned, let us underline that
the crucial role played by the highest log exponent corresponds to the concept
of monodromy filtration\footnote[1]{See paragraph 4.2
  page 40 of~\cite{christol-EMS}.}, which can be introduced even if one is
not totally sure that the linear differential operator is minimal order.

\vskip 0.5cm

$\bf Acknowledgments$
We thank A. Bostan and J-A Weil, for discussions on ODE's with algebraic solutions.
We thank A. Bostan for some p-curvature calculations.
We thank G. Christol, for monodromy filtration discussions.
We thank C. Koutschan for LGF discussions.

\vskip 0.5cm

\appendix

\section{Linear differential operators corresponding to Lattice Green functions}
\label{LGFAppend}

\subsection{MUM cases}

\subsubsection{Simple cubic  lattice Green functions in $\, d$ dimensions\\}
\label{LGFsc}

The lattice Green function of the $\, d$-dimensional simple cubic lattice is given
by the multiple integral of a rational function with numerator 1, and non factorizable
denominator depending on $ \, d$ variables, and on the parameter $ \, t$.

The diagonal $ \, 1/Q$, with polynomial $\, Q$ given by
\begin{eqnarray}
&&  \hspace{-0.98in}   \quad   \, \,    \quad  \quad  \qquad \quad \quad \quad
  Q\,  \, = \, \,\,\, \, 
  1\, \, \, \,\, - \sum_{j=1}^{d} x_j \,\,\, \, \,
  -  x_{d+1} \cdot \, \sum_{j=1}^{d} \,\, \prod_{i \ne j}^{d} x_i, 
\end{eqnarray}
depending on  $\, N_v =\,  d\, +1$ variables, reproduces the {\em simple cubic lattice}
of dimension $\, d$.
The diagonal of $\, 1/Q$ is annihilated by an irreducible
linear  differential operator of order $\, d$, {\em  having MUM}. 
The maximum power of the logarithmic formal solutions of the linear differential operators
up to $ \, d=\, 8$, and the corresponding rational functions follow the two conjectures
(\ref{conjmain}), (\ref{conjadditio}). 

\vskip .1cm

\subsubsection{The diamond lattice Green function\\}
\label{LGFdiam}

The LGF of the $\,  3$-dimensional diamond lattice is given by the diagonal of $\, 1/Q$, where
\begin{eqnarray}
&&  \hspace{-0.98in}   \quad   \quad  \quad  \quad   \quad   \quad   \quad   \quad 
  Q \,  \,= \,\, \,\,  \,  
  1 \,\,\,  \,
  -x y z u \cdot \, S(x,y,z) \cdot \,
     S\Bigl( {{1} \over {x}}, \, {{1} \over {y}}, \, {{1} \over {z}}\Bigr), 
\end{eqnarray}
with:
\begin{eqnarray}
&&  \hspace{-0.98in}   \quad   \quad  \quad  \quad   \quad  \quad  \quad     \quad   \quad 
  S(x,y,z)\,  \, = \, \,\, \,\,  \, 
  x \,  + {1 \over x} \, \, \, \,    +z \cdot \,  \left( y \, + {1 \over y} \right), 
\end{eqnarray}
With the fourth variable $\,  u$, the number of variables is $\,  N_v =\,  4$.
The corresponding order-three linear differential operator is irreducible, its differential Galois
group is included in the orthogonal group  $\, SO(3, \, \mathbb{C})$,
and its formal solutions with the highest log-power behave as $\,  \ln(t)^2$,
in agreement with $\,  N_v=\, 2\, +2 =\,  4\, $ variables.

\vskip 0.1cm

The LGF of the $4$-dimensional diamond lattice is given by the diagonal of $\,  1/Q$,
depending on $\, N_v=\, 5\, $ variables.
The corresponding order-four linear  differential operator is irreducible. Its
differential Galois group is (included) in the symplectic group $\, Sp(4, \, \mathbb{C} )$.
Its formal solution with the highest log-power behaves as $\,\ln(t)^3$,
in agreement with $\,  N_v=\, 3\, +2 =\,  5\,$ variables.

\vskip 0.1cm

Likewise, the LGF of the $5$-dimensional diamond lattice is given by the diagonal of $\,1/Q$,
which depends on $\,N_v= \,6 \, $ variables.
The corresponding order-five linear differential operator is irreducible. Its differential Galois group
is (included) in the orthogonal group $\, SO(5, \, \mathbb{C} )$,
and with the formal solution with the highest log-power at the origin
behaving as $\, \ln(t)^4$, in agreement with $\,  N_v=\, 4\, +2 =\,  6 \, $ variables.

All these examples are in agreement with the two conjectures (\ref{conjmain}) and (\ref{conjadditio}). 

\vskip 0.1cm

\subsubsection{On the connection between the LGF of the diamond and the simple cubic lattices\\}
\label{conndiasc}

There is a known relation~\cite{guttmann-2010,guttmann-prellberg-1993,glasser-montaldi-1993}
between the LGF of the
$d$-dimensional diamond lattice and the LGF of the $(d+1)$-dimensional simple cubic lattice.
The general term of the series, corresponding to the $d$-dimensional diamond lattice
is the square of the multinomial coefficient on 
$\, d+1\, $ indices. The general term of the series corresponding to the $\, d$-dimensional simple cubic lattice,
is the square of the multinomial coefficient on $\, d$ indices times the binomial $\,  2 n \choose n$.

\vskip 0.1cm

The origin of this binomial is considered by Guttmann in~\cite{guttmann-2010}. Let us
see how this appears on the diagonals
of $\, 1/Q$.
Call $ \,  Q(d+1, sc)$ the polynomial corresponding to the $ \, (d+1)$-dimensional simple cubic lattice.
Rewrite the polynomial $\, Q(d, dia)$ corresponding to the $\, d$-dimensional diamond lattice as
\begin{eqnarray}
\label{Qddimdia}
&&  \hspace{-0.98in}   \quad   \quad  \, \,   \quad   \quad   \quad  \quad   \quad   \quad   \quad 
  Q(d, dia) \,  \, \, = \,\, \, \, \,
  1\, \, \,\, - t \, \cdot \, S \cdot \, {\tilde S}, 
\end{eqnarray}
where ${\tilde S}$ is the same as the Laurent polynomial $\, S$, and
where each variable is changed to its reciprocal.
The variable $ \, t$ is the product of all the variables, i.e. the variables occurring in $ \,S$,
and an additional variable.
Changing in (\ref{Qddimdia}),  $\,t \, $ to $ \, 4\,t \, $ for rescaling purposes, 
the relation between the LGF of simple cubic and diamond lattices reads:
\begin{eqnarray}
&&  \hspace{-0.98in}   \quad    \quad  \quad  \, \,  \, \, \, \,   \quad   \quad   \quad 
{\rm Diag} \Bigl( {{1} \over {Q(d+1, sc) }} \Bigr)
   \,  \, \, = \,  \, \,\,\,
  {\rm Diag} \Big( {1 \over \sqrt{ Q(d, dia) } } \Bigr).
\end{eqnarray}

\vskip 0.1cm

{\bf Remark A.1.}
This is similar to the examples in sections \ref{algCONNrat1}, and \ref{algCONNrat2},
giving connection between the LGF of (simple cubic and face centred cubic) lattices,
and Calabi-Yau equations.
See also \ref{sqrtR2CY}.

\subsection{ Non-MUM examples: Face centred cubic lattice Green functions in $\, d$ dimensions}
\label{LGFfcc}

The lattice Green function of the $\, d$-dimensional face-centred cubic lattice
is given by the multiple integral of a rational
function with numerator 1, and non factorizable denominator depending on $ \, d$
variables.
For these  lattice Green we have {\em not} MUM (except for $\, d=2, 3, 4$). 

\vskip 0.1cm

The diagonal $ \, 1/Q$ where $\, Q$ is the polynomial
\begin{eqnarray}
&&  \hspace{-0.98in}   \quad  \quad  \quad   \quad  \quad  \quad 
Q(x_1, \, x_2, \, \cdots,  \, x_d,  \,x_{d+1} )
\nonumber \\
&&  \hspace{-0.98in}   \quad  \quad  \quad   \quad  \quad  \quad \quad  \quad 
\,\, = \,\, \,\, \, \,
1 \,\,  \,\,
  - \left( \prod_{j=1}^{d+1} x_j \right) \cdot \,
\sum_{j > i}^{d} \,\left(x_i + {1 \over x_i} \right) \cdot \, \left(x_j + {1 \over x_j} \right), 
\end{eqnarray}
depending on $\,N_v= \,d\,+1 \, $ variables, reproduces
the face centred cubic lattice Green function of dimension $ \, d$.

\vskip 0.1cm

From $\, d = \, 2 \,$ to $\, d = \, 12$, we denote the corresponding
linear differential operators (the subscript being the order), $\, G_2^{2Dfcc}$,
$\, G_3^{3Dfcc}$ \cite{joyce-1998},
$ \, G_4^{4Dfcc}$~\cite{guttmann-2010}, $ \, G_6^{5Dfcc}$~\cite{broadhurst-2009},
$ \, G_8^{6Dfcc}$~\cite{koutschan-2013}, $  \, G_{11}^{7Dfcc}$~\cite{2015-LGF-fcc7},
and ($G_{14}^{8Dfcc}$,  $ \, G_{18}^{9Dfcc}$,
$\, G_{22}^{10Dfcc}$, $\, G_{27}^{11Dfcc}$, $\, G_{32}^{12Dfcc}$)~\cite{2016-LGF-fcc8to12}.

\vskip 0.1cm

The linear differential operators, up to $ \, d = \, 9$, are known
to be irreducible~\cite{2015-LGF-fcc7,2016-LGF-fcc8to12}.
The linear differential operators, up to $\, d= \, 12$, have formal solutions
with the highest log-power at the  origin behaving as $ \, \ln(t)^n$.
All the linear differential operators (and their corresponding rational functions)
(see Table \ref{Ta:2}) are in agreement with
the two conjectures (\ref{conjmain}) and (\ref{conjadditio}).

\vskip 0.1cm
All the linear differential operators up to\footnote[1]{For $ \, d= \, 12$,
  the linear differential operator $ G_{32}^{12Dfcc}$ is known only modulo some primes.}
$ \, d = \, 11$ have {\em non globally bounded} formal solutions
around all the singularities $ \,t \ne \, 0$. There
is an exception for $ \,G_3^{3Dfcc}$ which has, at the singularity $ \, t= \, -1/4$,
a {\em globally bounded} formal solution in front of 
$\ln(t+1/4)^2$, i.e. the {\em same} maximum as around the singularity $t=0$.

\vskip 0.1cm

\begin{table}[htp]
  \caption{Minimal number of variables, order, maximum exponent of $\ln(t)^n$ of the formal
    solutions and differential Galois group for the LGF of the fcc lattice of dimension $d= \, 2, 3, \cdots, 12$.
 }

\label{Ta:2}
\begin{center}
\begin{tabular}{|c|c|c|c|c|c|c|c|c|c|c|c|}\hline
    $d$               &  2    &  3   &  4  &  5  &  6  &  7  &  8  &  9  & 10  & 11  &  12        \\ \hline 
    $N_v$           &  3   &  4  &  5  &  6  &  7  &  8  &  9  & 10  & 11  &  12   & 13     \\ \hline 
    {\rm Order}   &  2   &  3  &  4  &  6  &  8  &  11  &  14  & 18  & 22  &  27   & 32     \\ \hline 
    $n$               &  1   &   2  &  3  &  4  &  5  &  6  &  7  &  8  & 9 & 10  &  11     \\ \hline 
    Sp or SO       &  Sp   &  SO  &  Sp  &  SO  & Sp  &  SO  &  Sp  & SO  & Sp  &  SO   & Sp     \\ \hline 
 \end{tabular}
\end{center}
\end{table}

\vskip 0.1cm

\section{The minimum number of variables in the rational function}
\label{binomialVSfactorial}

For general coefficient of the series written as nested sums of
binomials, one may use the integral representation of the binomial
\begin{eqnarray}
\label{binomRepr}
&&  \hspace{-0.98in}   \quad   \quad  \quad \quad \quad  \quad \quad \quad \quad 
   { n \choose k} \,\, = \,\, \, \,
   {\frac{1}{2 \pi i}} \int_C {\frac{(1+z)^n}{z^k}} \, {\frac{dz}{z}},
\end{eqnarray}
to write down the rational function~\cite{2013-rationality-integrality-ising}. The calculations are
straithforward~\cite{2013-rationality-integrality-ising}, and one obtains, this way, the 
rational function with as many variables as binomials plus one more variable.
Furthermore, the denominator polynomial $\,Q$ will be in a non factorized form,
for more than one summation
in the general term of the series.

\vskip 0.1cm

Let us consider, for instance, the general term of the Calabi-Yau series
(number 16 in~\cite{TablesCalabi-2010})
\begin{eqnarray}
&&  \hspace{-0.98in}    \quad   \quad    \quad \quad \quad   \quad  
  CY_{16}   \,\, = \,\,\,\,
   \sum_{n=0}^{\infty}  \sum_{k=0}^{n}  \, \,  {2n \choose n} \cdot \, {n \choose k}^2
   \cdot \, {2k \choose k}\cdot \, {2n-2k \choose n-k} \cdot \,t^n
 \nonumber \\
 &&  \hspace{-0.98in}    \quad    \quad  \quad \quad   \quad   \quad \quad     \quad  \quad  
 \, = \,  \, \,\,\,\,
 1 \,\,  \, +8 \,t \,\, \,\, +168\,{t}^{2} \, \,+5120\,{t}^{3}
 \, \,\, \,\, + \, \cdots
\end{eqnarray}

Using the integral representation (\ref{binomRepr}), each binomial,
and the variable $\, t$,  brings one variable. One obtains
\begin{eqnarray}
&&  \hspace{-0.98in}  \quad  \, \,   \quad \quad  \quad \quad \quad \quad  \quad \quad \quad \quad
CY_{16}   \, \,=\,  \, \, \, {\rm Diag} \left( {\frac{1}{Q_1 \cdot \,Q_2}}\right), 
\end{eqnarray}
where:
\begin{eqnarray}
&&  \hspace{-0.98in}  \quad  \quad  \quad  \,\,  \,\, 
Q_1  \, = \,  \,\, \,\, 
 1 \, \, \, \,  -z_0 z_2 z_3 z_4 \cdot  \,  (1+z_1)^2 \cdot  \, (1+z_2)  \cdot  \, (1+z_3)  \cdot  \, (1+z_5)^2,
\nonumber \\
&&  \hspace{-0.98in}  \quad  \quad  \quad \,\,  \,\, 
Q_2  \, = \, \,\, \, \,
1 \,  \, \, \, - z_0 z_5 \cdot \, (1 +z_1)^2  \cdot  \,  (1 +z_2)  \cdot  \, (1+z_3)  \cdot  \, (1 +z_4)^2.
\end{eqnarray}
This diagonal of rational function representation of the Calabi-Yau $\, CY_{16}$
depends on six variables with a {\em factorizable} denominator.
This is a diagonal of rational function representation of $\, CY_{16}$, but
this simple nested sum of binomials {\em does not provide the minimal number of
variables representation} for the diagonal.

\vskip 0.1cm

Let us now show how the rational function $\, 1/Q$, in Table 1, is obtained for Calabi-Yau
$\, CY_{16}$. 

\vskip 0.1cm

The procedure amounts to casting the general term into the original form
of the {\em multinomial} theorem, instead of the previous nested sum of binomials form.
This means making the numerator with one factorial by introducing more summations.
Converting to factorials, and using the formula 
\begin{eqnarray}
&&  \hspace{-0.98in}  \quad  \quad  \quad \quad  \quad \quad \quad  \quad \quad \quad \,  
  {\frac{(2 n)!}{n!^4}} \, \, = \, \, \,\,\, 
  \sum_{ k=0}^{n} \, {\frac { 1}{  k! ^{2} \,   \left(n -k \right) ! ^{2} }}, 
\end{eqnarray}
in the general term of $ \, CY_{16}$, one
obtains\footnote[5]{This is the original form of the general term given
in Section 8.1 of~\cite{batyrev-stranten-1995}.}:
\begin{eqnarray}
\label{CY16facto}
&&  \hspace{-0.98in}  \quad  \quad  \quad  \quad  \quad  
CY_{16}   \, \, \, = \, \, \, \sum_{n, k, k_1, k_2} 
{\frac{(2 n)!}{k_1!^2\, \,   k_2!^2 \,\,   (k-k_1)!^2\, \,   (n-k-k_2)!^2}} \cdot \,t^n. 
\end{eqnarray}
The general term  carries one factorial in the numerator, and eight factorials in the denominator. 
There are, then, 8 monomials in the multinomial expansion of the unknown polynomial denominator
$\, \, Q \, \, = \,  \,\, 1 \,\,  \, -\left( T_1 \,+T_2 \, + \, \cdots \, +T_8 \right)$.

These monomials correspond to each factorial as 
\begin{eqnarray}
&&  \hspace{-0.98in}  \quad  \quad  \quad  \quad  \quad 
  T_1 \rightarrow \, \, k_1, \quad \,\quad  \, \, \,  \quad T_2 \rightarrow \,\,  k_1,
   \quad \quad \quad \, \, \, \, T_3 \rightarrow \,\, k_2,
   \quad \quad  \quad \, \,  \, \,T_4 \rightarrow \, \, k_2,
  \nonumber \\
&&  \hspace{-0.98in}  \quad  \quad  \quad  \quad  \quad 
  T_5 \rightarrow \, \, k -k_1,  \quad \quad \quad \quad  \quad \quad  \quad  \, \,
  T_6 \rightarrow \,\, k -k_1,
  \\
&&  \hspace{-0.98in}  \quad  \quad  \quad    \quad  \quad   
  T_7 \rightarrow \,\, n-k-k_2, \quad \quad  \quad \,   \quad \quad \,
  T_8 \rightarrow \,\, n -k -k_2
  \nonumber
\end{eqnarray}
and satisfy
\begin{eqnarray}
\label{condTj}
\fl \quad \quad \quad  \quad \, 
 {\frac{T_1 T_2}{T_5 T_6}} \,= \,  \, 1, \qquad \,\, {\frac{T_3 T_4}{T_7 T_8}} \, = \,1, 
\qquad  {\frac{T_5 T_6}{T_7 T_8}} \,= \, \,  1, \qquad \,\,  T_7 T_8 \,= \, t,  
\end{eqnarray}
which give:
\begin{eqnarray}
 &&  \hspace{-0.98in}  \quad  \quad  \quad  \quad 
  T_2 \,= \,  {\frac{t}{T_1}},  \qquad \,\,  T_4 \,=\, {\frac{t}{T_3}},
  \qquad \,\, T_6 \,=\,{\frac{t}{T_5}}, \qquad \,\,  T_8 \,=\,{\frac{t}{T_7}}.
\end{eqnarray}

The variable $\, t$ is the product of all the variables. There are four unfixed $ \, T_j$,
the number of variables should be greater, or equal, to 4.
Guided by our conjectures, let us, first, assume there are five variables, i.e. $\,t = \, x y z u v$.
Choosing $\, T_1 = \, x$, $\, T_3 =\, y$, $ \, T_5 =\, z \, $ and $\, T_7 =\, u$,
we obtain the polynomial $\, Q$ given in Table 1.
On may also choose $ \, T_1 = \, x$, $ \, T_3 = \, y$, $ \, T_5 = \, x u\, $
and $ \, T_7  = \, y v$,  to obtain
\begin{eqnarray}
&&  \hspace{-0.98in}   \quad  \quad \quad  \quad 
Q \,\, = \,\,\,\,\,
   1 \, \, \,\,\, - x \cdot \, \left( 1\, +u + u z +u z v  \right)
      \,\, \, \, - y \cdot \, \left(1\,+ v+v z+v z u \right), 
\end{eqnarray}
and the diagonal of $\,1/Q$, then identifies with $\, CY_{16}$.

Assume, now, that we start with 6 variables, i.e. $\,t = \,x y z u v w$,
and fix $\,T_1 =\, x$, $\,T_3 =\,y$, $ \, T_5 =\,z$ and $\,T_7 = \,u$.
One obtains the polynomial
\begin{eqnarray}
&&  \hspace{-0.98in}  \quad  \quad   \quad    \quad  
Q \,\,=\, \,\,\,\,
 1 \,\,\, \,  \,
   - \Bigl( x + y +z +u \, \, \,\, + v w \cdot \, \left(  x y z + y z u + z u x + u x y \right)  \Bigr), 
\end{eqnarray}
where it is clear that the product  $\, v w \, $ stands for {\em just one} variable. 

\vskip 0.1cm

{\bf Remark B.1.}
One should note that the conditions (\ref{condTj}) are satisfied with 4 variables giving 
$\, Q\, \, =\, \, \,  1 \, \,\,  - \left( x+y+z +u \, \, +x y z +y z u +z u x +u x y \right)$,
the diagonal of which {\em does not} identity with $\,CY_{16}$.
The corresponding linear differential operator is of order four, is irreducible, has his
differential Galois group included in $\, Sp(4, \, \mathbb{C} )$,
and the most logarithmic singular formal solution is in $\,\ln(t)^2$.

\vskip 0.1cm

{\bf Remark B.2.}
Assume that, in the numerator of (\ref{CY16facto}), one has $\, a^{k_1} \, (2 n)! \, \, t^n$,
the first condition in (\ref{condTj}) changes to 
\begin{eqnarray}
&&  \hspace{-0.98in}  \quad  \quad  \quad \quad   \quad  \quad   \quad     \quad \quad  
  {\frac{T_1 T_2}{T_5 T_6}} \,= \, \,  1
   \qquad  \longrightarrow \qquad \, \, 
 {\frac{T_1 T_2}{T_5 T_6}} \,= \, \, a
\end{eqnarray}
and the denominator polynomial $\,Q$ becomes:
\begin{eqnarray}
&&  \hspace{-0.98in}  \quad  \quad \quad     \quad \quad    \quad  
  1 \,\,\,   \,
    - \Bigl( x + y +z +u \, \, \,\,  + v \cdot \,\left( x z u +x y u+x y z +a\, y z u\right)  \Bigr), 
\end{eqnarray}

\vskip 0.1cm

{\bf Remark B.3.}
The procedure to obtain the rational function is straightforward when the numerator
of the general term of the series can be cast into only one factorial. It may happen
that this is not mandatory. Consider, for instance, the series
\begin{eqnarray}
 &&  \hspace{-0.98in}  \quad \quad  \,  \quad  \quad  \quad  \quad  \quad 
 \sum_{n=0}^{\infty}  \sum_{k=0}^{n} \,  { n \choose k}^3 \cdot \, t^n \,\, = \,\,\,\,
 \sum_{n=0}^{\infty}  \sum_{k=0}^{n} \,  \left(  {\frac{n!}{k! \, (n-k)! }}  \right)^3\cdot  \,t^n
 \nonumber \\
&&  \hspace{-0.98in} \,  \quad  \, \, \, \,    \quad  \quad  \quad  \quad  \quad  \quad  \quad  \quad 
    \,  = \, \,\,\,\,
   1\, \,\, +2\,t\,\, +10\,{t}^{2}\,\, +56\,{t}^{3}\,\, +346\,{t}^{4}
   \,\,\, \,\, \,  +\, \cdots
\end{eqnarray}
which is the generating function of sequence {\bf A} in Zagier's tables
of binomial coefficients sums (see p.~354 in~\cite{Zagier-2009}).
With the integral representation of the binomial, the rational function
is in the form $\,1/Q_1/Q_2$, and depends on {\em four} variables:
\begin{eqnarray}
&&  \hspace{-0.98in}  \quad \quad  \quad  \quad  \quad  \quad  
Q_1 \,  = \, \, \,\,
1 \,\, \,\, - z_0 \cdot \, (1\,+z_1) \cdot  \, (1\,+z_2) \cdot \, (1\,+z_3),
\nonumber  \\
&&  \hspace{-0.98in}  \quad \quad  \quad  \quad  \quad  \quad  
Q_2 \,  = \,\, \,\,
1 \,\,\,\,  - z_0 \, z_1\, z_2\, z_3 \cdot \, (1\,+z_1) \cdot \, (1\,+z_2) \cdot \, (1\,+z_3).
\end{eqnarray}

The procedure applied, on each binomial, gives the rational function $ \, 1/Q$,
dependent on $\, N_v= \, 3 \, $ variables, where 
\begin{eqnarray}
\quad   \,
Q \,\, = \,\, \,\,
  1\, \,\,\, - \left( x+y \right) \cdot \, \left( x+z \right) \cdot \, \left( y+z \right), 
\end{eqnarray}
or:
\begin{eqnarray}
\quad   
Q \,\, = \,\,  \,\,\,
1\, \, \,\, \, - \left( x + y + z  \, \,\, -4 \, x y z  \right). 
\end{eqnarray}
The corresponding order-two  linear differential operator has a differential Galois group
(included) in $ \, Sp(2, \, \mathbb{C} )$. Its formal solution
with the highest log-power behaves as $\,\ln(t)^1$, in agreement with
$\, N_v \, = \, 1\, +2 \, = \, 3\,$ variables.

\vskip 0.1cm

\section{The factors occurring in the $ \, \chi^{(n)}$'s of the Ising model}
\label{chin}

The linear differential operators corresponding to the $\, n$-particles
contributions to the magnetic susceptibility of the Ising model, namely the $\, \chi^{(n)}$'s,
and especially their decomposition in {\em products and direct sums of many factors}
are recalled below (the subscript denote the order of the linear differential operator): 
\begin{eqnarray}
\fl 
  \chi^{(1)}, \qquad \qquad L_1,
  \nonumber \\
\fl 
  \chi^{(2)}, \qquad \qquad L_2,
  \nonumber \\
\fl 
  \chi^{(3)}, \qquad \qquad L_7 \,= \,\, L_1 \oplus (Y_3 \cdot \, Z_2  \cdot \, N_1),
  \nonumber \\
\fl 
  \chi^{(4)}, \qquad \qquad L_{10} \,  \,=\, \,\,
  L_2 \oplus ({\tilde L}_4 \cdot \, L_4^{(4)}), \qquad \quad \quad L_4^{(4)}
  \,  \, =  \, \,  L_{1,3} \, \left( L_{1,2} \oplus  L_{1,1} \oplus \,  D_t \right), 
  \nonumber \\
\fl 
  \chi^{(5)}, \quad \quad \quad \quad \quad L_{33} \,  \,=\,  \,\,
  L_7 \oplus \, (L_5 \, L_{12} \cdot \,{\tilde L}_1) \cdot \,  \,
  \Bigl( V_2 \oplus \, (Z_2 \cdot \,N_1) \oplus \, (F_3 \cdot \,F_2 \cdot \,L_1^{s}) \Bigr),
  \nonumber \\
\fl 
  \chi^{(6)}, \quad \quad \, \, \,  L_{52} \,  \,=\,  \,
  L_{10} \oplus \, (L_6 \cdot  \, {\tilde  L}_{2} \cdot  \, L_{21}) \cdot
  \,\Bigl( \left( D_t \, -{{1} \over {t}} \right) \oplus L_3 \oplus
  \, L_4^{(4)} \oplus \, (L_4 \,{\tilde  L}_{3} \cdot \, L_2^{e}) \Bigr),
  \nonumber 
\end{eqnarray}
All the factors, occurring in the linear differential operators, have been shown to be
such that their differential Galois group is
either  symplectic or orthogonal (see~\cite{2014-Ising-SG} and references therein).

The blocks of factors (of order greater than 1) with unique factorization
(with their corresponding differential Galois group), are:
\begin{eqnarray}
\fl \qquad \quad
Y_3\, Z_2, \qquad \quad \,\, Y_3 \rightarrow  SO(3, \,\mathbb{C} ),
\qquad \quad \, Z_2 \rightarrow Sp(2, \, \mathbb{C} )
  \nonumber \\
\fl \qquad \quad
L_5\, L_{12}, \qquad \,\,\, \, L_5 \rightarrow  SO(5, \, \mathbb{C} ),
\qquad \quad\,  L_{12} \rightarrow Sp(12, \, \mathbb{C} ), 
  \nonumber \\
\fl \qquad \quad
F_3\, F_2, \qquad \quad \, \,F_3 \rightarrow  SO(3, \, \mathbb{C} ),
\qquad \quad \, F_{2} \rightarrow Sp(2, \, \mathbb{C} ), 
  \nonumber \\
\fl \qquad \quad
  L_6\,{\tilde  L}_{2} \, L_{21}, \quad \quad\,\,  L_6 \rightarrow  SO(6, \, \mathbb{C} ),
  \quad \quad \quad \, {\tilde  L}_{2} \rightarrow Sp(2, \, \mathbb{C} ),
  \quad \, \,  \, L_{21} \rightarrow  SO(21, \, \mathbb{C} ), 
  \nonumber \\
\fl \qquad \quad
  L_4\,{\tilde  L}_{3} \,L_{2}^{e}, \qquad \, \,L_4 \rightarrow  Sp(4, \,\mathbb{C}),
  \qquad \quad {\tilde  L}_{3} \rightarrow SO(3, \, \mathbb{C} ),
  \quad  \quad \, L_{2}^{e} \rightarrow  Sp(2, \,\mathbb{C}).
  \nonumber 
\end{eqnarray}

Each time we deal with a block of factors with
unique\footnote[1]{No direct sum factorization.} factorization,
the differential Galois groups of the 
linear differential operators inside the block, are, {\em in alternance},
 (included in) symplectic, or orthogonal, differential Galois groups.

\vskip 0.1cm

Let us note that we have shown~\cite{2013-rationality-integrality-ising}
that the $\, \chi^{(n)}$'s are actually diagonals of rational functions.
However, for each linear differential operator factor in the blocks, it is quite hard to find the
corresponding rational (or even algebraic) function whose diagonal is annihilated
by this linear differential operator factor.
Conjecture (\ref{conjmain}), can just be used to determine the minimum number
of variables in the rational function.
This should, also, be consistent with the fact that {\em only}
the series in front of $\,\ln(t)^n$, with $\, n$ the maximum exponent, are globally bounded.

\vskip 0.1cm

Consider, for instance, the order-twelve linear differential operator $\, L_{12} \, $ occurring
in the factorization of $\, L_{33}$, the linear differential operator
annihilating  $\, \chi^{(5)}$. The singularities of $\, L_{12}$, besides $\,t = \, \infty$,
are the roots of:
\begin{eqnarray}
  \label{singL12}
 &&  \hspace{-0.98in}  \quad \quad \quad 
  t \cdot \, \left( 4\,t -1 \right) \cdot \,  \left( 4\,t +1 \right) \cdot \,
  \left( t +1 \right) \left( 2\,t +1 \right) \cdot \,  \left( t -1 \right)
  \cdot \, \left( 4\,{t}^{2}+3\,t+1 \right)
    \nonumber \\
 &&  \hspace{-0.98in}  \quad \quad  \quad 
    \left( {t}^{2}-3\,t+1 \right) \cdot \, \left( 4\,{t}^{2}-2\,t-1 \right)
  \cdot \,  \left( 8\,{t}^{2}+4\,t+1 \right) 
  \cdot \,  \left( 4\,{t}^{3}-3\,{t}^{2}-t+1 \right)
    \nonumber \\
 &&  \hspace{-0.98in}  \quad \quad \quad 
    \left( 4\,{t}^{3}-5\,{t}^{2}+7\,t-1 \right)
    \cdot \, \left( 4\,{t}^{4}+15\,{t}^{3}+20\,{t}^{2}+8\,t+1 \right)   
   \, \,  \, = \,\,\,  \, 0.
\end{eqnarray}

Among the formal solutions of $\, L_{12} \, $ around the origin, there are
two series in front
of $\,\ln(t)^3$, (the exponent $\,n= \, 3 \, $ being the maximum):
\begin{eqnarray}
  \label{un}
&&  \hspace{-0.98in} \quad \quad \quad \quad \,
  {t}^{6} \,\,\, \, + {\frac {61533918683}{35574000}} \, {t}^{8} \, \,  \, \,
  + {\frac {3912998847493}{53361000}}  \,  {t}^{9} \,
  -{\frac {1448659961931211}{533610000}}\, {t}^{10} 
\nonumber  \\
&&  \hspace{-0.98in}  \quad   \quad \quad  \quad \quad  \quad \quad \,
  \,\,  \, \,\,
  -{\frac {718444385040931}{7623000}}\, {t}^{11}
  \, \, \,\, \, + \,\,  \cdots 
\end{eqnarray}
and 
\begin{eqnarray}
   \label{deux}
&&  \hspace{-0.98in}   \quad  \quad \quad  \quad \quad 
  {t}^{7} \, \, \, \,  -{\frac {13806647}{1386000}}\, \, {t}^{8} \,\, \,  \,
   - {\frac {190483891}{297000}}\,  \,  {t}^{9} \, - {\frac {16842849293}{2970000}} \,  {t}^{10} 
 \nonumber   \\
&&  \hspace{-0.98in}  \quad  \quad \quad \quad \quad  \quad \quad  \quad \quad 
 -{\frac {140703207221}{297000}}\, {t}^{11}
 \, \, \, \, \, + \,\,  \cdots
\end{eqnarray}
{\em Each series} (\ref{un}) and  (\ref{deux})  {\em is globally bounded}. According
 to conjecture (\ref{conjmain}), these two globally bounded series
 solutions of the linear differential operator $\, L_{12}$,
 {\em should be diagonal of rational functions}
of $ \, N_v= \, 3\, +2 = \, 5 \, $ variables.

Around the singularity $\,t =\, \infty = \, 1/s$, there are also two series in front of $\,\ln(s)^3$,
(the same value of the exponent), and these series are {\em also globally bounded}.
At the singularity $\,t= \, 1/4$, there is one series in front of $\,\ln(t \, -1/4)^3 \, $
among the formal solutions, and this series is  {\em actually globally bounded}.

Around each of the other singularities (with rational values), i.e.
$\, t = \, -1/4, \,  -1/2, \,  -1, \, 1$,
the maximum of power log in the formal solutions
is, respectively, $ \, 2, \, 1, \, 1, \, 1$, and all these series are {\em not globally bounded}.

For the singularities $\, t = \, t_s$, roots of the polynomials in (\ref{singL12})
of degree 2, 3 and 4, we only checked that, for each, 
the formal solutions behave at the most as $\,\ln(t-t_s)^1$.

\section{Other examples}
\label{split}

Here, we give some examples that contradict, at first sight,
the affirmation that the diagonal of $\, 1/Q\, $ with non factorizable 
(resp. factorizable) $\, Q$ over the rationals, is annihilated by an irreducible
(resp. factorizable) linear differential operator. 

The three rational functions $ \, 1/Q_j$ with $\, Q_j$ given by 
\begin{eqnarray}
  Q_1 \, \,=\,\,\,\,  1 \, \, \, \,  +(x+y+z \, \,  +x y + y z \, \, \,   -x^3 y z),
  \\
Q_2 \, \,=\, \,\,\, 1 \,\,\,  \, -(x+y+z  \, \,+x^4 y),  
\\
Q_3\,  \,=\, \,\, \,1 \,\, \,\,  -\left(x+y+z \,\, \, +x^2 \cdot \,  (y + z)   \right), 
\end{eqnarray}
rule out the irreducibility statement. 
The diagonal of the rational function $\, 1/Q_j$ is annihilated by an
order-four linear differential operator
$\, L_4^{(j)}$ which factorizes as a {\em direct sum} of 
two order-two linear differential operators ($L_2^{(j)}$ and $\, M_2^{(j)}$): 
\begin{eqnarray}
  \qquad \quad
 L_4^{(j)}\, \, = \, \, \, L_2^{(j)} \, \oplus \, M_2^{(j)}. 
\end{eqnarray}
The three examples follow exactly the same features. For the three examples, one has
\begin{eqnarray}
 &&  \hspace{-0.98in}  \quad  \quad \, \, \,   \quad 
  {\rm sol}(L_2^{(j)})\,  \, = \, \, \,
    {\rm Diag} \left( {1\, +x \over Q_j} \right), \qquad
    {\rm sol}(M_2^{(j)})  \,  \,= \,\,  {\rm Diag} \left( {1 -x \over Q_j} \right), 
\end{eqnarray}
and for the three examples, their solutions $\, {\rm sol}(L_2^{(j)})$ and $\, {\rm sol}(M_2^{(j)})$
can be  written as pullbacked hypergeometric function  
$\, {}_2F_1\left([1/12, 5/12],[1], \, \bullet \right) \,$ with an algebraic
prefactor (see~\cite{ABKM-2002.00789}, for $\, Q_1$).

\vskip 0.1cm

Note however, that when the polynomials $\, Q_j\,$ are considered with {\em  generic values}
of the coefficients in front of the monomials, the resulting
linear differential operators, annihilating the diagonal of $\, 1/Q_j$,
are of order four, and  are {\em irreducible}.

\vskip 0.1cm

Introducing a parameter $\, \mu$ in $\, Q_1$
\begin{eqnarray}
&&  \hspace{-0.98in}  \quad  \quad \, \, \,  \quad \quad \quad
\,  \, \, 
Q_1(\mu) \,\, \, = \,\, \, \,\, 
  1 \, \, \,\,  +(x+y+z \,\,  +x y + y z \, \, \, \, - \mu\cdot \,  x^3 y z), 
\end{eqnarray}
and using the method of factorization of linear  differential operators modulo primes
(see Section 4 in~\cite{2009-chi5} and 
Remark 6 in~\cite{2014-Ising-SG}), one finds that the linear differential operator annihilating
the diagonal 
$\, {\rm Diag}(1/Q_1(\mu))$ factorizes for only two 
values of $\, \mu$. The trivial $\, \mu=\, 0\, $ where $\, Q_1(0)\,  =\,  (1\, +x+z)\, (1+y)$,
and the particular value $\, \mu =\, 1$.

\vskip 0.1cm

The example with $ \, Q_1(\mu=1)\, $ is presented in~\cite{ABKM-2002.00789} and analyzed
via the notion of split Jacobian~\cite{kumar-2015}.

Note that either irreducible or factorizable in direct sum, the linear differential operators
$\,  L_2^{(j)}$, $\, M_2^{(j)}$ have their differential Galois groups
included in  symplectic  groups,
and their formal solution with the highest log-power behaves as $\, \ln(t)^1$, indicating 
a minimal number $\,\, N_v= \, 3 \, = \, 1 \, +2 \,\, $
of variables occurring in the rational function $\, 1/Q_j$.
This is also the case for the order-four irreducible differential operator corresponding
to the diagonal of $1/Q_1(\mu)$, for generic values of the parameter $\, \mu$.

\vskip 0.1cm

\section{Other examples of  diagonal with a factorization  of the denominator: $ \, Q = \, Q_1\,Q_2$ }
\label{otherfactofamillies}

\subsection{Generalization of example (\ref{Q69}) \\}
 
 Let us generalize the example given in (\ref{Q69}), with polynomials:
\begin{eqnarray}
  \label{Q69n}
&&  \hspace{-0.98in}  \quad \quad \quad \quad  \quad 
  {\tilde Q}_n \,\, \,= \, \,\, 
  \Bigl(1 \, -(x+y+z) \Bigr) \cdot \, (1 \, -y - z^n) \quad \quad \,\,\,\,\,\quad n \, = \, 3, \, 4.
\end{eqnarray}
For $\, n= \, 3$ the diagonal of $\,1/{\tilde Q}_3$ is annihilated by an order-eight linear
differential operator $\, L_8$  with the unique factorization
$\, L_8 \, = \, L_2 \cdot \, L_6$, where the order-two linear
differential operator $\, L_2$ is homomorphic to the order-two operator
annihilating the diagonal of $\, 1/\Bigl(1 \, -(x+y+z) \Bigr)$, and where the  ``dressing'' order-six linear
differential operator $\, L_6$,  has {\em no logarithmic} formal
solution around {\em all} the singularities, and where  all these formal 
solutions are globally bounded.  In fact it is quite hard to see that these
globally bounded series are  {\em algebraic series}. However, the calculation of the $p$-curvature of the 
linear differential operator $ \, L_6$ shows that these
$\, p$-curvatures are zero for $\, 11 \le p \le \, 73$.
For $\, n= \, 4$ the diagonal of $\,1/{\tilde Q}_4$ is annihilated by an order-ten linear
differential operator $\, N_{10}$  with the unique factorization
$\, N_{10} \, = \, N_2 \cdot \, N_8$, where the order-two linear
differential operator $\, N_2$ is homomorphic to the order-two operator
annihilating the diagonal of $\, 1/\Bigl(1 \, -(x+y+z) \Bigr)$, and where, again, the  ``dressing'' order-eight linear
differential operator $\, N_8$,  has {\em no logarithmic} formal
solutions around {\em all} the singularities, and where all these formal 
solutions are globally bounded. Again it is quite hard to see that these
globally bounded series are algebraic series. The calculation of the $p$-curvature of the 
linear differential operator $ \, N_8$ shows that these
$\, p$-curvatures are zero for $\, 17 \le p \le \, 73$, strongly suggesting  {\em algebraic solutions}. 

\subsection{A second familly}
\label{sectex}
Let us consider the polynomials similar to polynomials (\ref{Q69n}): 
\begin{eqnarray}
  \label{Qn}
&&  \hspace{-0.98in}  \quad \quad  \quad \quad 
  {\tilde Q}_n \,\, \,= \, \,\, 
  \Bigl(1 \, -(x+y+z) \Bigr) \cdot \, (1 \, \, -y \,z^n) \quad \quad \,\,\,\,\quad
  n \, = \,2, \, 3, \, 4.
\end{eqnarray}
The diagonal of $\,1/{\tilde Q}_n \, $ is, now, annihilated by a linear
differential operator $\, L_{2\,n\, +2}$  of order $\, 2\, n \, +2\, $ with the unique factorization
$\, L_{2\,n\, +2} \, = \, L_2 \cdot \, L_{2\,n}$, where  the order-two linear
differential operator $\, L_2$ is homomorphic to the order-two operator
annihilating the diagonal of $\, 1/\Bigl(1 \, -(x+y+z) \Bigr)$, and where the  ``dressing'' linear
differential operators $\, L_{2\,n}$, of order $\, 2\, n$,  have {\em no logarithmic} formal
solution around {\em all} the singularities, and where  all these formal 
solutions are globally bounded. Again, it is quite hard to
see that these series are algebraic series.
The calculation of the $p$-curvature of the 
linear differential operators shows that these
$\, p$-curvatures are zero, suggesting (according to Grothendieck-Katz
conjecture~\cite{Grothendieck-Katz}) that all these series are
 {\em algebraic series}.

\vskip 0.1cm

\section{Direct sum versus unique factorization for the diagonal of $\, 1/Q$, with $ \, Q = \, Q_1\,Q_2$}
\label{directsumtounique}

The examples of section \ref{facto}  dealt with a polynomial $\, Q$ that factorizes as $\,\, Q_1^{(c)} \, Q_2$, where
the polynomial $\,  Q_1^{(c)}\,$ contains {\em all} the variables, and the polynomials $\, Q_2\,$
has a smaller number of variables. Call $\, L_q$, the linear differential operator annihilating the diagonal
of $ \,  1/Q_1^{(c)}$. The linear differential operator, annihilating 
the diagonal of $ \, 1/ Q_1^{(c)}/Q_2 $, appears with a unique factorization
as $ \, N_q \cdot \, M_n$, where $ \, N_q$ is homomorphic with $\, L_q$.
There is no known relation between the linear differential operator, corresponding to $ \, 1/Q_2$,
and the "dressing" linear differential operator $ \, M_n$.

\vskip 0.1cm

\subsection{Direct sum}
Let us illustrate equation (\ref{foreword1}).  
Let us consider the situation where $\, Q$ factorizes as $ \, Q_1^{(c)} \, Q_2^{(c)}$,
and where both polynomials $ \, Q_1^{(c)}\,$ and $\,  Q_2^{(c)}\,$
carry all the variables. Call $\, L_q$, $L_n$ and ${\cal L}_m$ the linear differential operators
annihilating respectively, $ \, {\rm Diag}(1/Q_1^{(c)})$, $ \, {\rm Diag}(1/Q_2^{(c)})$
and $ \, {\rm Diag}(1/Q_1^{(c)}/Q_2^{(c)})$.
Since the polynomials $\, Q_1^{(c)}$ and $\, Q_2^{(c)}$ are on an equal footing in terms
of number of variables, one may expect the linear differential operator $\, {\cal L}_m \, $
to have a left factor homomorphic to $\, L_q$ as well as another left factor 
homomorphic to $\, L_n$. This means that $\, {\cal L}_m \, $ has a factorization in {\em direct sum}
between the homomorphic linear differential operators of $\, L_q \, $ and $\, L_n$.

\vskip 0.1cm

Take the example $ \, Q =  \, Q_1 \, Q_2\, $ where:
\begin{eqnarray}
&&  \hspace{-0.98in}  \quad   \quad \quad  \quad  \quad  
  Q_1 \,\, = \,\,\,\, 1 \,\, \, -(x +y +z),  \qquad  \quad
  Q_2 \,\,=\, \,\,\, 1\, \,\,  -(x+y\,+z^3). 
\end{eqnarray}
We have (the same) three variables occurring in both polynomials. Call $\,L_2$ (resp. $L_6$)
the order-two (resp. order-six) 
linear differential operator annihilating the diagonal of $\,1/Q_1$ (resp. $\,1/Q_2$).
The diagonal of $\,1/Q_1/Q_2 \, $ is annihilated by an order-ten linear differential operator that
factorizes in a {\em direct sum} as
\begin{eqnarray}
&&  \hspace{-0.98in}  \quad    \quad \quad 
   L_{10}\,  \,\, = \, \,\, \, (N_2 \oplus N_6) \cdot \,  N_1 \cdot \, L_1
  \,  \, \, = \, \, \, \,
   (N_2 \cdot \, N_1 \cdot \, L_1) \, \oplus \, (N_6 \cdot \, N_1 \cdot \, L_1), 
\end{eqnarray}
where $\, N_1$ and $\, L_1$ are order-one linear differential operators 
\begin{eqnarray}
&&  \hspace{-0.98in}  \quad  \,   \quad \quad \quad \quad \quad \quad 
  N_1 \cdot L_1 \, \, = \,\,\,
  \left(  D_t \,  + {1 \over 2 \, t} \right) \oplus \,\left(  D_t \,+ {1 \over 2 \cdot \, (1+t) } \right), 
\end{eqnarray}
with {\em algebraic} solutions $\,  t^{-1/2}$ and $\, (1+t)^{-1/2}$.
The linear differential operator $\, N_2$ (resp. $\, N_6$)
{\em is homomorphic with} $ \, L_2$ (resp. $ \, L_6$). Here, we have
a direct sum $\, N_2 \oplus \, N_6$:
there is no unique factorization. Both $\, N_2$ and $\, N_6$
are such that their differential Galois groups are  symplectic, and both have 
formal solutions with the highest log-power at the origin
behaving in $ \, \ln(t)^1$, in agreement with
 $ \, N_v = \, 1 \, +2 \, = \, 3\, $ variables for the rational function we started with.

\vskip 0.1cm

\subsection{From direct sum to unique factorization}
\label{fromdirectsumto}

Let us see how the direct sum {\em actually reduces to a simple product},
when one of the factors $\, Q_j$ has less variables. 
Take the rational functions $\, 1/Q_1 \, $ and $\, 1/Q_2 \, $ with the polynomials: 
\begin{eqnarray}
  \label{Q1Q2def}
 &&  \hspace{-0.98in}     \quad \,  \, 
    Q_1 \, \, = \, \, \, \, 1 \,\,\,\,  -(x+y+z),  \,   \, \, \,  \quad \quad 
       Q_2(\alpha) \, \,=\, \,\, \, 1 \, \,\, \, -(x+y \,  \, +\alpha \cdot \, z \, \,  +x y).
\end{eqnarray}
The diagonal $\,{\rm Diag}(1/Q_1)$ (resp. ${\rm Diag}(1/Q_2(\alpha))$) is annihilated
by an order-two linear differential operator $\, L_2^{(1)}$ (resp. $\,L_2^{(2)}$).
The diagonal of $\, 1/Q_1/Q_2(\alpha)$ 
\begin{eqnarray}
\label{Q1Q2alpha}
&&  \hspace{-0.98in}    \, \,   \quad \quad  \quad  \quad 
   {\rm Diag}\left( {\frac{1}{Q_1\, Q_2}}(\alpha) \right)
    \\
&&  \hspace{-0.98in}   \quad  \quad  \quad    \quad  \quad  \quad  \quad 
   \, \, = \,\,\,\, \, \, 
   1 \, \, \, \, +(13+14 \,\alpha) \cdot \,t \, \,\, \,   +(241+273\, \alpha +306 \,\alpha^2) \cdot \,t^2
   \, \, \, \, \, \,  + \, \cdots
   \nonumber
\end{eqnarray}
is annihilated by a (quite large\footnote[1]{The polynomial coefficients of $\, L_6$
  are of degree $\, 21$ in $\, t$ and degree $\, 22$  in $\, \alpha$. })
order-six linear differential operator $ \, L_6$, depending on $\, \alpha$,
and it factorizes {\em in a direct sum} as:
\begin{eqnarray}
  \label{DIRECTSUM}
 &&  \hspace{-0.98in}  \quad   \quad \quad 
  L_6   \,\, =  \, \, \,  \, \left( N_2^{(1)} \oplus \, N_2^{(2)} \right) \cdot \, H_2
    \,\,   \, =  \, \, \, \,
    (N_2^{(1)} \cdot \, H_2) \oplus  \, (N_2^{(2)} \cdot \, H_2). 
\end{eqnarray}
The order-two linear differential ``dressing'' operator $ \, H_2\, $
has two {\em algebraic} solutions:
\begin{eqnarray}
&&  \hspace{-0.98in}  \,   \quad 
 {\rm sol}(H_2)  \, \, = \, \,\,
   {\frac{1}{\sqrt{t} \cdot \, \sqrt{p_2}}} \cdot \, \left( const. \, A^{1/4} \, \,
   + const. \, (t \,  \, -\alpha \, +1) \cdot \, A^{-1/4}  \right), 
 \quad \quad \quad \hbox{where:} 
 \nonumber  \\
&&  \hspace{-0.98in}  \quad  \quad   \quad  \quad  \quad  
 A \, \,=\, \,\,\,
 p_1 \, \, \, + 4 \cdot \, (2 \alpha-3) \cdot \, \sqrt{\alpha-1} \cdot \,  \sqrt{t} \cdot \,  \sqrt{p_2},
  \quad \quad \quad \quad \quad \quad  \hbox{where:} 
    \nonumber \\
&&  \hspace{-0.98in}  \quad  \quad   \quad   \quad  \quad  
p_1\,  \,=\,\,\,\,
t^2 \, \,\,
      +2 \cdot \,  (\alpha-1) \cdot \,  (4 \alpha-5)  \cdot \, (4 \alpha-7) \cdot \, t \,\, \,  +(\alpha-1)^2,
   \nonumber \\
 &&  \hspace{-0.98in}  \quad  \quad   \quad \quad  \quad  
 p_2 \, \,=\,\,\, \,  t^2\,\,\,
     +2  \cdot \, (\alpha-1)  \cdot \, (8 \alpha^2-24 \alpha+17) \cdot \, t \,\,\,  +(\alpha-1)^2.
\end{eqnarray}
The order-two linear differential operator $\, N_2^{(1)}$ (resp. $\, N_2^{(2)}$) {\em is homomorphic with the
linear differential operator} $ \, L_2^{(1)}$ (resp. $\, L_2^{(2)}$). The linear differential
operator $ \, L_2^{(1)} \, $ does not depend on $ \, \alpha$,
{\em but the intertwiner} $\, W_1^{(1)} \, $  {\em does}:
\begin{eqnarray}
  &&  \hspace{-0.98in}  \quad  \quad \quad \quad \quad \quad \quad \quad \, \,\, 
     N_2^{(1)}(\alpha) \cdot \,  W_1^{(1)}(\alpha)
     \, \,\, = \, \,\,  \, {\tilde W}_1^{(1)}(\alpha) \cdot \, L_2^{(1)}. 
\end{eqnarray}

\vskip 0.1cm

From the direct-sum decomposition (\ref{DIRECTSUM}), the diagonal of
$\, 1/Q_1/Q_2(\alpha) \,$ which is annihilated by the linear
differential operator $\, L_6$, reads:
\begin{eqnarray}
&&  \hspace{-0.98in}  \, \,    
 {\rm Diag}\left( {\frac{1}{Q_1\, Q_2(\alpha)}}(\alpha) \right) \, = \, \, \,\,
 {\rm sol}\left( N_2^{(1)}(\alpha) \cdot \, H_2(\alpha) \right) \,\,\,
 + {\rm sol}\left( N_2^{(2)}(\alpha)  \cdot \, H_2(\alpha) \right).
\end{eqnarray}
For $\, \alpha = \, 0$ the analytical solution (at $ \, t= \, 0$) of
$\, N_2^{(2)}(\alpha) \cdot \, H_2(\alpha)$ is the analytical solution (at $ \, t= \, 0$) of
$ \, H_2(\alpha)$ for $\, \alpha = \, 0$, namely the algebraic series 
\begin{eqnarray}
  \label{solH2}
&&  \hspace{-0.98in}  \quad 
  {\cal A} \, \, \,  = \,  \,
\,\,\,\,  1 \,\,\, +21 \,t \,\,\, +561 \,t^2 \,\, \,+16213\,{t}^{3} \, \, +487521\, t^4 \, \, +15015573\, t^5
\,\, \,  +\, \cdots
\end{eqnarray}
{\em together with  the constant function}.
The other solution of $ \, H_2(\alpha)$, for $\, \alpha = \, 0$, is not analytic at $\, t= \, 0$:
\begin{eqnarray}
&&  \hspace{-0.98in}  \quad  \quad  \quad  \quad \quad 
    S_2 \, = \,\, \,
    t^{-1/2} \cdot \, \left( 1\, \,\,  +13\,t\,\,  \, +321\,{t}^{2}\,  \, +8989\,{t}^{3}
    \, \,\, \,   + \, \cdots \right). 
 \end{eqnarray}

The analytical solutions (at $ \, t= \, 0$) of $\, N_2^{(1)}(\alpha) \cdot \, H_2(\alpha) \, $
for $\, \alpha = \, 0$ are, the analytical solution (at $ \, t= \, 0$) of
$ \, H_2(\alpha)$ for $\, \alpha = \, 0$, namely (\ref{solH2}), together
with the series:
\begin{eqnarray}
  \label{solN1H2}
  &&  \hspace{-0.98in}   \quad  \quad    \quad  \quad    \quad  \quad
 {\cal S} \, \, \,  = \,  \,
\,\,\,\,   t \, +40\, t^2 \, +1400\, x^3 \, +47110\, t^4 \, +1560328\, t^5
  \, \, + \, \, \, \cdots 
\end{eqnarray}
 The diagonal series  (\ref{Q1Q2alpha}) for $\, \alpha= \, 0$ reads: 
\begin{eqnarray}
 &&  \hspace{-0.98in}   \quad  \quad  \quad  
  1\,\, \, +13\,t \,\,  \, +241\,t^2 \, \, \, + 5013\,t^3 \, \, +110641\, t^4 \, \,
  +2532949\, t^5 \, \,\, + \,\cdots
\end{eqnarray}
which is nothing but the following linear combination of (\ref{solH2}) and (\ref{solN1H2}):
\begin{eqnarray}
 &&  \hspace{-0.98in}   \quad  \quad  \quad  \quad  \quad  
  {\rm Diag}\left( {\frac{1}{Q_1 \cdot \,  Q_2(\alpha)=0}} \right)
  \, \, \, = \,\, \,\,\, \, 
    {\cal A}  \,\,  -8 \cdot \,  {\cal S}.
\end{eqnarray}
This series is, thus, annihilated by {\em only} the linear
differential operator $\, N_2^{(1)} \cdot \, H_2$, for $\, \alpha= \, 0$,
 {\em with no need of the linear differential operator} $\, N_2^{(2)} \cdot \, H_2\, $
for  $\, \alpha= \, 0$.

\vskip .1cm 

 {\bf Remark F.1.} Note that the ``spurious'' linear differential operator $ \, N_2^{(2)} \cdot \, H_2$,
 actually factorizes as a direct sum for $\, \alpha =\, 0$
 \begin{eqnarray}
 &&  \hspace{-0.98in}   \quad  \quad  \quad \quad  \quad   \quad  
   N_2^{(2)}\cdot  \, H_2 (\alpha=0) \,\, \, = \,\,\, \,
      D_t \,  \oplus \,  \, \Bigl(L_1 \cdot \, H_2 (\alpha=0)\Bigr), 
 \end{eqnarray}
 where the order-one linear differential operator $\, L_1$ annihilates
 a rational solution.

\vskip 0.1cm


\section{Rational function from four variables to three variables as a parameter varies}
\label{rat4to3}

The example in section \ref{anotherdiag}  shows that, depending on the number
(four or three) of variables on which the diagonal is performed, 
we obtain, in the formal solutions of the correspondig linear differential operators,
the power 2 (i.e. $\ln(t)^2$) or the power 1  (i.e. $\ln(s)^1$)
in agreement with conjecture (\ref{conjmain}).

Here, we consider an example (depending on one parameter) where the diagonal
is performed on all the four variables, and the result,
for some value of the parameter ($b= \, 2$ in \ref{F3sub} below),
will indicate that we deal, in fact, with only three variables
in agreement with the power $ \, \ln(t)^1$ obtained in the formal solutions.

\vskip 0.1cm  

Let us consider the diagonal of the following rational function of {\em four} variables
$\, x, \, y,\,  z, \, u$, and a parameter $\, b$
\begin{eqnarray}
\label{diagab}
 &&  \hspace{-0.98in}     \quad \quad \,  \quad  \quad  
 {\rm Diag}\left( {\frac{1}{ Q_1\, Q_2 }} \right)
 \, \, = \,\, \,\,\, 
    1 \, \,\, \, +20 \cdot \, ( b \, +2) \cdot \, t \,\,\, \, +756 \cdot \,
  \Bigl( b^2 \, +3\, b \, +6 \Bigr) \cdot \, t^2 
\nonumber  \\
 &&  \hspace{-0.98in}   \quad \quad   \, \, \qquad \qquad \quad
  +34320 \cdot \,
  \Bigl(b^3 \, +4\, b^2\,  +10\, b\,+ 20 \Bigr) \cdot \, t^3
  \, \, \,\,\,  + \,\,  \cdots
\end{eqnarray}
where:
\begin{eqnarray}
 &&  \hspace{-0.98in}    \quad  \quad \quad \, \,
Q_1 \, \,= \,\,  \, 1 \,\, \,  \, -x \, -y \, -z \, - \, u,
\qquad \quad 
Q_2  \,\,= \, \, \, 1  \,\,  \, \, -x \, -y \, -b \cdot \, z. 
\end{eqnarray}
The power series (\ref{diagab}) is annihilated (for generic values of $\,b$)
by an order-six linear differential operator with the unique 
factorization
\begin{eqnarray}
  \label{L6b}
  &&  \hspace{-0.98in}  \quad   \quad  \quad \quad  \quad    \quad  \quad   \quad   \quad \quad
L_6(b) \, \, = \, \, \,
     L_3(b)  \cdot \, L_2(b) \,  \cdot \, L_1
\end{eqnarray}
where the order-one linear differential operator,  $\, L_1$, reads $\, D_t \, +1/2/t$.

The order-three linear differential operator $ \, L_3(b) \, $ has the following
hypergeometric solution:
\begin{eqnarray}
\label{L3bsol}
  &&  \hspace{-0.98in}   \quad \quad \quad \quad\quad \quad
{\frac {1}{t^2 \cdot \, (1 -b \, \, \, +64 \,b^2\cdot \, t)}} \cdot \,
  _3F_2\Bigl([{{3}\over {2}}, \, {{7}\over {4}}, \, {{9}\over {4}}], \,  [1,  \, 2],
     \, \, 256 \cdot  \, t  \Bigr).
\end{eqnarray}
The formal series solutions of $\, L_3(b)$ have, at most, a $\, \ln(t)^2$.

The order-two linear differential operator\footnote[2]{This order-two
linear differential operator is {\em not} homomorphic with the order-two
linear differential operator annihilating $\, 1/Q_2$.}
$ \, L_2(b) \, $ has the following $\, _2F_1$
hypergeometric solution:
\begin{eqnarray}
\label{L2bsol}
  &&  \hspace{-0.98in}   \quad \quad \quad \quad  \quad \quad \quad \quad \quad \quad
     {{1} \over {t}} \cdot \, _2F_1\Bigl([{{3}\over {4}}, \, {{5}\over {4}}], \,  [1],
        \, \, {{ 64 \cdot \, b^2} \over {b-1  }} \cdot  \, t  \Bigr).
\end{eqnarray}
The formal series solutions of $\,  L_2(b) \, $ have, at most, a $\, \ln(t)^1$
logarithmic power.

The formal series solutions of $\, L_6(b) \, $ have, at most, a $\, \ln(t)^2$
logarithmic power. This is in agreement with 
conjecture (\ref{conjmain}), and the fact that  (\ref{diagab})
corresponds to a rational function of $\, 2\, +2\, = \, 4\, $  variables.

\vskip 0.1cm

Let us introduce the  order-three linear differential operator $\, U_3$, annihilating 
\begin{eqnarray}
\label{diagQ1}
 &&  \hspace{-0.98in}   \quad  \quad  \quad  \quad  \quad  
{\rm Diag}\left( {\frac{1}{ Q_1 }} \right) \,\,= \,\,\,\,\, 
1 \,\, \, +24\,t \,\,\, \, +2520 \,{t}^{2}\,\,\,  +369600\,{t}^{3}
  \,  \,\, \,+ \,\,\, \cdots
\end{eqnarray}
This  order-three linear differential operator  reads:
\begin{eqnarray}
\label{U3}
 &&  \hspace{-0.98in}     \quad  \quad   \quad  \quad  \quad   \quad  
U_3 \, \, = \, \, \, \, \,
 (1 \, -256\, t) \cdot \, t^2\, \cdot \,  D_t^3
    \, \,  \,\,
  +3 \,\cdot \, (1 \, -384\,  t) \cdot \, t \,  \cdot \,  D_t^2
\nonumber \\
  &&  \hspace{-0.98in}    \quad   \quad  \quad    \quad  \quad  \quad  \quad   \quad   \quad  \quad    
  \, +(1 \, -816 \, t) \cdot \, D_t \, \,\,\, -24 
\end{eqnarray}
The formal series solutions of $\,U_3$ have, at most, a $\, \ln(t)^2\,$
logarithmic power, in agreement with 
conjecture (\ref{conjmain}) and the fact that  $\,1/Q_1$ is  a rational function of four variables.
This order-three linear differential operator $\,U_3$ is, in fact, the symmetric square
of an order-two  linear differential operator.  The diagonal (\ref{diagQ1})
reads:
\begin{eqnarray}
\label{diagabSOL}
  &&  \hspace{-0.98in}   \quad \quad \quad \quad 
 _3F_2\Bigl([{{1} \over{4}}, \, {{2} \over{4}}, \, {{3} \over{4}}], \, [1, \, 1], \,   256\, t  \Bigr) 
 \, \, = \, \, \,
     \Bigl( \, _2F_1\Bigl([{{1} \over{8}}, \, {{3} \over{8}}], \, [1], \,   256\,  t  \Bigr) \Bigr)^2.
\end{eqnarray}
The  order-three linear differential operator $\, L_3(b)$, in the factorization given in (\ref{L6b}),
is {\em actually homomorphic} to the
order-three linear differential operator $\, U_3$.

\vskip 0.1cm

All the results above are for the generic values of the parameter $\,b$.
In the sequel we consider the situation with the values of $\,b$ for which the series (\ref{diagab})
will be annihilated by a linear differential operator of order less than six.  
These values of $\,b$ are\footnote[1]{These values of $\,b$ can be obtained using the 
  method of factorization of linear differential operators
 modulo primes (see Section 4 in~\cite{2009-chi5}
and Remark 6 in~\cite{2014-Ising-SG}). 
They can also be obtained by reducing the four singularities
$\,t = \, \infty,\, 0, \,1/256, (b-1)/64/b^2 $ of $\, L_6(b)$
to the three singularities
$\,t =  \infty, 0, 1/256 $.}  $\,b=\, 0$, $\,b=\, 1$ and $\,b=\, 2$.

\vskip 0.1cm

\subsection{The  $\,b = \, 0\, $ case}
For the value of the parameter $\,b= \, 0$, the factorization of the order-six
linear differential operator $\,L_6(b)$ becomes
\begin{eqnarray}
\label{L6b0}
&&  \hspace{-0.98in}     \quad   \quad  \quad \quad   \quad   \quad \quad \quad
 L_6(b=0) \, \, \, = \, \,  \, \,
 \Bigl(\Bigl(  D_t +{1 \over t} \Bigr)  \, \cdot \, D_t  \Bigr)
    \, \oplus \,  \Bigl( N_3 \, \cdot \, L_1 \Bigr).
\end{eqnarray}
For  $\,b= \, 0 \, $ the series (\ref{diagab}) is annihilated by the order-four linear
differential operator $ \, N_3 \,\cdot\,  L_1$,
whose solution (analytic at $t\, = \, 0$) reads:
\begin{eqnarray}
  &&  \hspace{-0.98in}   \, \quad \quad \quad \quad \quad \quad
 {\rm sol}(N_3 \cdot \, L_1 ) \,  \,  \, = \,  \, \,
 \, _4F_3\Bigl([{{1} \over{2}}, \, {{1} \over{2}}, \, {{3} \over{4}},\, {{5} \over{4}}], \,
 [1, \, 1,\, {{3} \over{2}}], \,   256\, t  \Bigr). 
\end{eqnarray}
The formal solutions, at the origin, of $ \, N_3 \,\cdot L_1 \,$
carry the maximum exponent $ \, \ln(t)^2$,
indicating, according to conjecture
(\ref{conjmain}), that we deal with a rational function with $\, 4 \, = 2\, +2 \, $ variables.

\vskip 0.1cm

\subsection{The  $\,b= \, 1$ case}
For the value of the parameter $\,b= \, 1$, the factorization of the order-six linear
differential operator $\,L_6(b) \, $ reads:
\begin{eqnarray}
\fl \qquad
  \label{L6b1}
 &&  \hspace{-0.98in}     \quad  \quad \quad   \quad  \quad  \quad \quad \quad \quad
    L_6(b=1) \, \, = \, \, \,
    \Bigl( D_t \, +{1 \over 2 t} \Bigr)\, \oplus \,
    \Bigl( D_t\,  +{3 \over 4 t} \Bigr)\, \oplus \,  \Bigl( D_t \, +{5 \over 4 t} \Bigr)\, 
 \oplus \,  M_3. 
\end{eqnarray}
For $\,b= \, 1$ the series (\ref{diagab})  is annihilated by the order-three linear
differential operator $ \, M_3$, with the 
(analytic at $\, t= \, 0$) solution: 
\begin{eqnarray}
  &&  \hspace{-0.98in}   \quad \quad \quad \quad \quad \quad  \quad \quad 
 {\rm sol}(M_3 ) \, \,= \, \,\,
   _3F_2\Bigl([{{1} \over{2}}, \, {{3} \over{4}},\, {{5} \over{4}}], \, [1, \, 2], \,   256\, t  \Bigr). 
\end{eqnarray}

The formal solutions of $\, M_3$, at the origin,  carry the maximum exponent $\ln(t)^2$
indicating, according to conjecture
(\ref{conjmain}), that we deal with a rational function with $\, 2 \, +2 \, = \, 4\, $
variables. This rational function actually reads (with $ \, t=\, x y z u$):
\begin{eqnarray}
\ &&  \hspace{-0.98in}   \,\, \, \, 
  { \left( 1-320\,t \right) \cdot \, Q_1^2 \, \, \,
    -\left( 1-384 \,t \right) \cdot \, Q_1 \, \, \,  -2 \cdot \,(1-256\,t )
 \cdot \,  \left( x^2+y^2+z^2+u^2 \right)  \over 
32 \cdot \, t \cdot \, Q_1^3}. 
\end{eqnarray}

\vskip 0.1cm

\subsection{The $\,b=\, 2$ case}
\label{F3sub}
For  $\,b= \, 2 \, $ the factorization of the order-six linear
differential operator $\, L_6(b) \, $ becomes:
\begin{eqnarray}
  \label{L6b2}
  &&  \hspace{-0.98in}     \quad  \quad   \quad   \quad \quad \quad \quad \quad\quad
L_6(b=2) \, \, = \, \, \, \,  V_3 \, \oplus  \,  \Bigl( L_2(b=2) \,\cdot L_1 \Bigr).
\end{eqnarray}
The series (\ref{diagab}), for the parameter $\,b= \, 2$, is annihilated by the order-three
linear differential operator $\,  L_2(b=2) \,\cdot \, L_1$
with the hypergeometric solution:
\begin{eqnarray}
\label{solL2b2}
  &&  \hspace{-0.98in}   \quad \quad \quad  \quad \quad  \quad \quad 
 {\rm sol}(L_2(b=2) \,\cdot L_1 ) \,= \,\, \, 
 _3F_2\Bigl([{{1} \over{2}}, \, {{3} \over{4}},\, {{5} \over{4}}],
     \, [1, \, {{3} \over{2}}], \,   256 \, t  \Bigr)
  \\
  &&  \hspace{-0.98in}   \quad \qquad \qquad \quad  \quad \quad  \quad \quad  \,
  \,=\,\, \,  \, 1 \,\,  \, +80\, t \,\, \,  +12096\, {t}^{2} \,  \, +2196480\, {t}^{3}
     \,\, \,  \, \,  +\, \,  \cdots
 \nonumber 
\end{eqnarray}
The formal solutions at the origin of  $ \,  L_2(b=2) \,\cdot L_1$
carry the maximum exponent $ \, \ln(t)^1$
indicating, according to conjecture
(\ref{conjmain}), that we deal with a rational function with $\, 3 \, = \, 1\, +2\, $
variables, while the rational function
we started with, was dependent on four variables. 

The power series (\ref{solL2b2}) should, then,  be the diagonal of a rational function
depending on {\em only three} variables.

Introducing the polynomial
\begin{eqnarray}
&&  \hspace{-0.98in}   \quad \quad \quad \quad \quad \quad \quad \quad \quad \quad
Q_3 \, \,=\, \, \, \,  1  \, \, \,\,\,  -x \, -y \, \,\, \, -b \cdot \,u,  
\end{eqnarray}
the partial fraction form of $\, 1/Q_1/Q_2\, $ in the variable $\,z$,  reads for $\,b= \, 2$:
\begin{eqnarray}
  \label{partfrac}
&&  \hspace{-0.98in}   \quad \quad \quad \quad \quad\quad \quad \quad \quad \quad \quad
 {\frac{1}{Q_1 \,Q_2}}\,  \,= \, \,\,
        - {\frac{1}{Q_1 \,Q_3}} \, \,\, +{\frac{2}{Q_2 \,Q_3}}.
\end{eqnarray}
The diagonal of $\, 1/Q_1/Q_2$,   and the diagonal of $\,1/Q_1/Q_3$,  are identical, by the symmetry
$\, (z, \, u) \,  \rightarrow \, (u, \, z)$.
We end up with
\begin{eqnarray}
 &&  \hspace{-0.98in}   \quad \quad \quad \quad \quad \quad \,\,\,
  {\rm Diag}\left( {\frac{1}{ Q_1\, Q_2 }} \right) \,  \,= \,  \,  \,
  {\rm Diag}\left( {\frac{1}{ Q_2\, Q_3 }} \right)
  \\
 &&  \hspace{-0.98in}   \quad  \qquad \qquad \qquad \qquad
  \, = \, \, \,  {\rm Diag}\left( {\frac{1}{ \left( 1  \, \, -x \, -y \, -2 \,z \right) \cdot \,
  \left( 1  \, \, -x \, -y \, -2 \,u \right)  }} \right). 
  \nonumber
\end{eqnarray}
It remains to show that this diagonal depends, in fact,
{\em on only three variables} instead of four variables:
\begin{eqnarray}
 &&  \hspace{-0.98in}   \quad \quad \quad \quad  \,\,\,\,  
 {\rm Diag}\left( {\frac{1}{ Q_2\, Q_3 }} \right) \,\, = \,\,  \,
 {\rm Diag}\left( {1 \over (1 \,\,  -x-y)^2}
 \cdot \, \sum_{i, j} {\frac{(2 z)^i\, (2u)^j}{ (1\,\,  -x-y)^{i+j} }}  \right)
  \nonumber \\
 &&  \hspace{-0.98in}   \quad \quad \quad \qquad \qquad 
\,=\,\,\,
    {\rm Diag}\left( {1 \over (1 \,\,  -x-y)^2}
    \cdot \, \sum_{k=0}^{\infty} {\frac{(4\, z\,u)^k}{ (1 \,\,  -x-y)^{2 k} }}  \right). 
\end{eqnarray}
The sum on the index $\,k$ gives:
\begin{eqnarray}
 &&  \hspace{-0.98in}   \quad \quad  \quad \quad  \quad   \quad   \quad   \quad
 {\rm Diag}\left( {\frac{1}{ Q_2\, Q_3 }} \right) \,  \,=\, \,   \,  
{\rm Diag}\left( {\frac{1}{ (1\, \, -x-y)^2 \, \,  \, - 4 \,z\,u }}  \right).   
\end{eqnarray}
As far as the diagonal is concerned, the product $\,z \,u$ stands for {\em only one} variable.
The diagonal given in (\ref{diagab}) for $\,b = \, 2$, which is the series given in (\ref{solL2b2})
is the diagonal of a rational function depending on {\em three} variables, as the power of the logarithm
in the formal solutions of $\,  L_2(b=2) \,\cdot \,  L_1\, $ indicates.

\vskip 0.1cm

{\bf Remark G.1.}
In the factorization of $\, L_6(b)$ for $\,b= \, 2$, given in  (\ref{L6b2}), the
linear differential operator $ \, L_2(b=2) \,\cdot \, L_1 \, $ annihilates the
series (\ref{diagab}) for the value $\,b= \, 2$. Therefore, the
linear differential operator $\,V_3\, $ becomes ``spurious''.
The analytical solution (at the origin) of the differential operator $\, V_3$ reads
\begin{eqnarray}
\label{solV3}
&&  \hspace{-0.98in}   \quad \quad \quad \quad \quad \quad \quad \quad \quad
{\rm sol}(V_3)\,  \,= \, \, \,  \, 
 _3F_2\Bigl([{{2} \over{4}}, \, {{3} \over{4}}, \, {{5} \over{4}}], \, [1, \, 1], \,   256 \, t  \Bigr),  
\end{eqnarray}
to be compared with the solution (\ref{diagabSOL}) of the linear differential operator $\, U_3$.
The linear differential operators  $\,U_3 \, $ and $\,V_3\,$ are {\em actually  homomorphic}
\begin{eqnarray}
&&  \hspace{-0.98in} \quad \quad \quad  \quad \quad  
  (1\, -256\, t) \cdot \, t^2  \cdot \,  V_3 \cdot  \, \Bigl( t\cdot \, D_t \,  \, \,  +{1 \over 4} \Bigr)
  \nonumber \\
 &&  \hspace{-0.98in}   \quad \quad \quad \quad \quad  \quad \quad    \quad \quad  
  \, \,= \, \, \, \,
\Bigl( t \cdot \, D_t\,\,   +{5 \over 4}  \Bigr)  \cdot \, (1\, -256\, t) \cdot \, t^2  \cdot \,  U_3,
\end{eqnarray}
which shows (using the relation\footnote[1]{For three variables we had  equation (\ref{intwQnQ})
  in section \ref{power}.} (\ref{homIntert1}) in \ref{classratfun}, here $\, a \, = \, 4$)   
that (\ref{solV3}) is in fact:
\begin{eqnarray}
 &&  \hspace{-0.98in}   \quad \quad \quad \quad \quad \quad \quad \quad \quad \quad \quad
{\rm sol}(V_3)\,  \, = \, \, \, 
 {\rm Diag}\Bigl( {\frac{1}{ Q_1^2 }} \Bigr).
\end{eqnarray}
At the value of $\,b=\, 2$, even if the linear differential operator $ \, V_3$ is spurious
with respect to the series (\ref{diagab}) for $\,b= \, 2$, 
the linear differential operator $\,L_6(b)$  has kept, for $\,b=\, 2$,
some "memory" of the rational function $ \, 1/Q_1$, through the spurious operator $ \, V_3$.

\vskip 0.1cm

\section{Square root of rational functions versus Calabi-Yau equations}
\label{sqrtR2CY}

 As in sections 7.2 and 7.3, we give, here, two more examples where
the diagonal of square root of a rational function
produces a Calabi-Yau equation.

\subsection{The square root of the rational function of the LGF 4-D simple cubic}

Recall the rational function $\, 1/Q$,  depending on 5 variables,
and corresponding to the lattice Green function of
the 4-dimensional simple cubic lattice:
\begin{eqnarray}
\label{G1_Q1}
&&  \hspace{-0.98in}  \quad \quad\,   \quad \quad
Q_1  \, \,= \,  \, \,\, \, 
   1 \, \, \, \,
   - \Bigl(x + y + z + u \, \,  \,  \,  + v \cdot \, (x y z + x y u + x z u + y z u) \Bigr),
  \\
&&  \hspace{-0.98in} \quad  \quad   \quad \quad \, \, 
{\rm Diag} \Bigl( {{1} \over { Q_1}} \Bigr) \,  \,  \,  \, \, 
   \, = \,  \,\,   \, \,
 1\, \, \,+8\,t\,\, +168\,{t}^{2}\,\, +5120\,{t}^{3}\,\, +190120\,{t}^{4}
  \, \, \,\, \, + \, \, \cdots
\nonumber 
\end{eqnarray}
Considering, now, the diagonal of the {\em square root}
of this rational function (\ref{G1_Q1})
\begin{eqnarray}
\label{Qhalf}
&&  \hspace{-0.98in}  \quad \quad \,   \, \,
  Q_2 \,  \,\, = \,  \, \,
   \Bigl( 1 \,  \,  \,\,
   -  \Bigl(x + y + z + u \, \,  \,  \,  + v \cdot \, (x y z + x y u + x z u + y z u) \Bigr) \Bigr)^{1/2}, 
\\
&&  \hspace{-0.98in} \quad  \quad \, \,  \, \,
{\rm Diag} \Bigl( {{1} \over { Q_2}} \Bigr) \,  \, = \,  \,  \, \, \,
1 \, \, \, \, +3\,t \,\,\, +{\frac {735}{16}}\,{t}^{2} \,\,\, +1155\,{t}^{3}
\,\, \, +{\frac {152927775}{ 4096}}\,{t}^{4}
   \,\, \,\, \, \, + \, \, \cdots
\nonumber 
\end{eqnarray}
one obtains an annihilating irreducible order-five linear differential operator $\,L_5$, with
a differential Galois group included in the orthogonal group $\, SO(5, \, \mathbb{C})$,
and with $ \, \ln(t)^4$ highest log-power formal solution at the origin. According to
conjecture (\ref{conjmain}), a rational function $ \, 1/Q_3$, 
{\em depending on $\, 6= 4 \, +2$ variables} should exist. With the
introduction of an extra variable $\, w$, it actually reads:
\begin{eqnarray}
&&  \hspace{-0.98in}  \quad  \quad   \quad  \quad  \quad  \quad    \quad  \quad    \quad    \quad  \quad    \quad  
  Q_3  \, \, \, = \, \,  \,\,\, Q_1 \, \,\, \,  - {w^{1/2} \over 4}.
\end{eqnarray}
The diagonal of the {\em algebraic} function $\, 1/Q_2\, $  identifies with the diagonal
of the {\em rational} function $\, 1/Q_3$.

The linear differential operator $\,L_5$ has MUM, and is actually the exterior square of
an order-four linear differential operator, 
$  L_5  \, = \,  {\rm ext}^2 \left( L_4 \right)$. The linear differential operator $\, L_4$
also has MUM, at $ \, t = \, 0$, the indicial exponents being four
times $ \, 1/2$.
Let us introduce the linear differential operator $\,N_4$:
\begin{eqnarray}
 &&  \hspace{-0.98in}  \quad  \quad   \quad  \quad  \quad  \quad    \quad 
    N_4 \, \, \, = \, \,\,   \,
    L_4 \cdot  \, \sqrt{t} \cdot \left(1 \, -16\,t \right)^{1/4} \cdot \,  \left(1 \, -64\,t \right)^{1/4}. 
\end{eqnarray}
With the scaling $\, t \rightarrow  \, 16\,t$, the linear differential operator
$\,N_4$ reads (with $\theta = \,  t\, D_t$):
\begin{eqnarray}
 &&  \hspace{-0.98in}   \qquad 
N_4 \, \, = \, \, \,
{\theta}^{4} \, \, \,
  -4  \cdot \, t   \cdot \,  \left( 960\,{\theta}^{4} +640 \,{\theta}^{3} +574\,{\theta}^{2}+254\,\theta +41 \right)
  \nonumber  \\
 &&  \hspace{-0.98in}   \qquad \quad \quad
  +16  \cdot \, t^2  \cdot \, \left( 356352\,{\theta}^{4}+451328\,{\theta}^{2}
  +475136\,{\theta}^{3}+199424\,\theta+34257 \right)
  \nonumber \\
 &&  \hspace{-0.98in}   \qquad \quad \quad
  -2^{12} \cdot \, t^3  \cdot \, \left( 1003520\,{\theta}^{4}+2007040\,{\theta}^{3}
  +2098048\,{\theta}^{2}+1043328\,\theta+198453 \right)
  \nonumber \\
 &&  \hspace{-0.98in}   \qquad \quad \quad
  +2^{22} \cdot \, t^4  \cdot \, \left( 356352\,{\theta}^{4}+950272\,{\theta}^{3}
  +1126912\,{\theta}^{2}+618752\,\theta+124913 \right)
  \nonumber  \\
 &&  \hspace{-0.98in}   \qquad \quad \quad
  -2^{37} \cdot \, t^5   \cdot \, \left( 2\,\theta+1 \right)  \cdot \,
  \left( 960\,{\theta}^{3}+2720\,{\theta}^{2}+2854\,\theta+987 \right)
  \nonumber \\
 &&  \hspace{-0.98in}   \qquad \quad \quad
    +2^{48}\cdot \,t^6  \cdot \,  \left( 2\,\theta+1 \right)  \cdot \,   \left( 2\,\theta+3 \right)  \cdot \,
    \left( 4\,\theta+3 \right) \cdot \,  \left( 4\,\theta+5 \right). 
\end{eqnarray}

\vskip 0.1cm

To be of Calabi-Yau type, the linear differential operator $ \, N_4\, $ must satisfy
some conditions~\cite{Almk-Zud-2006,Enck-Strat-2006}.
The differential Galois group of $N_4$ is included in $ \, Sp(4, \, \mathbb{C} )$, the
linear differential operator $ \, N_4$ has MUM, and at the infinity, the indicial 
exponents $\, 1/2, \, 3/4, \, 5/4, \, 3/2$ are such that
$\, 1/2 \, +3/2 = \, 3/4 \, +5/4 \, = \, 2 \, $
is a rational.
Morever, the coefficients of the power series given  below in (\ref{theS0}), and the
instanton numbers given below in (\ref{instan}), should be integers.

\vskip 0.1cm

The formal solutions, at the origin, of the linear differential operator $\,N_4$ are
\begin{eqnarray}
\label{theS0}
S_0 &=& \, \, 1 \, \, \, +164\,t \,\,\,   +66972\,{t}^{2} \,\,  \, +38050160\,{t}^{3}
\,\, \, \,   + \, \, \cdots
 \nonumber  \\
  S_1 &=& \,\, \, S_0 \cdot \, \ln(t) \,\,  + S_{1,0} 
  \\
  S_2 &=&\, \, {1 \over 2}\cdot \,S_0 \cdot \, \ln(t)^2 \,\,  +S_{1,0}\cdot \, \ln(t) \, \,  + S_{2,0} 
 \nonumber \\
  S_3 &=& \,\, {1 \over 6} \cdot \,S_0 \cdot \, \ln(t)^3\,\,\,
 +{1 \over 2} \cdot\, S_{1,0} \cdot \, \ln(t)^2 \,\, + S_{2,0} \cdot \, \ln(t)\, \,\,   + S_{3,0}
  \nonumber 
\end{eqnarray}
with:
\begin{eqnarray}
  S_{1,0} \,\, = \, \,\,\, 
  360\,t \,\,  \,+182484 \,{t}^{2} \,\,\,  +111758064\,{t}^{3}
 \,  \, \, \,\, + \, \,\cdots
  \nonumber \\
  S_{2,0} \, \,= \,\,\,\, 
  -128\,t \,\,\,  -12768\,{t}^{2} \,\, \, +{\frac {36154528}{9}}\,{t}^{3}
  \,\,  \,  \, + \,\,\,  \cdots
  \nonumber \\
  S_{3,0} \,\, = \, \,\,\, 
  256\,t \,\,\,  +98560\,{t}^{2} \,\, \,  +{\frac {1479864064}{27}}\,{t}^{3}
  \,\,\,\,  \, + \,\,\,\,  \cdots
  \nonumber 
\end{eqnarray}

Let $\, z = \, S_1/S_0$, the nome $\,q$, defined as $\, q = \, \exp (z)$, has the expansion
\begin{eqnarray}
\label{nome1}
q  \,\,= \, \,\, \,\, t \,\,\, \, +360\,{t}^{2} \,\, \, +188244\,{t}^{3} \,\,\,  +119619168\,{t}^{4}
  \,\,\, \,   \, + \, \,\, \cdots
\end{eqnarray}
and the mirror map reads:
\begin{eqnarray}
\label{mirror1}
  t  \,\, = \, \,  \,\, q \,\, \, \, -360\,{q}^{2} \,\,\,  +70956\,{q}^{3} \,\,\,   -14059968\,{q}^{4}
 \,  \,  \, \,\, + \, \,\,  \, \cdots
\end{eqnarray}
The Yukawa coupling defined by
\begin{eqnarray}
&&  \hspace{-0.98in}  \, \quad \quad \quad \quad \quad \quad \quad \quad \quad \quad \quad 
K(q)\, \, = \,\, \,{\frac{d^2}{dz^2}} \left( {\frac{S_2}{S_0}} \right), 
\end{eqnarray}
reads:
\begin{eqnarray}
\label{Yuk1}
&&  \hspace{-0.98in}  \, \,  \quad
K(q)  \,\, = \,  \, \,
  1 \, \, -128 \,q \, \,-41984 \,{q}^{2} \,\,  -13919744 \,{q}^{3} \,\,  -4141162496 \,{q}^{4}
  \, \,\,  + \, \, \, \cdots
\end{eqnarray}
The Yukawa coupling, expanded in a Lambert series, reads
\begin{eqnarray}
  &&  \hspace{-0.98in}  \, \quad  \quad  \quad \quad  \quad \quad \quad \quad \quad 
     K(q) \, \, = \,\,  \,\, \,
     1 \,\,\, +\, \sum_{j=1}^\infty \, n_j \, \, {\frac{j^3\, q^j}{1-q^j}}, 
\end{eqnarray}
and gives the "instantons numbers" $ \, n_j,  \,\, j= \, 1, \,2, \,\,\, \cdots$
\begin{eqnarray}
\label{instan}
 &&  \hspace{-0.98in}  \, 
-128, \, -5232, \, -{\frac {1546624}{3}},\, -64705008, \, -7960717440, \, -1089730087792, \, \, \, \cdots
\end{eqnarray}
where, with $ \, n_0 = \, 3$, the numbers $ \, n_0\, n_j$ {\em are actually integers}.

\vskip 0.1cm

The linear differential operator $\, N_4\, $ satifies the Calabi-Yau type conditions,
and especially the series given in
(\ref{theS0}), (\ref{nome1}), (\ref{mirror1}) and (\ref{Yuk1}) have {\em integer} coefficients.

\vskip .1cm 

\subsection{The square root of the rational function of the LGF 4-D body centred cubic}
\label{LFG4Dbody}

Let us consider the rational function $\, 1/Q_1$, where the polynomial denominator $\, Q_1\, $ reads:
\begin{eqnarray}
  \label{Q1def}
&&  \hspace{-0.98in}   \qquad \qquad \quad \quad \quad
 Q_1 \, = \, \,  \,\,
 1  \, \,\, -\left(x +z \right) \cdot \, \left(1 +y \right) \cdot\, \left(1 +u \right) \cdot \, \left(1 +v \right). 
\end{eqnarray}
This is the multivariate polynomial corresponding to the Calabi-Yau number 3 (see Table \ref{Ta:1}).
The diagonal of $\, 1/Q_1 \, $ also identifies with the LGF
of the 4-D body centred cubic lattice~\cite{guttmann-2009}.

Considering, the diagonal of the reciprocal of the {\em square root} of the polynomial function (\ref{Q1def}),
  one obtains:
\begin{eqnarray}
&&  \hspace{-0.98in}  \quad  \quad \quad \quad \, \, 
{\rm Diag}\left( {1 \over  \sqrt{Q_1} } \right)  \, \, = \,\, \, \, \,
{}_5F_4\left( [{1 \over 2}, {1 \over 2}, {1 \over 2}, {1 \over 4}, {3 \over 4}],
\, [1, 1, 1,1], \,  \,  256 \,t \right)
\nonumber \\
 &&  \hspace{-0.98in}  \qquad  \quad \qquad \quad \quad \quad 
=  \, \,   \, \, \,
    1 \, \,  \,  \,+6 \,t \,\,  \, +{\frac {2835}{8}}\,{t}^{2} \, \,  \,+{\frac {144375}{4}}\,{t}^{3}
  \,  \, \,  \,\, + \,  \, \cdots
\end{eqnarray}
This series is annihilated by an irreducible order-five linear differential operator $\, L_5$, which has
a differential Galois group included in the orthogonal group $\, SO(5, \,  \mathbb{C})$.
The operator $\, L_5\, $ has a $ \, \ln(t)^4\, $ highest log-power formal solution
at the origin. According to
conjecture (\ref{conjmain}), a rational function $ \, 1/Q_w$, 
{\em depending on $\, 6 \, = \,  4 \, +2 \, $ variables}, should exist. It actually reads
(with an extra variable $\, w$):
\begin{eqnarray}
&&  \hspace{-0.98in}  \quad  \quad   \quad \quad \quad \quad  \quad  \quad  \quad   \quad  \quad    \quad  
  Q_w  \, \, = \,  \,\,\,\, Q_1 \, \, \,   - {w^{1/2}. \over 4}.
\end{eqnarray}
The diagonal of $\, 1/\sqrt{Q_1} $ actually identifies with the diagonal of $\, 1/Q_w$.

The order-five linear differential operator $\,L_5\, $ has MUM, and is actually the exterior square
of an order-four linear differential operator, 
$ \, L_5  \, = \, \,  {\rm ext}^2 \left( L_4 \right)$. The linear differential operator $\, L_4$
has MUM, at $\, t= \, 0$, the indicial exponents being four
times $ \, 1/2$.

Let us introduce the linear differential operator $\,N_4$:
\begin{eqnarray}
&&  \hspace{-0.98in}  \quad  \quad  \quad \quad  \quad \quad  \quad  \quad  \quad   \quad  
 N_4 \,\, = \,\,\,\,
 L_4 \cdot \, \sqrt{t} \cdot \, \left(1\, -256\,t \right)^{1/4}. 
\end{eqnarray}
With the scaling $\, t \,  \rightarrow  \, 16\,t$, the linear differential operator
$\,N_4$ reads (with $ \, \theta = \,  t\,D_t$):
\begin{eqnarray}
&&  \hspace{-0.98in}  \qquad  \, \, \quad \quad
N_4  \, \,=\, \,  \, 
{\theta}^{4} \,  \, 
-16 \cdot  \,t \cdot \,
\left( 758 \,{\theta}^{4} +512\,{\theta}^{3}+440\,{\theta}^{2}+184\,\theta\, +25 \right)
  \nonumber  \\
&&  \hspace{-0.98in}  \qquad \quad \quad \quad \quad \quad
  +2^{12} \cdot \, t^2 \cdot   \, \left( 12288\,{\theta}^{4}+16384\,{\theta}^{2}
  \, +13056\,{\theta}^{3} \, +3840\,\theta \, +401 \right)
  \nonumber \\
&&  \hspace{-0.98in}  \qquad \quad\quad  \quad \quad \quad
  -2^{24} \cdot \, t^3 \cdot \, \left( 8\,{\theta}+3 \right)^2
  \cdot \,  \left( 8\,{\theta}+5 \right)^2. 
\end{eqnarray}

\vskip 0.1cm

Similarly to the previous example, one obtains, from the formal solutions at the origin of $\,N_4$,
 the nome
\begin{eqnarray}
  \label{nome2}
&&  \hspace{-0.98in}  \quad \quad\,  \, \, \, 
   q  \, \,= \,  \, \, \exp (z) \, = \,\,  \,\,
t \,\, \, +1344\,{t}^{2} \,\, \, +2906016\,{t}^{3} \,\,  +7605501952\,{t}^{4}
 \,  \, \, + \, \, \cdots
\end{eqnarray}
 the mirror map
\begin{eqnarray}
  \label{mirror2}
  &&  \hspace{-0.98in} \quad \quad \quad \quad \,  \, \, \, 
  t \,  \, = \,  \, \, \,
  q \,\, \,\,-1344 \,{q}^{2} \,\,\, +706656 \,{q}^{3} \,\,\,  -215652352\,{q}^{4}
  \, \,\,\,  + \, \, \,  \cdots
\end{eqnarray}
and the Yukawa coupling
\begin{eqnarray}
\label{Yuk2}
 &&  \hspace{-0.98in}  
K(q)  \, \, = \, \,  \, \, 
  1\, \, \, -736\,q\, -836832\,{q}^{2}\, -726599680\,{q}^{3}\, -579941553376\,{q}^{4}
  \, \,\,   +\,  \,  \cdots
\end{eqnarray}
giving the instantons numbers, $n_1$, $n_2$, $\cdots \,$ {\em as integers}:
\begin{eqnarray}
 &&  \hspace{-0.98in}  \qquad \quad
  -736, \,\,\,\, -104512,\,\,\,\, -26911072,\, \,\,\,-9061573696,\,\,\,\, -3547993303456,\,\,
  \nonumber \\
 &&  \hspace{-0.98in}  \qquad  \qquad  \quad  \quad \quad \quad \quad
  -1530399290794816, \, \, \,\, \, \, \,\, \,   \,  \cdots
\end{eqnarray}

\vskip .1cm

\section{A Class of rational functions}
\label{classratfun}

The expansion of the rational function $\, 1/Q$, where the polynomial $\, Q\, $ is given by
\begin{eqnarray}
&&  \hspace{-0.98in}  \quad  \quad  \quad \quad \quad  \quad   \quad   \quad   \quad   
Q  \, \, = \, \, \, \, \,
1 \, \, \, \, - \left( T_1 \, + T_2 \, + \cdots \, +T_n \right), 
\end{eqnarray}
where $ \, T_j$ are monomials, can be expanded in a multinomial sum:
\begin{eqnarray}
 &&  \hspace{-0.98in}  \quad  \quad  \quad \quad  \quad \quad    \, \, 
 {{1} \over {Q}} \,  \, = \, \,\,\,
 \sum \,\,   {\frac{(k_1 +k_2 +\cdots \, + k_n)!}{k_1! \,  k_2! \, \cdots \, k_n!}}
\cdot  \, T_1^{k_1} \,T_2^{k_2} \,\cdots \, T_n^{k_n}. 
\end{eqnarray}
The expansion of $ \, 1/Q^k$ reads
\begin{eqnarray}
&&  \hspace{-0.98in}  \quad  \quad  \quad \quad    \, \, 
  {{1} \over {Q^k}} \,  \, = \,\, \,\, \,
  \sum \,\, {\frac{\left(k\right)_{k_0}}{k_0!}} \cdot \,
   {\frac{(k_1\, +k_2\, +\cdots\,  +k_n)!}{k_1!\,  k_2! \, \cdots\,  k_n!}}
   \cdot \, T_1^{k_1} \,T_2^{k_2} \,\cdots \, T_n^{k_n}, 
\end{eqnarray}
where $ \, \left(k\right)_{k_0}$ is the Pochhammer symbol, and where
$ \,\,  k_0\,  =\, \, k_1\, +k_2\,\,  +\,  \cdots \,  +k_n$.

The expansion can be written:
\begin{eqnarray}
&&  \hspace{-0.98in}  \quad  \quad  \,  \quad  
{{1} \over {Q^k}}  \,  \, = \,\,\,\, 
 \sum \,\, {1 \over (k-1)!}\,
 {\frac{(k_1\, +k_2 \, +\cdots\, + k_n\, +k-1)!}{k_1! \,  k_2! \,\cdots\,  k_n!}}
   \cdot \, T_1^{k_1} \, T_2^{k_2} \,\cdots \,  T_n^{k_n}. 
\end{eqnarray}
Let us fix $ \, k= \, 2$, the expansion of $ \, 1/Q^2\, $ reads
\begin{eqnarray}
 &&  \hspace{-0.98in}    \quad   \quad \quad  \quad   \, \, \,\, \,  \,   
{{1} \over {Q^2}}  \,  \, = \,\,\,\,  \,   
\sum \,\,   {\frac{(k_1\, +k_2\, +\cdots \, + k_n\, +1)!}{k_1! \, k_2! \, \cdots \, k_n!}}
\cdot \, T_1^{k_1} \, T_2^{k_2} \,\cdots \, T_n^{k_n}, 
\end{eqnarray}
or
\begin{eqnarray}
&&  \hspace{-0.98in}   \quad      
   {{1} \over {Q^2}}  \,  \, =\, \,
   \\
&&  \hspace{-0.98in}   \, \,   \, \,   \quad        \quad      
   \,  \, =\, \,  \,
 \sum \,\,  (k_1\, +k_2\, +\, \cdots \, + k_n\, +1) \cdot \,
   {\frac{(k_1\, +k_2\, +\cdots\, + k_n)!}{k_1! \,  k_2! \,  \cdots \,  k_n!}}
   \cdot \, T_1^{k_1} \, T_2^{k_2} \,\cdots \,  T_n^{k_n},
   \nonumber 
\end{eqnarray}
to be compared with the expansion of $\, 1/Q$.

The diagonal of the rational function amounts to distributing each exponent $ \, k_j$
on the variables occurring in the monomials $ \, T_j$,
and constraining the resulting exponent on each variable to the
running index $ \, p$ of the sum. This gives the following system of linear equations
\begin{eqnarray}
  \quad  \quad 
  \alpha_1 \,k_1 \,+ \alpha_2 \, k_2 \,\, +\, \cdots
  \,\, = \,\, \, p, 
  \nonumber \\
  \quad  \quad  
  \beta_1 \,k_1\, + \beta_2 \, k_2 \,\, +\, \cdots
  \,\, = \,\, \, p, 
  \nonumber \\
   \quad  \quad 
\vdots
\end{eqnarray}
where $\,p$ (or $\,2p$, ...) is the running index of the series, and where
$ \, (\alpha_j,\,  \beta_j, \, \cdots )$
are the exponents of the variables $\, (x,\, y,\, \cdots)$
occurring in the monomial $\,T_j$. There are as many equations as variables,
and (in general) less than the number of monomials.

\vskip 0.1cm

Solving the system of equations fixes some $ \, (k_j)$ in terms of $p$,
and some other non fixed indices which are labelled as the set  $ \, \{k_r\}$. The
first factor in the sum of the expansion of $ \, 1/Q^2 \, $ becomes
\begin{eqnarray}
\label{sumkj}
  (k_1\, +k_2\, +\, \cdots \, + k_n \, +1)
  \, \,  \, = \, \,\,\,\,
       a \, p \,\, \, + g(\{k_r\}) \,\,\, +1, 
\end{eqnarray}
where $ \, a\, $ is a rational number, and where $ \, g(\{k_r\})$ is a notation for
the expression containing the remaining non fixed indices $ \, \{k_r\}$.
The diagonal of $ \, 1/Q\,$ and $ \, 1/Q^2\, $ read respectively: 
\begin{eqnarray}
&&  \hspace{-0.98in}  \quad   \quad   \quad \quad \quad \quad\quad
 {\rm Diag} \Bigl( {{1} \over {Q}} \Bigr)\,  \, = \, \, \, \,\,
  \sum_{p=0}^{\infty} \, C(p, \{k_r\}) \cdot  \, (x y \cdots )^p, 
\end{eqnarray}
\begin{eqnarray}
&&  \hspace{-0.98in}  \quad    \quad  \quad \quad
   {\rm Diag} \Bigl( {{1} \over {Q^2}} \Bigr)
\, \, = \,\, \,\,
\sum_{p=0}^{\infty}  \, \Bigl( a\, \, p \,\, + g(\{k_r\}) \, + 1 \Bigr)
\cdot  \,C(p, \{k_r\}) \, (x y \cdots )^p.   
\end{eqnarray}
For $\,g(\{k_r\}) = \, 0$, one obtains the relation (with $\, t = \, x y \cdots $):
\begin{eqnarray}
\label{homIntert1}
&&  \hspace{-0.98in}  \quad  \quad  \quad   \quad \quad \quad  \quad  \quad
  {\rm Diag}  \Bigl( {{1} \over {Q^2}} \Bigr)
   \,  \, = \, \, \, \Bigl( a \cdot \, t \cdot \, D_t \,\, + 1 \Bigr)
     \cdot  \,   {\rm Diag}\Bigl( {{1} \over {Q}} \Bigr).
\end{eqnarray}

\vskip 0.1cm

The condition $\, g(\{k_r\})=\, 0$ happens {\em when there are as many variables as monomials}.
Consider the case when the number of monomials exceeds
the number of variables (e.g. 3 variables and 4 monomials).
The solving on the indices $ \, (k_1, \,k_2, \, k_3, \,k_4)$, involves the
$ \, 3 \times \, 4 \,$ matrix $ \, A$:
\begin{eqnarray}
  \quad \quad \quad \quad 
  \alpha_1 \quad \alpha_2 \quad \alpha_3 \quad \alpha_4
  \nonumber \\
   \quad \quad \quad \quad 
  \beta_1 \quad \beta_2 \quad \beta_3 \quad \beta_4
  \nonumber \\
   \quad \quad \quad \quad 
  \gamma_1 \quad \gamma_2 \quad \gamma_3 \quad \gamma_4
  \nonumber 
\end{eqnarray}
Call $ \, A^{(j)}$ the $ \, 3 \times \, 3$ matrix obtained by discarding the
$j$-th column of the matrix $A$.
The indices $ \, (k_1, k_2, k_3)$ are obtained in terms of $\, k_4$ and of the running index
$ \, p$, with $ \, {\rm det}(A^{(4)}) \ne \, 0$. 
The sum (\ref{sumkj}) becomes:
\begin{eqnarray}
&&  \hspace{-0.98in}  \quad \,\, \,
  (k_1 \, + k_2 \, +k_3 \, +k_4 \, +1)
\, \,\, = \,\, \, \,\,
\,   a \, p \, \, \,  \,\,
  + k_4 \cdot \, \sum_{j-1}^4  \, (-1)^{j} \cdot \, {\frac{{\rm det} (A^{(j)})}{{\rm det} (A^{(4)})}}
\, \,\, \,  \,   +1. 
\end{eqnarray}
The condition for the homomorphism (\ref{homIntert1}) with order-one intertwinner to occur reads:
\begin{eqnarray}
&&  \hspace{-0.98in}  \quad    \quad  \quad  \quad  \quad \qquad \quad \quad \quad   \, 
  \sum_j  \,  (-1)^{j} \cdot \, {\rm det}(A^{(j)})
  \, \, \, = \, \, \, \, 0. 
\end{eqnarray}

\vskip 0.1cm

{\bf Remark I.1.}
In the case of three variables and five monomials, the sum corresponding to (\ref{sumkj}) reads
\begin{eqnarray}
&&  \hspace{-0.98in}  \quad \quad 
  (k_1 \, +k_2\,+k_3\,+k_4\, +k_5\, \, +1)
  \\
&&  \hspace{-0.98in}  \quad   \quad \, \, \,  \,  \,  \,
   = \,\,\,   \, \,
  a\, p \, \, \,\, + 1 \, \, \, \, 
+k_4  \cdot \, \sum_{j \ne 5}^5  \, (-1)^{j} \cdot \, {\frac{{\rm det} (A^{(j,5)})}{{\rm det} (A^{(4,5)})}} \,
  \, \,  \,
+ k_5 \cdot \, \sum_{j \ne 4}^5  \, (-1)^{j} \cdot \, {\frac{{\rm det} (A^{(j,4)})}{{\rm det} (A^{(4,5)})}}, 
  \nonumber
\end{eqnarray}
where $ \, A^{(j, i)}$ is the $ \, 3 \times \, 3$ matrix obtained by discarding the $j$-th
and the $i$-th column of the matrix $ \, A$, 
with $ \, {\rm det}(A^{(4,5)}) \ne \,  0$. This gives two conditions:
\begin{eqnarray}
\fl \quad \quad
  \sum_{j \ne 4}  \,  (-1)^{j} \cdot \, {\rm det}(A^{(j,4)}) \,= \, \, 0,
  \quad \quad {\rm and} \quad \quad \, \, \, 
\sum_{j \ne 5}  \,  (-1)^{j} \cdot \, {\rm det}(A^{(j,5)}) \,= \,\, 0.
\end{eqnarray}

\vskip 0.1cm

{\bf Remark I.2.}
These conditions occur for the rational functions $ \, 1/Q$
with polynomial $\, Q\, $ of the Calabi-Yau equations in Table 1. We checked, for some periods
of rational integrals given in~\cite{lairez-2014},
that the simple intertwinner $ \,\, \, a \cdot \, t \, D_t \, \, +1 \,\, $ occurs.

\end{document}